\documentclass[%
 aip,
 amsmath,amssymb,
 reprint,%
]{revtex4-1}

\usepackage{mathrsfs}
\usepackage{graphicx}
\usepackage{dcolumn}
\usepackage{bm}

\usepackage[utf8]{inputenc}
\usepackage[T1]{fontenc}
\usepackage{mathptmx}
\usepackage{etoolbox}
\usepackage{adjustbox}

\usepackage{color}

\usepackage{ulem} 

\makeatletter
\def\@email#1#2{%
 \endgroup
 \patchcmd{\titleblock@produce}
  {\frontmatter@RRAPformat}
  {\frontmatter@RRAPformat{\produce@RRAP{*#1\href{mailto:#2}{#2}}}\frontmatter@RRAPformat}
  {}{}
}%
\makeatother
\begin{document}

\preprint{AIP/123-QED}

\title[]{Rotationally Equivariant Super-Resolution of Velocity Fields in Two-Dimensional Fluids Using Convolutional Neural Networks}
\author{Yuki Yasuda}
 \altaffiliation[]{Global Scientific Information and Computing Center, Tokyo Institute of Technology,\\2-12-1 Ookayama, Meguro-ku, Tokyo 1528550, Japan}
 \email{yasuda.y.aa@m.titech.ac.jp}

\author{Ryo Onishi}
 \altaffiliation[]{Global Scientific Information and Computing Center, Tokyo Institute of Technology,\\2-12-1 Ookayama, Meguro-ku, Tokyo 1528550, Japan}
 \email{onishi.ryo@gsic.titech.ac.jp}

\date{\today}

\begin{abstract}
This paper investigates the super-resolution (SR) of velocity fields in two-dimensional fluids from the viewpoint of rotational equivariance. SR refers to techniques that estimate high-resolution images from those in low resolution and has lately been applied in fluid mechanics. The rotational equivariance of SR models is defined as the property in which the super-resolved velocity field is rotated according to a rotation of the input, which leads to the inference covariant to the orientation of fluid systems. Generally, the covariance in physics is related to symmetries. To clarify a relationship to symmetries, the rotational consistency of datasets for SR is newly introduced as the invariance of pairs of low- and high-resolution velocity fields with respect to rotation. This consistency is sufficient and necessary for SR models to acquire rotational equivariance from large datasets with supervised learning. Such a large dataset is not required when rotational equivariance is imposed on SR models through weight sharing of convolution kernels as prior knowledge. Even if a fluid system has rotational symmetry, this symmetry may not carry over to a velocity dataset, which is not rotationally consistent. This inconsistency can occur when the rotation does not commute with the generation of low-resolution velocity fields. These theoretical suggestions are supported by the results from numerical experiments, where two existing convolutional neural networks (CNNs) are converted into rotationally equivariant CNNs and the inferences of the four CNNs are compared after the supervised training.
\end{abstract}

\maketitle

\section{\label{sec:Introduction}Introduction}

In recent years, neural networks (NNs) have been actively applied to the field of fluid mechanics.\cite{Brunton2020, Duraisamy2021, Vinuesa2021, Brunton2022} In these applications, the physical validity of NNs in inference is important in both theory and practice. One method of enhancing the validity involves explicitly incorporating physics laws such as the Navier--Stokes equations into NNs. These networks are referred to as physics-informed neural networks\cite{Raissi2019} and have been extensively studied recently.\cite{Kashinath2021, Cai2022} Most physics laws are described with vectors, which are often processed as tuples of scalars such as colors in images.

The difference between scalar and vector is mathematically formulated via the covariance in geometry.\cite{Schutz1980, Nakahara2003} The covariance describes the change in the components of a tensor (e.g., scalar or vector) under certain coordinate transformations. In other words, by changing the direction of observation, it is possible to determine whether a numerical array is a vector having a direction or a tuple of scalars having no direction. These transformation laws can be regarded as a part of the definition of scalars and vectors.\cite{Dirac1996} 

By incorporating both covariance and physics law into NNs, they can be applied to data on various coordinate systems without compromising accuracy and physical validity. The covariance ensures that scalars and vectors are geometrically invariant and the forms of physics equations are invariant with respect to coordinate systems. If the physical validity is measured by the residual of a covariant physics equation, the value of this residual is independent of certain coordinate systems. Therefore, the residual is sufficiently small in any of these systems once it becomes nearly zero in training.

Super-resolution (SR) has not been well investigated from a geometric point of view. SR refers to methods of estimating high-resolution images from those with low resolution and has been studied in computer vision as an application of NNs.\cite{Dong2014, Ledig2017SRGAN, Wang2018ESRGAN, Ha2019, Anwar2020} The success of such NNs has resulted in an increased number of studies that focus on fluid-related SR: idealized turbulent flows in two\cite{Deng2019, Fukami2019, Maulik2020, Wang2020, Fukami2020, Fukami2021, Wang2021DSCMS} and three dimensions,\cite{Fukami2020, Fukami2021, Liu2020, Bode2021, Kim2021, Bao2022physics} Rayleigh–B\'{e}nard convection,\cite{Jiang2020} smoke motions in turbulent flows,\cite{Xie2018, Werhahn2019, Bai2020} flows in blood vessels,\cite{Ferdian2020, Sun2020, Gao2021} sea surface temperature,\cite{Ducournau2016, Maulik2020} and atmospheric flows.\cite{Cannon2011, Vandal2017, Rodrigues2018, Onishi2019, Leinonen2020, Stengel2020, Wang2021GMD, Wu2021, Yasuda2022} Physics laws such as the continuity equation can be taken into account as prior knowledge, rendering super-resolved flows more accurate and physically valid.\cite{Wang2020, Bode2021, Gao2021, Bao2022physics} However, studies investigating SR models in terms of geometry are scarce, and the invariance of physics equations has yet to be fully exploited.

\begin{figure}[htbp]
    \includegraphics[width=8.5cm]{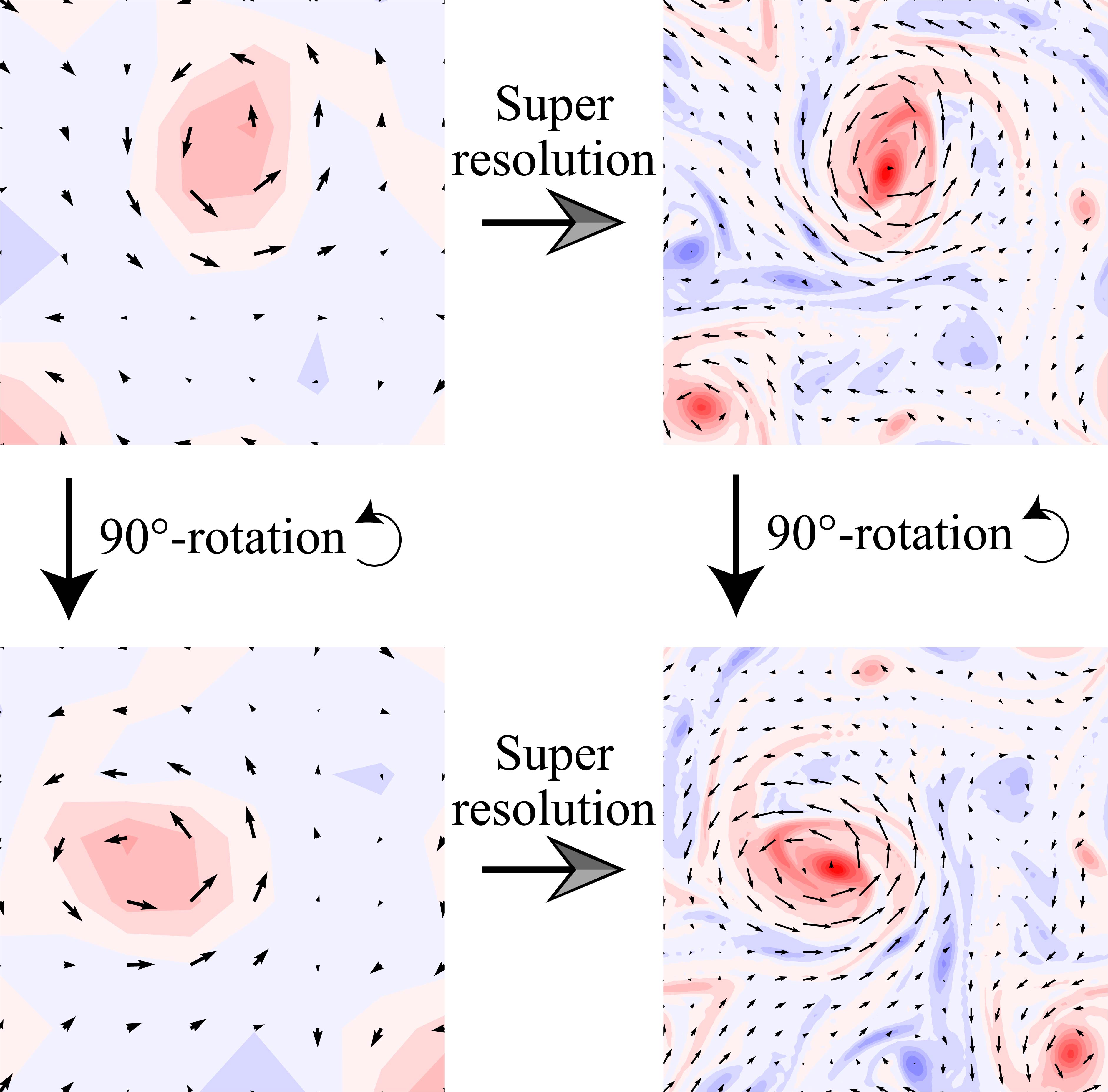}
    \caption{\label{fig:schematic-equivariance} Schematic for the equivariance of super-resolution. If a super-resolution model is equivariant to rotation, the rotation is commutative with the super-resolution. The colors represent a scalar field, and the arrows represent a vector field.}
\end{figure}

The covariance of super-resolved fields is satisfied by imposing the equivariance on SR models. The equivariance of a function is generally defined as the property in which the output is transformed according to a transformation of the input.\cite{Weiler2019, Bronstein2021} For instance, if an SR model is equivariant to rotation, rotating the input results in the output being rotated in the same manner (Fig. \ref{fig:schematic-equivariance}). Therefore, if the input is a vector field, the output satisfies the same transformation law, assuring that the output is a geometric vector field and the SR model applies to input in any orientation.

The equivariance is related to the symmetry of fluid systems. Symmetry in physics is described with Lagrangian.\cite{Landau1975, Salmon1998} If a Lagrangian is invariant under certain coordinate transformations, the system is said to be symmetric and the governing equations are unchanged. In deep learning, the term ``symmetry'' is used in a broader sense:\cite{Bronstein2021} if a certain property of a system or object is unchanged under a transformation, the system or object is said to have symmetry. For instance, if an SR model is equivariant to rotation (Fig. \ref{fig:schematic-equivariance}), the output preserves the covariance of vectors. Thus, in the broader sense, this SR model has rotational symmetry. The terms of covariance, equivariance, and symmetry are clarified in Section \ref{sec:covariance-equivariance-consistency}.

Various equivariant convolutional neural networks (CNNs) have been developed in recent years. Originally, the CNNs\cite{LeCun1989, Krizhevsky2012} are equivariant to translation. For instance, when an object in an image is translated, the output feature is translated in the same manner. CNNs with other types of equivariance (e.g., rotation) have been studied.\cite{Cohen2016, Cohen2017, Marcos2017, Worrall2017, Zhou2017, Schutt2017, Bekkers2018, Sosnovik2021} These CNNs are understood within unified frameworks describing the equivariant CNNs.\cite{Kondor2018, Cohen2019, Weiler2019} The theories and techniques concerning equivariant CNNs have been developed not only for two-dimensional images but also for graph data\cite{Fuchs2020} and three-dimensional volumetric\cite{Weiler2018} and point cloud data.\cite{Thomas2018}  These studies are considered a part of geometric deep learning.\cite{Bronstein2021, Atz2021}

In computer vision, equivariance has occasionally been utilized in SR. The equivariance can be incorporated into loss functions\cite{Chen2021a, Chen2021b} or discriminators of generative adversarial networks (GANs).\cite{Lee2021} In both methods, an image generator acquires the equivariance from training data. Few studies directly impose the equivariance on image generators. Xie et al.\cite{Xie2020} developed an SR model with a self-attention module. In their module, the point-wise convolution is utilized to construct key, query, and value, resulting in equivariance to any pixel permutation. Although their module is equivariant to various transformations, it cannot be applied to vector fields.

In fluid mechanics, several studies have employed equivariant NNs for the modeling of Reynolds stress tensors,\cite{Ling2016, Zhou2022, Han2022, Pawar2022} the surrogate modeling of Navier--Stokes equations,\cite{Wang2021, Suk2021} the estimation of flow over multiple particles,\cite{Siddani2021} and the estimation of higher-rank tensors of fluid stresses.\cite{Gao2020RotEqNet} Those studies considered the physics symmetries using equivariant NNs and reported that the imposed equivariance enhances the accuracy or robustness of NNs in inference. To the authors' knowledge, however, the effect of including the equivariance has not been examined in terms of the fluid SR.

This study aims to investigate the super-resolution of velocity fields in terms of rotational equivariance. To highlight the essence, idealized two-dimensional flows are super-resolved. We clarify relationships among the covariance, equivariance, and symmetry from a theoretical point of view. To support the obtained theoretical suggestions, numerical experiments were conducted. Two existing CNNs\cite{Fukami2019, Bode2021} for fluid SR were converted into rotationally equivariant CNNs by utilizing a theory of equivariant deep learning.\cite{Weiler2019} The inferences of the CNNs were compared after the supervised training.

The remaining paper is organized as follows. Section \ref{sec:covariance-equivariance-consistency} provides the theoretical discussion. Section \ref{subsec:equivariant-neural-networks} describes all CNNs used in this study. Section \ref{sec:methodology} describes the methods of training and evaluating the CNNs. Section \ref{sec:results} analyzes the results. Section \ref{sec:conclusions} presents the conclusions. The source code used in this study is available on the GitHub repository.\cite{Yasuda2022GH}

\section{\label{sec:covariance-equivariance-consistency} Covariance, equivariance, and symmetry in super-resolution}

\subsection{\label{subsec:covariance} Rotational covariance of scalars and vectors}

The two-dimensional Euclidean space is considered with the Cartesian coordinate system: $\bm{x} = (x, y)^{\rm T} \in \mathbb{R}^2$. Contravariant and covariant vectors are identified because of the orthonormal basis of the Cartesian coordinates. There are two types of transformations in the tensor analysis: active and passive transforms. In the former, scalars and vectors are transformed, whereas the coordinates remain fixed. The latter adopts the opposite treatment. The active transform is employed following the previous studies on equivariant CNNs.\cite{Cohen2019, Weiler2019}

The rotation is first formulated for a position vector $\bm{x} = (x, y)^{\mathrm{T}}$:
\begin{eqnarray}
\mathscr{R}_{\theta} \bullet \bm{x} &=& 
    \underbrace{\begin{pmatrix}
       \cos\theta & -\sin\theta \\
       \sin\theta & \cos\theta
    \end{pmatrix}}_{= R_{\theta}}
    \begin{pmatrix}
       x \\
       y
    \end{pmatrix}\; \label{eq:def-rotation},
\end{eqnarray}
where $\theta$ is a rotation angle and $R_\theta$ is the corresponding rotation matrix. The left-hand side denotes the action of rotation in abstract form, and the right-hand side is the matrix representation for $\bm{x}$. Generally, the matrix representation is determined by the object on which the rotation acts.

The covariance\cite{Schutz1980, Nakahara2003} with respect to rotation is described for a scalar field $\omega(\bm{x})$ and a vector field $\bm{v}(\bm{x})$ as follows:
\begin{subequations}
\label{eq:covariance-whole}
\begin{eqnarray}
    \mathscr{R}_{\theta} \bullet \omega(\bm{x}) &=& \omega(R^{-1}_{\theta} \bm{x})\;, \label{eq:covariance-scalar} \\
    \mathscr{R}_{\theta} \bullet \bm{v}(\bm{x}) &=& R_{\theta} \bm{v}(R^{-1}_{\theta}\bm{x})\;, \label{eq:covariance-vector}
\end{eqnarray}
\end{subequations}
Only the referred position is changed for a scalar as in (\ref{eq:covariance-scalar}), whereas both the referred position and components are changed for a vector as in (\ref{eq:covariance-vector}). The change in the vector components describes the rotation of its direction. Both scalars and vectors satisfy the linearity. The transformation law, i.e., the covariance (\ref{eq:covariance-whole}), distinguishes a vector from a tuple of scalars.

\subsection{\label{subsec:equivariance} Rotational equivariance of super-resolution models}

The equivariance\cite{Weiler2019, Bronstein2021} is generally defined as the commutativity of a function $f$ with a set of transformations. Consider a function $f:\;{F}^{(\rm in)} \mapsto {F}^{(\rm out)}$, where ${F}^{(\rm in)}$ (${F}^{(\rm out)}$) is an input (output) tensor field. The function $f$ is equivariant if it satisfies
\begin{equation}
    f (\mathscr{R}_{\theta} \bullet {F}^{(\rm in)}) = \mathscr{R}_{\theta} \bullet f({F}^{(\rm in)})\;. \label{eq:def-equivariance}
\end{equation}
The rotation $\mathscr{R}_{\theta}$ composes a group, and (\ref{eq:def-equivariance}) must hold for all elements of that group. For instance, if we consider a set of discrete rotations for every 120 degrees, (\ref{eq:def-equivariance}) must hold for all of those rotations. The invariance is a special case of the equivariance: $f(\mathscr{R}_{\theta} \bullet {F}^{(\rm in)}) = f({F}^{(\rm in)}) = {F}^{(\rm out)}$, where the output is independent of $\mathscr{R}_{\theta}$.

We argue that the equivariance of SR models is a sufficient condition for the covariance of super-resolved velocity fields. An SR model $f$ is assumed to be equivariant (\ref{eq:def-equivariance}); $\bm{v}_{\rm LR}$ is an input velocity field in low resolution (LR) and $\bm{\hat{v}}_{\rm HR}$ is the corresponding output in high resolution (HR). Both $\bm{v}_{\rm LR}$ and $\bm{\hat{v}}_{\rm HR}$ are snapshots at the same instantaneous time. When the input is transformed as $\mathscr{R}_{\theta} \bullet \bm{v}_{\rm LR}$, the SR model yields $f (\mathscr{R}_{\theta} \bullet \bm{v}_{\rm LR})$, which is equal to $\mathscr{R}_{\theta} \bullet f (\bm{v}_{\rm LR}) =\mathscr{R}_{\theta} \bullet \bm{\hat{v}}_{\rm HR}$. Clearly, the output $\bm{\hat{v}}_{\rm HR}$ is transformed in the same manner as the input $\bm{v}_{\rm LR}$ (Fig. \ref{fig:schematic-equivariance}), ensuring that the output is a covariant, geometric vector field.

\subsection{\label{subsec:data-consistency} Rotational consistency in dataset for super-resolution}

We propose that the rotational consistency in datasets is necessary and sufficient for SR models to acquire rotational equivariance through training. A similar discussion can be made for another transformation such as reflection. In the following, we consider supervised learning and assume that a dataset is so large that a trained SR model is sufficiently accurate.

Rotational consistency in datasets is newly introduced here as the invariance of pairs of LR and HR velocities with respect to rotation. Suppose a dataset $D$ and any pair in it $(\bm{v}_{\rm LR}, \bm{v}_{\rm HR})$. The dataset $D$ is rotationally consistent if it contains the transformed pair $(\mathscr{R}_{\theta} \bullet \bm{v}_{\rm LR}, \mathscr{R}_{\theta} \bullet \bm{v}_{\rm HR})$:
\begin{equation}
    {}^\forall(\bm{v}_{\rm LR}, \bm{v}_{\rm HR}) \in D \rightarrow (\mathscr{R}_{\theta} \bullet \bm{v}_{\rm LR}, \mathscr{R}_{\theta} \bullet \bm{v}_{\rm HR}) \in D. \label{eq:def-data-consistency}
\end{equation}
The consistency (\ref{eq:def-data-consistency}) is not necessarily related to physics. For instance, this concept can be applied to a dataset of pairs of LR and HR photos.

An SR model learns rotational equivariance if a training dataset is rotationally consistent. This dataset contains a velocity pair $(\bm{v}_{\rm LR}, \bm{v}_{\rm HR})$ and the rotated pair $(\mathscr{R}_{\theta} \bullet \bm{v}_{\rm LR}, \mathscr{R}_{\theta} \bullet \bm{v}_{\rm HR})$. Since the SR model is sufficiently accurate, it gives $f(\bm{v}_{\rm LR}) \approx \bm{v}_{\rm HR}$ and $f(\mathscr{R}_{\theta} \bullet \bm{v}_{\rm LR}) \approx \mathscr{R}_{\theta} \bullet \bm{v}_{\rm HR}$. Both equations imply $f(\mathscr{R}_{\theta} \bullet \bm{v}_{\rm LR}) = \mathscr{R}_{\theta} \bullet f(\bm{v}_{\rm LR})$, i.e., the rotational equivariance (\ref{eq:def-equivariance}).

Conversely, a training dataset is rotationally consistent if a trained SR model is rotationally equivariant. Consider an LR velocity $\bm{v}_{\rm LR}$ and the rotated velocity $\mathscr{R}_{\theta} \bullet \bm{v}_{\rm LR}$. The SR model yields $f(\bm{v}_{\rm LR}) = \bm{\hat{v}}_{\rm HR}$, where the inference is denoted by a hat. Using the equivariance (\ref{eq:def-equivariance}), we obtain $f(\mathscr{R}_{\theta} \bullet \bm{v}_{\rm LR}) = \mathscr{R}_{\theta} \bullet f(\bm{v}_{\rm LR}) = \mathscr{R}_{\theta} \bullet \bm{\hat{v}}_{\rm HR}$. Since the model accuracy is sufficiently high, the inference is close to the corresponding ground truth. Thus, the dataset contains pairs similar to $(\bm{v}_{\rm LR}, \bm{\hat{v}}_{\rm HR})$ and $(\mathscr{R}_{\theta} \bullet \bm{v}_{\rm LR}, \mathscr{R}_{\theta} \bullet \bm{\hat{v}}_{\rm HR})$, implying the rotational consistency (\ref{eq:def-data-consistency}). 

In summary, the above discussion suggests that the rotational consistency in datasets is necessary and sufficient for SR models to acquire rotational equivariance through training. Such a learning process is not required for rotationally equivariant CNNs,\cite{Weiler2019, Cohen2019} where the equivariance is imposed as prior knowledge through the weight sharing of convolution kernels. This fact indicates that the acquisition of rotational equivariance is not necessarily the result of training. The imposed and learned equivariance are compared in Section \ref{sec:results}.

The above discussion does not necessarily assume a global rotation and holds for local rotations. It may be rare for datasets to contain exactly the same pair as the rotated one. SR models usually infer outputs locally from inputs such as through convolution. These SR models can learn the local equivariance from datasets, which are consistent for local rotations. This point may be different from image classification, in which the inference is based globally on input. Weiler et al.\cite{Weiler2018CVPR} showed that the CNN did not learn rotational equivariance in an image classification task, while it learned when the rotational data augmentation was employed. In Sections \ref{subsec:results-decaying-turbulence} and \ref{subsec:results-barotropic-instability}, we demonstrate that the ordinary CNNs learn rotational equivariance without data augmentation. The difference from the result of Weiler et al.\cite{Weiler2018CVPR} may be attributed to the locality of SR. In the following discussion, the global rotation is considered, while the same results hold for local rotations.

\subsection{\label{subsec:example-data-inconsistency} Breaking of rotational consistency in dataset}

A dataset may not be rotationally consistent even if the fluid system has rotational symmetry. In other words, the rotational symmetry of fluid systems is not a sufficient condition for the rotational consistency of datasets. The term ``symmetry'' is used here in the sense of physics;\cite{Landau1975, Salmon1998} i.e., when a fluid system has rotational symmetry, the Navier–-Stokes equations and boundary conditions are unchanged under rotation and a rotated velocity field is a solution. We provide an example where the rotational consistency is broken due to a method of generating LR velocity.

\begin{figure}[htbp]
    \includegraphics[width=8.5cm]{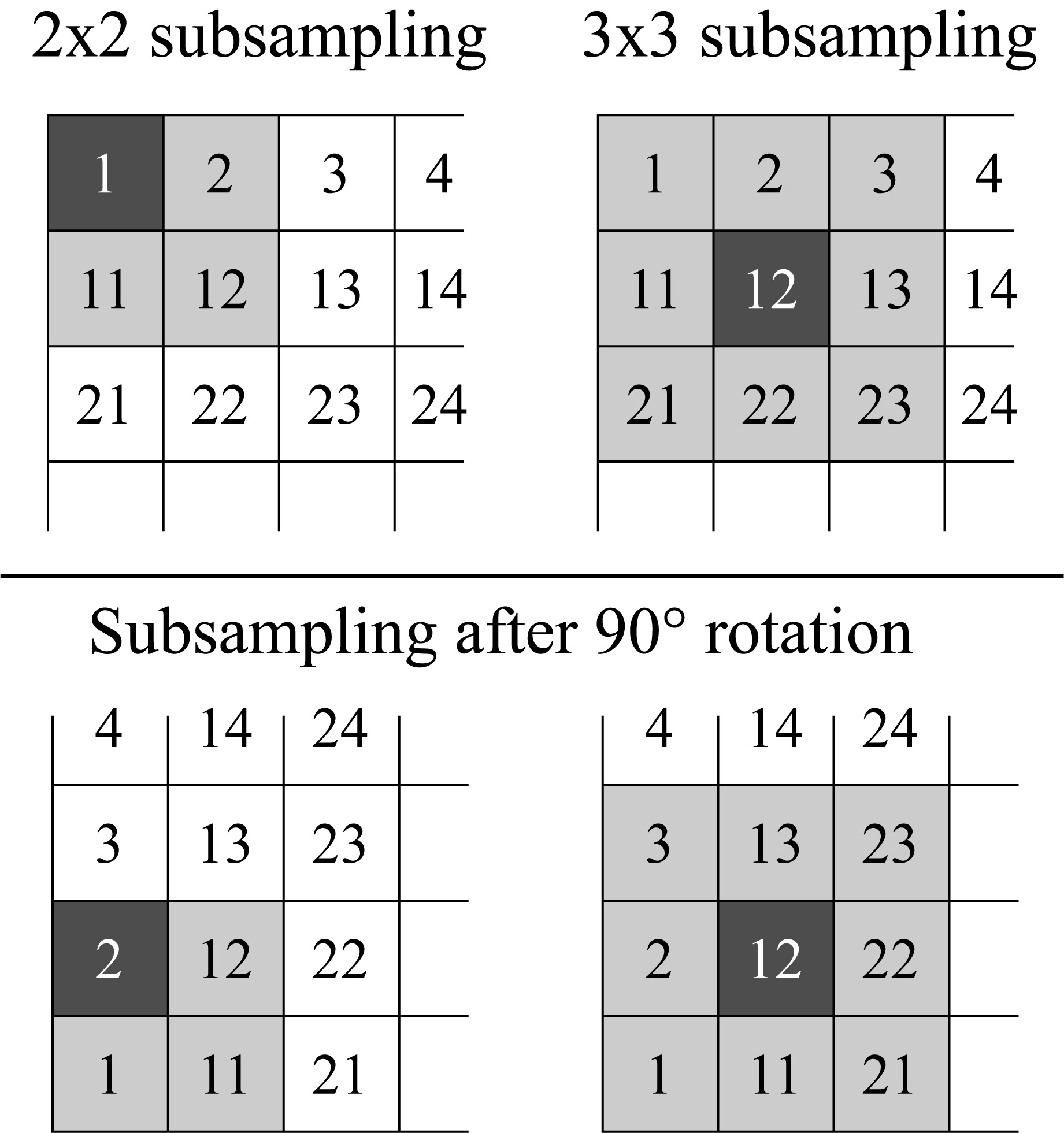}
    \caption{\label{fig:schematic-commutative-subsampling} Schematic showing the commutative property of subsampling with rotation. (left) The upper left pixels (black tiles) are sampled in the $2 \times 2$ windows. (right) The central pixels are sampled in the $3 \times 3$ windows regardless of the orientation.}
\end{figure}

Let us consider that LR images are generated by subsampling HR images. Figure \ref{fig:schematic-commutative-subsampling} is a schematic that compares the subsampling at intervals of $2\times2$ and $3\times3$ pixels. When the interval is $2\times2$, the LR image is varied through the orientation of the HR image, whereas it is not when the interval is $3\times3$. The pairing between LR and HR images depends on the orientation when the interval is $2\times2$ pixels.

The above schematic indicates that the subsampling at even intervals does not commute with the rotation of 90 degrees. This statement is formulated by
\begin{equation}
    d(\mathscr{R}_{90^\circ} \bullet{\bm{v}_{\rm HR}}) \ne \mathscr{R}_{90^\circ} \bullet d(\bm{v}_{\rm HR})\;, \label{eq:commutative-subsampling}
\end{equation}
where $d$ is the subsampling operation, $\mathscr{R}_{90^\circ}$ is the 90-degree rotation (\ref{eq:covariance-vector}), and $\bm{v}_{\rm HR}$ is an HR velocity field. The input LR velocity is given by $\bm{v}_{\rm LR} = d(\bm{v}_{\rm HR})$. In contrast, the subsampling at odd intervals is commutative with the 90-degree rotation, as indicated in Fig. \ref{fig:schematic-commutative-subsampling}.

The rotational consistency in a dataset is broken by the non-commutativity of subsampling (\ref{eq:commutative-subsampling}), even if the fluid system is rotationally symmetric. Suppose an HR velocity $\bm{v}_{\rm HR}$ and the rotated velocities $\mathscr{R}_{\theta} \bullet \bm{v}_{\rm HR}$ ($\theta = 90^\circ, 180^\circ, 270^\circ$), all of which are snapshots of actual flows because of the rotational symmetry. For simplicity, we focus on $\theta = 90^\circ$. The LR velocities are generated as $\bm{v}_{\rm LR} = d(\bm{v}_{\rm HR})$ and $\bm{v}^\prime_{\rm LR} = d(\mathscr{R}_{90^\circ} \bullet \bm{v}_{\rm HR})$. Importantly, $\bm{v}^\prime_{\rm LR}$ is not equal to $\mathscr{R}_{90^\circ} \bullet \bm{v}_{\rm LR}$ [$= \mathscr{R}_{90^\circ} \bullet d(\bm{v}_{\rm HR})$]. The resultant dataset contains $({\bm{v}_{\rm LR}},{\bm{v}_{\rm HR}})$ and  $(\bm{v}^\prime_{\rm LR}, \mathscr{R}_{90^\circ} \bullet \bm{v}_{\rm HR})$, but not $(\mathscr{R}_{90^\circ} \bullet \bm{v}_{\rm LR}, \mathscr{R}_{90^\circ} \bullet \bm{v}_{\rm HR})$. This fact indicates that the rotational consistency (\ref{eq:def-data-consistency}) is broken.

According to Section \ref{subsec:data-consistency}, the breaking of the rotational consistency in datasets indicates that SR models do not learn rotational equivariance. In such a case, if the equivariance is forcefully imposed on an SR model through weight sharing, the accuracy in SR is undermined. These points are demonstrated in Section \ref{sec:discussions}.

\subsection{\label{subsec:symmetry-data-consistency} A relationship between symmetry in fluid systems and consistency in datasets}

It may be natural to consider that the symmetries of a system are reflected in its dataset and an NN learns the symmetries from the dataset;\cite{Decelle2019, Wetzel2020, Ha2021, Krippendorf2021, Desai2022} however, this does not always hold in SR. The pairing of LR and HR data needs to be investigated. In fact, a dataset can be made rotationally inconsistent even if the fluid system has rotational symmetry (Section \ref{subsec:example-data-inconsistency}). Moreover, a converse example can be provided by applying data augmentation. The rotational consistency (\ref{eq:def-data-consistency}) is satisfied in an augmented dataset by rotating all LR and HR pairs even if the fluid system is not rotationally symmetric. In such a case, a rotated velocity field is not regarded as an actual flow, although it is contained in the dataset.

For further discussion, it may be necessary to restrict data generation methods. We propose two conditions:
\begin{description}
    \item[C1] HR velocity fields are directly generated from fluid experiments or simulations.
    \item[C2] LR velocity fields are generated from the HR velocity fields by an operation $p$ that commutes with rotation $\mathscr{R}_{\theta}$.
\end{description}
Under C1 and C2, a dataset is rotationally consistent if the fluid system has rotational symmetry. The first condition eliminates the possibility of modifying rotational symmetry through data augmentation. Because the fluid system has the symmetry, a sufficiently large dataset contains an HR velocity field and the rotated one. The second condition eliminates a case in which the pairing between LR and HR velocities depends on the orientation as in Section \ref{subsec:example-data-inconsistency}. For any pair $(\bm{v}_{\rm LR}, \bm{v}_{\rm HR})$, the rotated HR velocity $\mathscr{R}_{\theta} \bullet \bm{v}_{\rm HR}$ is contained in the dataset and it is tied to $\mathscr{R}_{\theta} \bullet \bm{v}_{\rm LR}$. The LR velocity obtained from $\mathscr{R}_{\theta} \bullet \bm{v}_{\rm HR}$ is given by $p(\mathscr{R}_{\theta} \bullet \bm{v}_{\rm HR})$, which is equal to $\mathscr{R}_{\theta} \bullet p(\bm{v}_{\rm HR}) = \mathscr{R}_{\theta} \bullet \bm{v}_{\rm LR}$ owing to the commutativity of $p$. Thus, the rotated pair $(\mathscr{R}_{\theta} \bullet \bm{v}_{\rm LR}, \mathscr{R}_{\theta} \bullet \bm{v}_{\rm HR})$ is contained and the dataset is rotationally consistent.

In summary, under C1 and C2, the rotational symmetry of a fluid system is reflected in its dataset; that is, the dataset is rotationally consistent. In this case, an SR model learns the rotational equivariance if the data size is sufficiently large (Section \ref{subsec:data-consistency}). Further, if the SR model becomes equivariant, the output of super-resolved velocity satisfies the covariance of vectors with respect to rotation (Section \ref{subsec:equivariance}).

The above conditions, C1 and C2, may be too strict. The previous studies reported that data augmentation is effective in reducing test errors\cite{Marcos2017, Fuchs2020, Siddani2021} or for CNNs to learn rotational equivariance.\cite{Weiler2018CVPR} Condition C1 eliminates those benefits through data augmentation. Moreover, the LR and HR data may be generated from separate fluid simulations.\cite{Wang2021, Kim2021} In such a case, condition C2 cannot be assumed because LR data are not generated from HR data. Recently, Desai et al.\cite{Desai2022} defined the symmetries of datasets as the invariance of probability density functions of data samples. They demonstrated that a GAN can discover the symmetries of a dataset in particle physics. In future work, the definition of symmetries in SR may be extended by using probability density functions to relax condition C2 for considering a case where LR and HR data are separately generated.

\section{\label{subsec:equivariant-neural-networks}CNNs used in this research}

\subsection{\label{subsec:two-cnns-for-sr}Two existing CNNs for super-resolution of fluids}

Numerical experiments were conducted to support the theoretical suggestions described in the previous section. Two types of CNNs were used to confirm that the results are not strongly dependent on the network architecture: the hybrid downsampled skip-connection/multi-scale model (DSC/MS)\cite{Fukami2019} and the residual in residual dense network (RRDN).\cite{Bode2021} Both CNNs were used for achieving single image super-resolution; the input was a snapshot of LR velocity, and the output was the HR velocity at the same instantaneous time.

\begin{figure*}[htbp]
    \includegraphics[width=17cm]{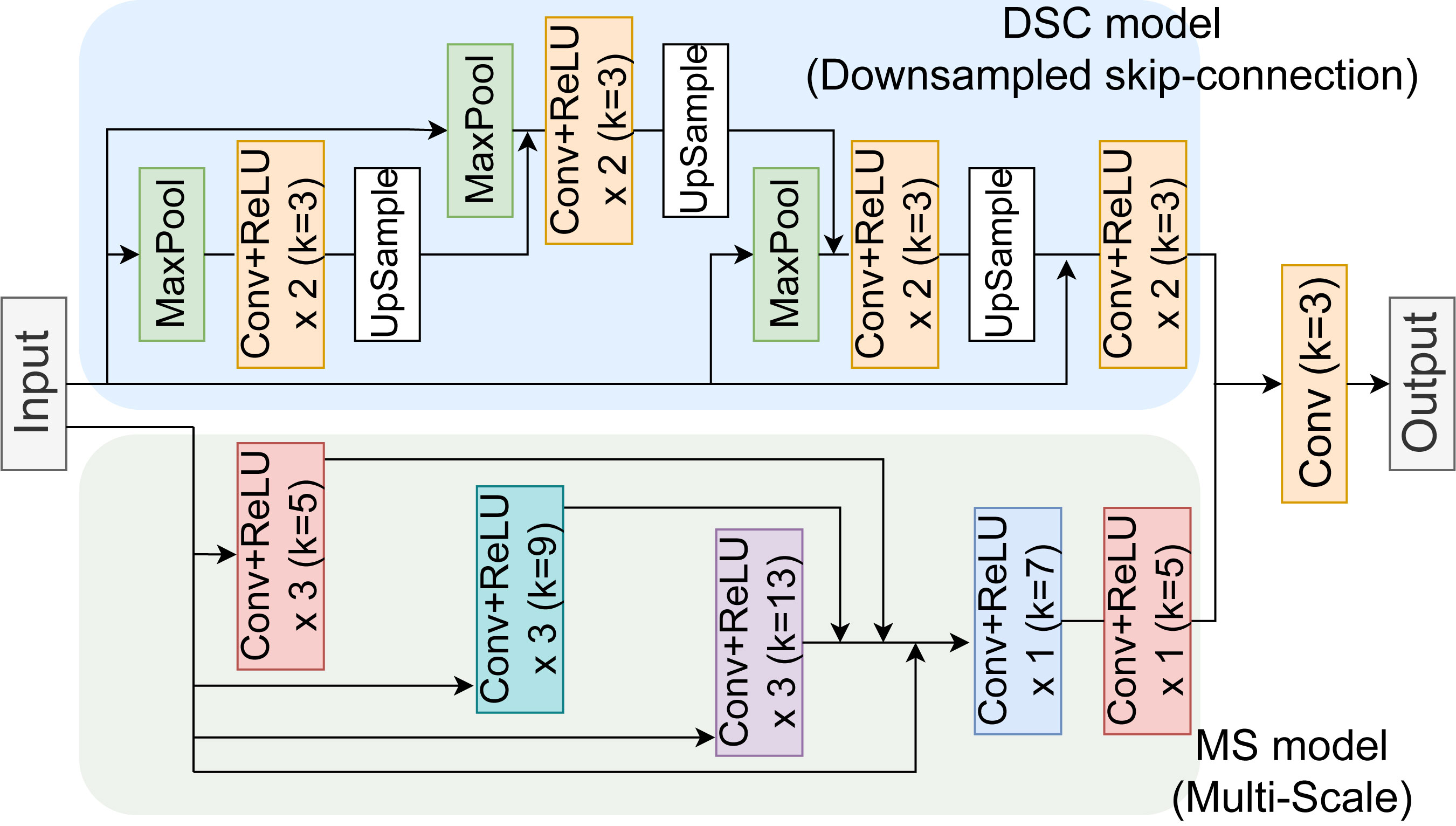}
    \caption{\label{fig:DSC/MS} Network architecture of the hybrid DSC/MS model (Downsampled Skip-Connection and Multi-Scale). The architecture follows the CNN proposed by Fukami et al.\cite{Fukami2019} All convolutions are two-dimensional. The symbol $k$ denotes the spatial size of the kernel. Linear interpolation is used in upsampling.}
\end{figure*}

The hybrid DSC/MS model (hereafter, DSC/MS) was proposed in pioneering work on fluid SR by Fukami et al.\cite{Fukami2019} They combined their downsampled skip connection (DSC) model with the multi-scale (MS) model.\cite{Du2018} Figure \ref{fig:DSC/MS} shows the network architecture of DSC/MS. Multi-scale signals in fluid fields can be captured by downsampling and skip connections. The DSC/MS can super-resolve both velocity and vorticity fields in two dimensions.\cite{Fukami2019} In subsequent studies, DSC/MS was applied to three-dimensional velocity,\cite{Fukami2020, Fukami2021} and spatio-temporal SR was achieved by successive use of DSC/MS in space and time.\cite{Fukami2021} Additionally, a deeper version of DSC/MS has been proposed.\cite{Wang2021DSCMS} Although DSC/MS was proposed relatively early, it is still among the most important CNNs in fluid SR.

\begin{figure*}[htbp]
    \includegraphics[width=17cm]{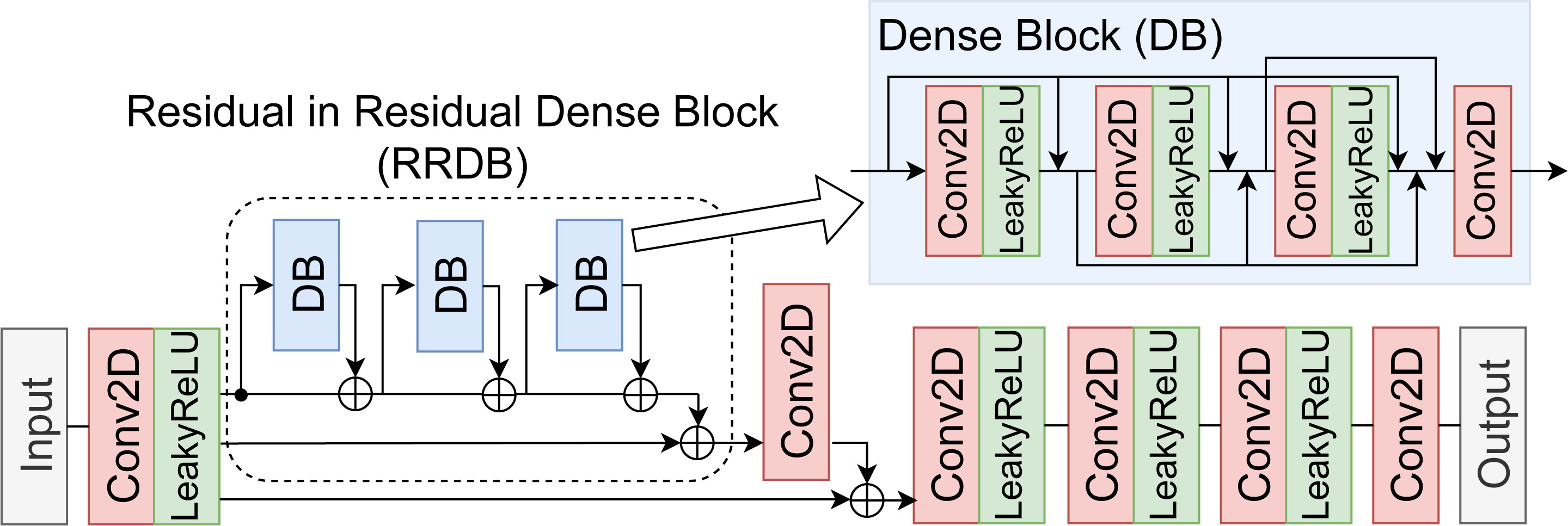}
    \caption{\label{fig:RRDN} Network architecture of the RRDN (Residual in Residual Dense Network). The architecture emulates that of the generator of the GAN proposed by Bode et al.\cite{Bode2021} All kernels in the convolution layers have a spatial size of $3\times3$.}
\end{figure*}

The RRDN was proposed by Bode et al.\cite{Bode2021} as the generator of their physics-informed GAN. They employed the RRDN for the SR of three-dimensional velocity. A network similar to the RRDN has been applied to two-dimensional passive scalar fields.\cite{Wang2020} The RRDN utilizes the residual in residual dense block (RRDB),\cite{Zhang2018RDN, Wang2018ESRGAN} which extracts multi-scale structures in fluid fields with dense skip connections. Figure \ref{fig:RRDN} shows the network architecture of the RRDN. The RRDB, the core module of the RRDN, has been utilized for SR in fluid mechanics\cite{Wang2020, Bode2021} as well as in computer vision.\cite{Zhang2018RDN, Wang2018ESRGAN} Thus, it is necessary to examine rotationally equivariant CNNs having RRDBs.

\subsection{\label{subsec:construction-equivariant-cnns}Construction of rotationally equivariant CNNs}

The DSC/MS and RRDN were converted into rotationally equivariant CNNs by replacing each layer with its equivariant counterpart. Mathematically, if each layer is rotationally equivariant, the entire network is equivariant as well.\cite{Weiler2019, Bronstein2021} The rotationally equivariant versions are referred to as the Eq-DSC/MS and Eq-RRDN. The DSC/MS and RRDN are equivariant to translation, while the Eq-DSC/MS and Eq-RRDN are equivariant to translation and rotation. The DSC/MS and RRDN are referred to as the original and conventional CNNs, while the Eq-DSC/MS and Eq-RRDN are referred to as the rotationally equivariant and $C_N$-equivariant CNNs. The meaning of $C_N$ is explained below. In the following, the replacement of convolution layers is discussed, followed by a description of the other layers.

The rotationally equivariant convolution is implemented by the theory and software library of Weiler and Cesa.\cite{Weiler2019} Specifically, the rotational equivariance is achieved by sharing the weights of convolution kernels along the azimuthal and channel directions. Compared with ordinary convolution, this weight sharing reduces the number of learnable parameters. The theory is briefly reviewed in Appendix \ref{sec-app:theory-CNN}.

The regular representation\cite{Gilmore2008, Zee2016} was adopted in all rotationally equivariant convolutions. It is necessary to determine the matrix representation of rotation, which is a hyper-parameter.\cite{Weiler2019} In the regular representation, the rotation is formulated as a permutation along the channel direction. There are three reasons for adopting the regular representation. First, it comprises all irreducible representations, i.e., all minimal representations.\cite{Gilmore2008, Zee2016} Second, CNNs using the regular representation have exhibited high accuracy in image classification.\cite{Cohen2016, Cohen2017, Marcos2017, Weiler2019} Third, each channel can be transformed nonlinearly and independently when the regular representation is used.\cite{Cohen2016, Weiler2019} In contrast, when another representation is employed, each channel may not be transformed independently by a nonlinear function such as ReLU. A function such as Norm-ReLU\cite{Weiler2019} is necessary, as it acts on the vector norm and preserves the vector direction. A limitation such as in Norm-ReLU may reduce the expressive power of CNNs and may not be suitable for SR.

In practice, vector fields are discretized in space, and the rotation acting on them is also discretized. Mathematically, rotations proportional to $360/N$ degrees are described by $C_N$, the cyclic group\cite{Gilmore2008, Zee2016} of order $N$ ($N \in \mathbb{N}$). CNNs equivariant to any rotation of $C_N$ are referred to as $C_N$-equivariant. This study focuses on rotations in multiples of 90 degrees. These rotations are described by $C_4$ or its subgroup $C_2$: $C_4$ consists of 0-, 90-, 180-, and 270-degree rotations, and $C_2$ consists of 0- and 180-degree rotations. Here, the identity map is referred to as the 0-degree rotation.

All convolution layers in the original CNNs were replaced with the $C_N$-equivariant layers, while maintaining the number of channels to avoid increasing the calculation time. For instance, the regular representation of $C_4$ is composed of the 4 elements. If both input and output of a convolution layer have 64 channels, the $C_4$-equivariant convolution employs 16 sets of features in the regular representation. The other hyper-parameters such as kernel size were the same as in the original convolution layer.

The implementation of the other layers was as follows. A max-pooling layer was replaced with the norm max-pooling\cite{Marcos2017, Weiler2019} when the input was a vector field, whereas it remained unchanged when the input was described with the regular representation. In all cases, the pooling operation is equivariant.\cite{Cohen2016, Marcos2017, Weiler2019} Data resizing was performed via linear or bicubic interpolation, both of which are equivariant because the weights of interpolation depend only on the absolute difference in the $x$ or $y$ coordinate.\cite{Key1981} The skip connection needs no replacement\cite{Wang2021} because it is an addition operation between features having the same type. A channel-wise nonlinear function such as ReLU or leaky ReLU needs no replacement because it is equivariant under the regular representation, as discussed above. In the input and output layers, no conversion was required because a vector such as velocity is a feature in the irreducible representation\cite{Gilmore2008, Zee2016} of order 1 and is decomposed with the regular representation.\cite{Weiler2019}

\section{\label{sec:methodology}Methods}

Supervised learning was performed using pairs of velocity fields in HR and LR. The HR velocity data were generated from numerical simulations of fluid dynamics. The LR velocity data were obtained by applying the subsampling or average pooling to the HR velocity fields. The source code used in this study is available on the GitHub repository.\cite{Yasuda2022GH}

\subsection{\label{subsec:fluid-simulations}Fluid dynamics simulations}

Two types of fluid dynamics simulations were conducted: freely decaying turbulence on a square flat torus and barotropic instability on a periodic channel. The rotational symmetries of the two fluid systems are different. Thus, the dependence of SR on rotational symmetry can be examined by comparing the results.

The governing equations for all experiments are as follows:
\begin{subequations}
\label{eq:2DNS-whole}
\begin{eqnarray}
    \frac{\partial \omega_z}{\partial t} + u \frac{\partial \omega_z}{\partial x} + v \frac{\partial \omega_z}{\partial y} &=& -\nu (-\Delta)^{n_{\nu}} \omega_z\;, \label{eq:2DNS-time-evolution} \\
    \omega_z &=& \Delta \psi = \frac{\partial v}{\partial x} -\frac{\partial u}{\partial y}\;, \label{eq:2DNS-vorticity} \\
    \left(u, v\right) &=& \left(-\frac{\partial \psi}{\partial y}, \frac{\partial \psi}{\partial x}\right)\;, \label{eq:2DNS-velocity}
\end{eqnarray}
\end{subequations}
where $t$ denotes time, $\nu$ is a positive real number, and $n_{\nu}$ is a positive integer. The stream function $\psi$ is derived by solving the Poisson equation (\ref{eq:2DNS-vorticity}) with the vorticity $\omega_z$, and the velocity components $(u, v)$ are obtained from $\psi$ in (\ref{eq:2DNS-velocity}). In two-dimensional Euclidean space, velocity $(u, v)$ is a vector field, while vorticity $\omega_z$ is a scalar field. In three dimensions, $\omega_z$ is the vertical ($z$) component of the vorticity vector, and the subscript $z$ denotes this fact. Note that large-scale flows in the ocean or atmosphere are described by similar equations,\cite{Vallis2017} i.e., quasi-geostrophic equations; the only difference is that the Poisson equation is three-dimensional and includes the depth or altitude dimension.

The governing equations (\ref{eq:2DNS-whole}) were numerically solved with the spectral method and the fourth-order Runge-Kutta method. A software library, ISPACK,\cite{ispack2015} was employed to perform spectral calculations. A discrete Fourier transform was employed along a periodic direction, while a discrete sine/cosine transform was along a direction bounded by two walls.

The freely decaying turbulence experiment was conducted following Refs.~\onlinecite{Fukami2019, Taira2016}. Here, $\nu = 10^{-2}$ and $n_{\nu} = 1$ were set, leading the governing equations (\ref{eq:2DNS-whole}) to the two-dimensional Navier--Stokes equations. A doubly periodic boundary condition was imposed on the flow domain $[0, 2\pi] \times [0, 2\pi]$. This domain, which is a square flat torus, has a rotational symmetry of 90 degrees. The grid size was $128 \times 128$ and determined a truncation wavenumber of 42 for alias-free simulations. The initial condition was given at random while keeping the energy spectrum of $E(k) = a \times k \exp(-k^2 / k_0^2)$, where $k_0 = 26.5$ and $k$ is the magnitude of the wavenumber vector. The Reynolds number ${\rm Re}$ was controlled by $a$: ${\rm Re} \approx 80.0$ for the training data ($a=9$) and ${\rm Re} \approx 93.3$ for the test data ($a=12$). Here, ${\rm Re}$ is defined by ${\rm Re} = U^\ast L^\ast / \nu$, where $U^\ast$ is the velocity scale given by the square root of the spatially averaged initial energy and $L^\ast$ is the initial integral length scale.\cite{Taira2016} Subsequently, while varying the random initial condition, 1,000 simulations were run for training, and 100 were run for testing. In each simulation, 10 snapshots of velocity were sampled between $1.1 \le t \le 2.0$ at intervals of $0.1$. Thus, the training and test datasets comprise 10,000 and 1,000 snapshots, respectively. Results similar to those shown below were obtained from the halved test dataset, implying that the 1,000 snapshots are sufficient for estimating generalization errors.

The barotropic instability\cite{Vallis2017} was simulated in a periodic channel. Here, $\nu = 10^{-30}$ and $n_{\nu} = 8$ were set, resulting in the governing equations (\ref{eq:2DNS-whole}) being the two-dimensional Euler equations with hyper-viscosity. The channel domain is $[0, 2\pi] \times [-\pi/2, \pi/2]$, where the $x$ direction is periodic and the $y$ direction is bounded by the two walls [$\psi = 0 \;\;(y = \pm \pi/2)$]. This domain exhibits rotational symmetry of 180 degrees. The grid size was $128 \times 64$ and determined a truncation wavenumber of $42$ for alias-free simulations. The initial condition was given by a superposition of an unstable laminar flow and random perturbations. The laminar flow has a shear region of
\begin{equation}
    u = \begin{cases}
        U & (\frac{l}{2} \le y \le \frac{\pi}{2})\;, \\
        U + 2U \frac{(y - \frac{l}{2})}{l} & (-\frac{l}{2} < y < \frac{l}{2})\;, \\
        -U & (-\frac{\pi}{2} \le y \le -\frac{l}{2})\;, \label{eq:initial-cond-barotropic-instability}
    \end{cases}
\end{equation}
where the shear was controlled by speed $U$ and width $l$. The speed was set at $U = 0.25$ (positive shear) or $-0.25$ (negative shear). The numbers of experiments with $U = 0.25$ and $-0.25$ were kept the same. The width was set at $l=0.45$ for training and $0.40$ for testing. Each initial perturbation had a random phase and a constant amplitude of $10^{-4}$. Subsequently, 800 and 80 simulations were run for training and testing, respectively, while varying the random phases. In each simulation, 14 snapshots of velocity were sampled between $31 \le t \le 70$ at intervals of $3$. Thus, the training and test datasets comprise 11,200 and 1,120 snapshots, respectively. Results similar to those shown below were obtained from the halved test dataset, implying that the 1,120 snapshots are sufficient for estimating generalization errors.

\subsection{\label{subsec:data-preparation} Data preparation}

The LR snapshots of velocity were obtained by applying the subsampling or average pooling to the HR snapshots, which were generated from the fluid simulations. Subsampling or average pooling has been employed in previous studies on fluid SR.\cite{Vandal2017, Xie2018, Fukami2019, Deng2019, Onishi2019, Leinonen2020, Liu2020, Stengel2020, Fukami2021, Kim2021, Wang2021DSCMS, Yasuda2022}

The LR velocity data were obtained in four steps. First, a scale factor $s$ was determined. When $s=2$, for instance, an LR image is twice as coarse as the HR image. Second, an HR snapshot of size $H \times W$ was resized to $H^{\prime} \times W^{\prime}$, where $H^{\prime}$ ($W^{\prime}$) was the multiple of $s$ closest to $H$ ($W$). Third, coarse-graining was performed on the resized HR snapshot by using subsampling or average pooling. In subsampling, pixel values were extracted at intervals of $s \times s$. In average pooling, each non-overlapping area of $s \times s$ was replaced with its mean. Fourth, the obtained snapshot of $H^{\prime}/s \times W^{\prime}/s$ was resized to $H \times W$. Bicubic interpolation\cite{Key1981} was used in all cases of resizing mentioned above. The above algorithm was applied to $u$ or $v$ separately because each velocity component can be added or subtracted independently in the Cartesian coordinates.

The commutativity with rotation is noted here. As discussed in Section \ref{subsec:example-data-inconsistency}, subsampling with an odd $s$ commutes with the rotation in multiples of 90 degrees, while it does not when $s$ is even. The average pooling commutes with the rotation regardless of the parity of $s$ because the averaging operation does not depend on the orientation of images.

In preprocessing, velocity component $u$ or $v$ was scaled with the common parameter $\sigma$ such that 99.9\% of scaled values were within the range of $[-1,1]$: $u \rightarrow u/\sigma$ and $v \rightarrow v/\sigma$. This scaling is isotropic; hence, it preserves the rotational symmetry of velocity.

\subsection{\label{subsec:training-neural-networks}Implementation of CNNs}

The CNNs were implemented with PyTorch\cite{pytorch2019} 1.8.0 and e2cnn\cite{Weiler2019} 0.2.1. The latter is a software library for the equivariant deep learning of two-dimensional images.

As for the $C_N$-equivariant CNNs, the hyper-parameter $N$ was determined such that it reflected the rotational symmetry of the fluid system: $N = 4$ in the decaying turbulence and $N = 2$ in the barotropic instability experiment. The former is rotationally symmetric at every 90 degrees and the latter at 180 degrees. The effect of changing $N$ is investigated in Appendix \ref{subsec-app:dependency-cn}.

The Adam optimizer\cite{Kingma2015} was used coupled with the loss function of mean squared error. This loss is invariant to rotation. The mini-batch size for the DSC/MS and Eq-DSC/MS was 100, while that for the RRDN and Eq-RRDN was 32. The learning rate ranged between $1.0 \times 10^{-3}$ and $1.0 \times 10^{-4}$ and was generally the same for the DSC/MS and Eq-DSC/MS and for the RRDN and Eq-RRDN. Each training was terminated by early stopping with a patience parameter of 30 epochs, where 30\% of the training data was used for the validation.

Data augmentation was not employed because of the following two reasons. First, data augmentation is not necessary for $C_N$-equivariant CNNs because they take rotational equivariance into account as prior knowledge.\cite{Kondor2018, Cohen2019, Weiler2019} Second, data augmentation may impose inappropriate symmetries that the original data do not possess. According to Section \ref{subsec:symmetry-data-consistency}, if data augmentation is not employed and the LR data generation commutes with the rotation, the rotational symmetry of fluid systems is reflected in datasets. Thus, without data augmentation, only the commutativity of LR data generation needs to be considered.

\subsection{\label{subsec:evaluation-scores}Evaluation metrics of CNNs}

The following three metrics were utilized to evaluate the CNNs: norm error ratio, energy spectral error, and equivariance error ratio. An SR model, implemented by a CNN, is denoted by $f: \bm{v}_{\rm LR}^{(i)} \rightarrow \bm{\hat{v}}_{\rm HR}^{(i)}$, where $\bm{v}_{\rm LR}^{(i)}$ is a snapshot of LR velocity, $\bm{\hat{v}}_{\rm HR}^{(i)}$ is the super-resolved velocity, and $i$ is an index in the test dataset. The ground truth of HR velocity is denoted by $\bm{v}_{\rm HR}^{(i)}$ without a hat.

The norm error ratio (NER) is a score for pixel-wise errors:
\begin{equation}
    NER = \frac{1}{M} \sum_i \left[ \frac{ \sum_{\rm space} \| \bm{\hat{v}}_{\rm HR}^{(i)} - \bm{v}_{\rm HR}^{(i)} \| }{\sum_{\rm space}\|\bm{v}_{\rm HR}^{(i)}\|} \right] \;, \label{eq:NER}
\end{equation}
where $\|\cdot\|$ represents the Euclidean norm and $M$ is the size of the test dataset.

The energy spectral error (ESE) was introduced by Wang et al.\cite{Wang2021} to evaluate the validity of spatial patterns. The ESE is equal to the root-mean-squared error of the logarithm of the energy spectra between the inference and ground truth:
\begin{equation}
    ESE = \sqrt{\frac{1}{k_{\max} M}\sum_{i,k}\left[\log \hat{E}^{(i)}(k) - \log E^{(i)}(k)\right]^2}\;,\label{eq:ESE}
\end{equation}
where $k_{\max}$ is the maximum of $k$, and $\hat{E}^{(i)}(k)$ and $E^{(i)}(k)$ are the energy spectra of the inference and ground truth, respectively. The ESE is interpreted as the error of spatial correlations through the Wiener–-Khintchine theorem.

The equivariance error\cite{Wang2021} measures the equivariance of any model and is calculated without referring to the ground truth. The equivariance error ratio (EER) is defined by
\begin{equation}
    EER = \frac{1}{M} \sum_i \left[ \frac{\sum_{\text{central region}} \| f(\mathscr{R}_{\theta}\bullet\bm{v}_{\rm LR}^{(i)}) -  \mathscr{R}_{\theta}\bullet f(\bm{v}_{\rm LR}^{(i)}) \|}{\sum_{\text{central region}} \| \mathscr{R}_{\theta}\bullet f(\bm{v}_{\rm LR}^{(i)}) \|} \right]\;, \label{eq:EER}
\end{equation}
The summation inside the brackets was taken over the central $48 \times 48$ pixels to avoid extrapolation effects such as zero padding. If an SR model is rotationally equivariant, the EER is nearly zero. The EER (\ref{eq:EER}) measures the rotational equivariance. If the rotation $\mathscr{R}_{\theta}$ is replaced with another transformation in (\ref{eq:EER}), the corresponding EER can be defined. The NER and ESE are referred to as the test errors and are distinct from the EER because the NER and ESE are based on the ground truth while the EER is not.

\section{\label{sec:results}Results and Discussion}

\subsection{\label{subsec:results-decaying-turbulence}Freely decaying turbulence experiment}

This subsection compares the $C_4$-equivariant CNNs with the original ones in the freely decaying turbulence experiment. We discuss the SR of the LR velocity generated by subsampling. Similar results were obtained for the LR velocity by average pooling. The scale factor $s$ was fixed to odd numbers: $s = 5$, $9$, or $13$. The SR with an even $s$ is examined in Section \ref{sec:discussions}.

\begin{figure*}[htbp]
    \includegraphics[width=17cm]{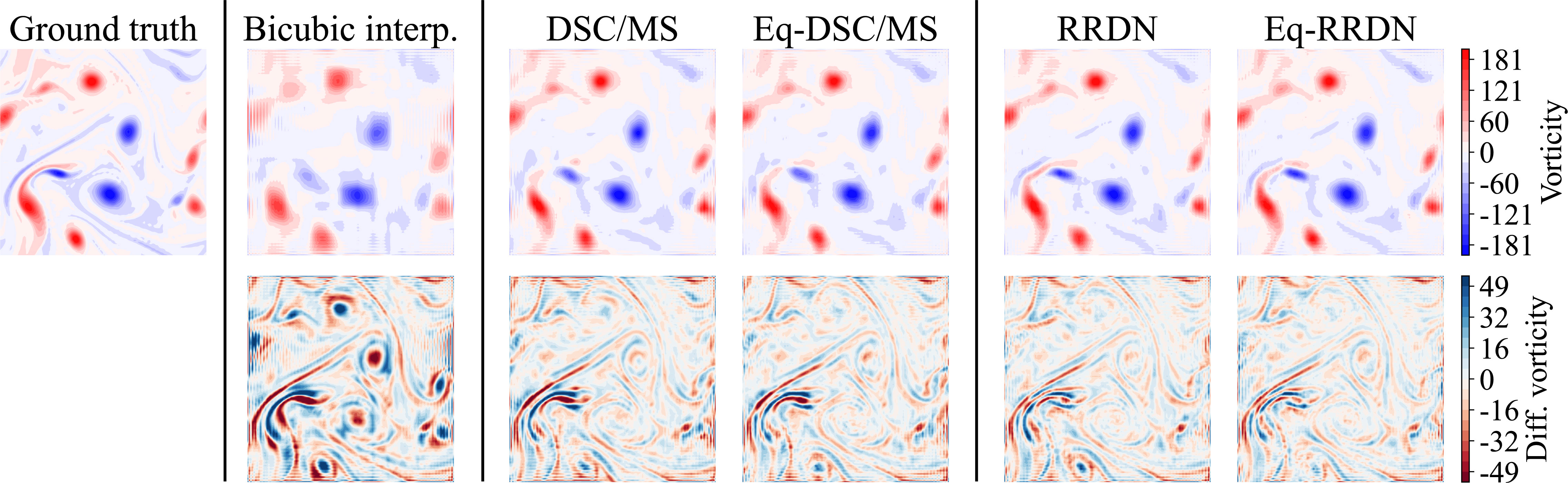}
    \caption{\label{fig:sr-decaying-turbulence} An example of the vorticity calculated from the super-resolved velocity in the decaying turbulence. The low-resolution velocity was generated by subsampling with the scale factor $s=9$. ``Bicubic interp.'' stands for bicubic interpolation. The bottom row shows the difference from the ground truth.}
\end{figure*}

The $C_4$-equivariant CNNs reproduced HR velocity fields similar to those of the original CNNs. Figure \ref{fig:sr-decaying-turbulence} shows an example of the vorticity calculated from the super-resolved velocity when $s = 9$. The finite difference operation deriving the vorticity emphasizes the small-scale structure in the super-resolved velocity. Regarding the baseline of bicubic interpolation, the shape of each vortex is ambiguous, and the difference from the ground truth is large. All four CNNs accurately reproduce the small-scale structure. The snapshots of vorticity by Eq-DSC/MS and Eq-RRDN are nearly identical to those by DSC/MS and RRDN, respectively.

\begin{figure*}[htbp]
    \includegraphics[width=8.5cm]{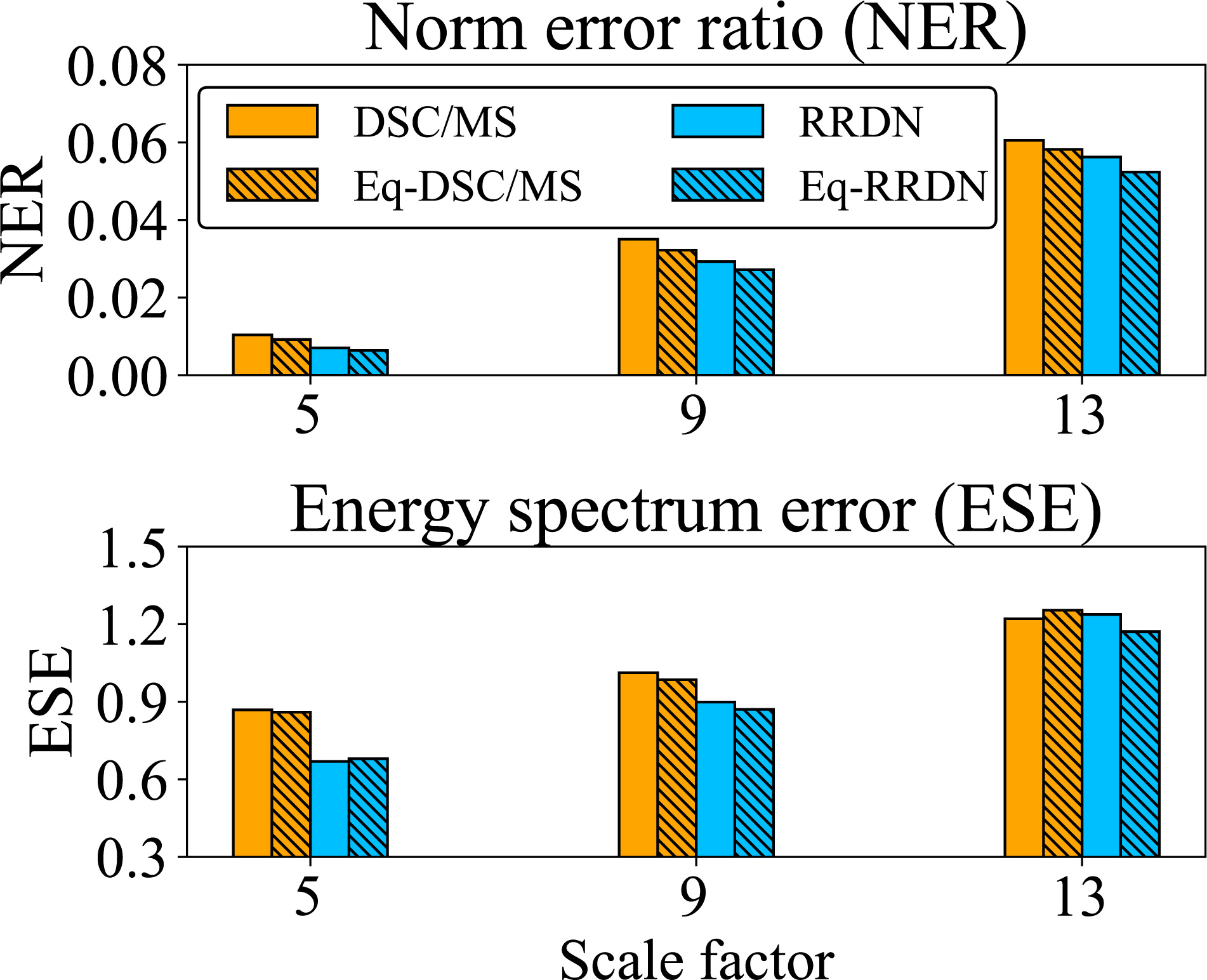}
    \caption{\label{fig:test-errors-decaying-turbulence} Test errors of the super-resolved velocity in the decaying turbulence. The input low-resolution velocity was generated by subsampling. The norm error ratio (NER) and energy spectrum error (ESE) are defined by (\ref{eq:NER}) and (\ref{eq:ESE}), respectively.}
\end{figure*}

The test errors of the $C_4$-equivariant CNNs were comparable to those of the original CNNs. Figure \ref{fig:test-errors-decaying-turbulence} shows the NER (\ref{eq:NER}) and ESE (\ref{eq:ESE}) at $s = 5$, $9$, and $13$. Both errors increase as $s$ becomes large, likely because more small-scale components are removed from the input LR velocity and SR becomes more difficult. At each $s$, the test errors of Eq-DSC/MS and Eq-RRDN are approximately equal to or slightly smaller than those of DSC/MS and RRDN, respectively. Further, the test errors of RRDN and Eq-RRDN tend to be smaller than those of DSC/MS and Eq-DSC/MS. This result may be attributed to the fact that the RRDN and Eq-RRDN are deeper and have more learnable parameters than the DSC/MS and Eq-DSC/MS, as discussed below.

\begin{figure*}[htbp]
    \includegraphics[width=17cm]{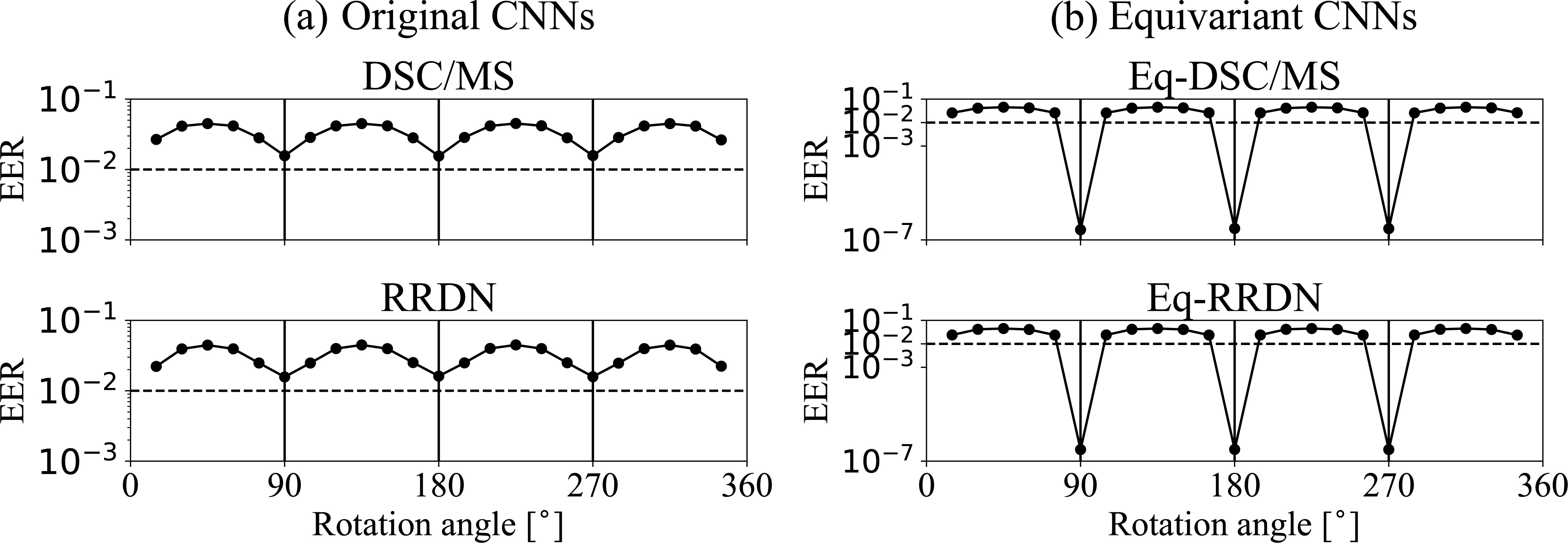}
    \caption{\label{fig:equiv-errors-decaying-turbulence} Equivariance error ratio (EER) against the rotation angle in the decaying turbulence. The EER is defined by (\ref{eq:EER}). The input low-resolution velocity was generated by subsampling with the scale factor $s=9$. The intervals of rotation angles are 15${}^{\circ}$, where 0${}^{\circ}$ and 360${}^{\circ}$ are omitted. The horizontal dashed lines show $EER=10^{-2}$.}
\end{figure*}

The original CNNs learned rotational equivariance from the training data, which explains why the original and $C_4$-equivariant CNNs show comparable accuracy in SR. Figure \ref{fig:equiv-errors-decaying-turbulence} shows the EERs (\ref{eq:EER}) at the various rotation angles when $s=9$. The Eq-DSC/MS and Eq-RRDN show relatively small EERs ($\sim 10^{-7}$) at the multiples of 90 degrees [Fig. \ref{fig:equiv-errors-decaying-turbulence}(b)], confirming the imposed $C_4$-equivariance as prior knowledge. The EERs shown by DSC/MS and RRDN at the multiples of 90 degrees are smaller than those shown at the other angles [Fig. \ref{fig:equiv-errors-decaying-turbulence}(a)]. These EERs are approximately $10^{-2}$ (dashed lines) and on the same order of magnitude as the NERs in Fig. \ref{fig:test-errors-decaying-turbulence}. Thus, it is difficult to distinguish between errors due to the SR and the rotation.

According to Sections \ref{subsec:data-consistency} through \ref{subsec:symmetry-data-consistency}, the rotational symmetry of the decaying turbulence system is reflected in the rotational consistency of the dataset because the subsampling of an odd number $s$ commutes with the 90-degree rotation. This consistency indicates that ordinary CNNs learn the $C_4$-equivariance if the dataset is sufficiently large. The result of Fig. \ref{fig:equiv-errors-decaying-turbulence} supports these theoretical suggestions.

Learning the $C_4$-equivariance implies that the number of effective parameters is reduced. Table \ref{tab:num-learnable-params} compares the numbers of all trainable parameters in the CNNs. The DSC/MS (RRDN) has approximately 7.8 (6.0) times more parameters than the Eq-DSC/MS (Eq-RRDN). This difference is due to the weight sharing imposed on the rotationally equivariant kernels.\cite{Weiler2019} The DSC/MS and RRDN acquired the $C_4$-equivariance through training; thus, their kernel weights would have become closer to those of Eq-DSC/MS or Eq-RRDN. Recently, it has been pointed out that the effective degrees of freedom of NNs are much smaller than the total number of trainable parameters;\citep{Maddox2020, Neyshabur2017} however, the reasons for that reduction are not quite clear. Our result implies that an NN can learn the symmetry of a system, reducing the effective degrees of freedom.

\begin{table}[htbp]
    \caption{\label{tab:num-learnable-params}Numbers of trainable parameters in the CNNs.}
    \begin{ruledtabular}
    \begin{tabular}{lr}
        Model name&Number of trainable parameters\\
        \hline
        DSC/MS&141,298\\
        $C_4$-equivariant DSC/MS&18,564\\
        \hline
        RRDN&1,256,770\\
        $C_4$-equivariant RRDN&209,552\\
    \end{tabular}
    \end{ruledtabular}
\end{table}

\begin{figure}[htbp]
    \includegraphics[width=8.5cm]{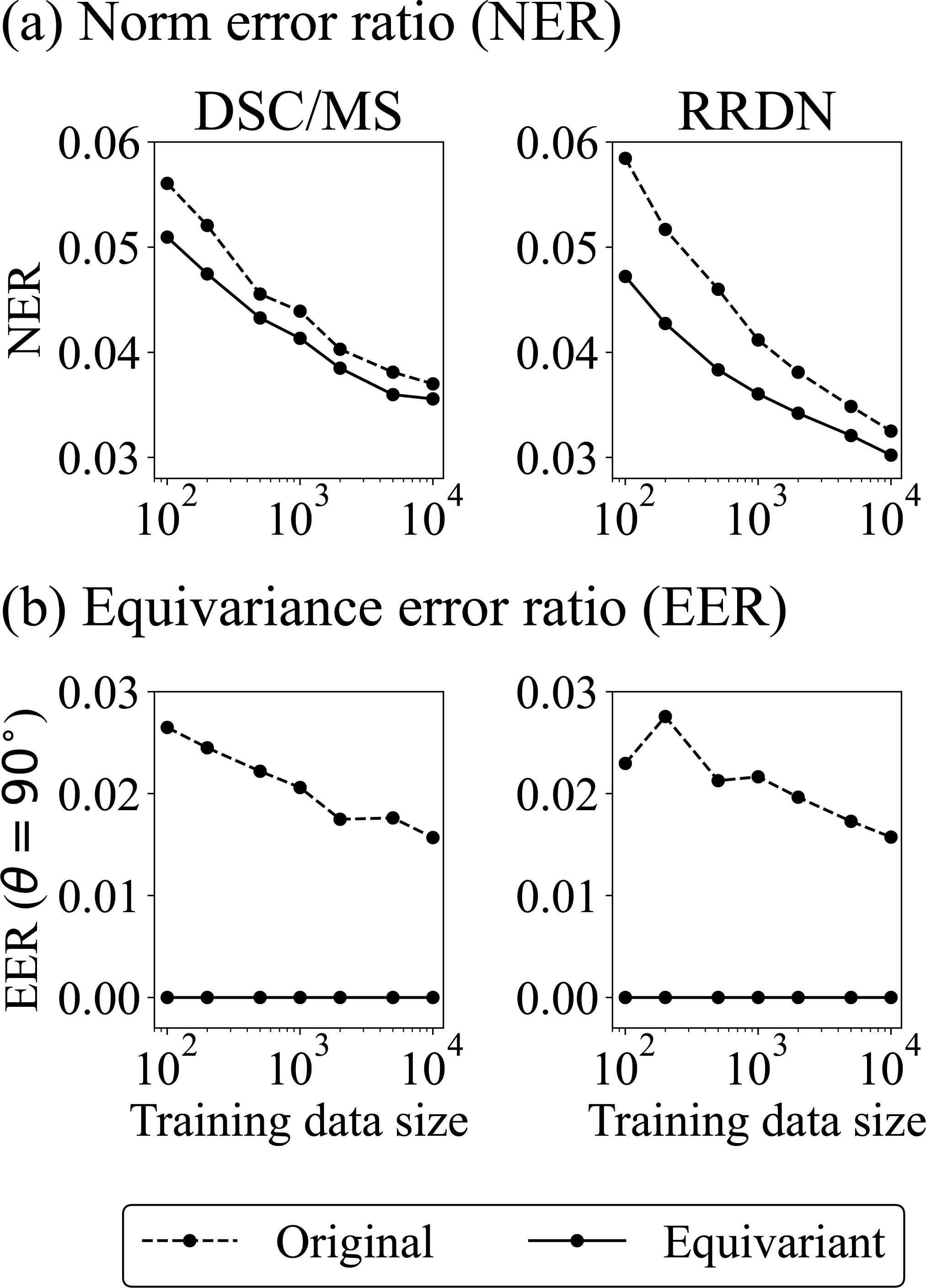}
    \caption{\label{fig:dependency-data-size} (a) Norm error ratios (NERs) and (b) equivariance error ratios (EERs) against the various sizes of the training data in the decaying turbulence. The NER and EER are defined by  (\ref{eq:NER}) and (\ref{eq:EER}), respectively. The EER of the 90-degree rotation was evaluated. The input low-resolution velocity was generated by subsampling with the scale factor $s=9$ in all cases. The training data size means the number of snapshots used in optimizing the CNNs, where 30\% of the data was used for the validation.}
\end{figure}

When the training data size was decreased, the NER and EER increased for the original CNNs. Figure \ref{fig:dependency-data-size} shows the NERs (\ref{eq:NER}) and EERs (\ref{eq:EER}) at the various sizes of the training data, where the EER at 90 degrees was evaluated. For the DSC/MS and RRDN (dashed lines), the reduction in NER is associated with that in EER. This result implies that the acquisition of the $C_4$-equivariance contributes to reducing the test error or vice versa.

The test error was reduced more by imposing the $C_4$-equivariance, as the training data size became small. This imposition was achieved by sharing the weights in the convolution kernels, resulting in the EER being nearly zero for Eq-DSC/MS and Eq-RRDN [Fig. \ref{fig:dependency-data-size}(b)]. The NERs of the $C_4$-equivariant CNNs (solid lines) are smaller than those of the original (dashed lines) [Fig. \ref{fig:dependency-data-size}(a)]. This difference in NER tends to be larger as the data size decreases. Moreover, the NER difference between Eq-RRDN and RRDN is larger than that between Eq-DSC/MS and DSC/MS. Thus, the conversion to rotationally equivariant networks is more effective for the RRDN than for the DSC/MS. Those results can be explained by the fact that the RRDN has many more trainable parameters than the DSC/MS (Table \ref{tab:num-learnable-params}). The learning of the $C_4$-equivariance means that more kernel weights take similar values along the azimuthal and channel directions; or equivalently, the kernel pattern becomes similar to the pattern from the weight sharing of the $C_4$-equivariant kernels.\cite{Weiler2019} This learning process may require more data as a CNN becomes deeper, i.e., an increase in trainable parameters. The imposition of rotational equivariance can eliminate the need for such a large dataset. The previous studies\cite{Weiler2018, Weiler2019, Siddani2021} reported that rotationally equivariant CNNs are trainable and show higher accuracy than ordinary CNNs when the training dataset is small. Our result is consistent with theirs; in addition, we reveal that such equivariant CNNs may be effective when the original CNNs are deep.

\subsection{\label{subsec:results-barotropic-instability}Barotropic instability experiment}

This subsection discusses the barotropic instability experiment. The results are similar to those of the decaying turbulence in the previous subsection. The main difference stems from the 180-degree rotational symmetry in the periodic channel. The $C_2$-equivariant CNNs were compared with the original ones. The LR velocity obtained by subsampling is super-resolved here by employing the RRDN and Eq-RRDN. Similar results were obtained for the LR velocity by average pooling. We also confirmed that the results are not strongly dependent on the network architecture, namely the RRDN or DSC/MS.

\begin{figure*}[thbp]
    \includegraphics[width=17cm]{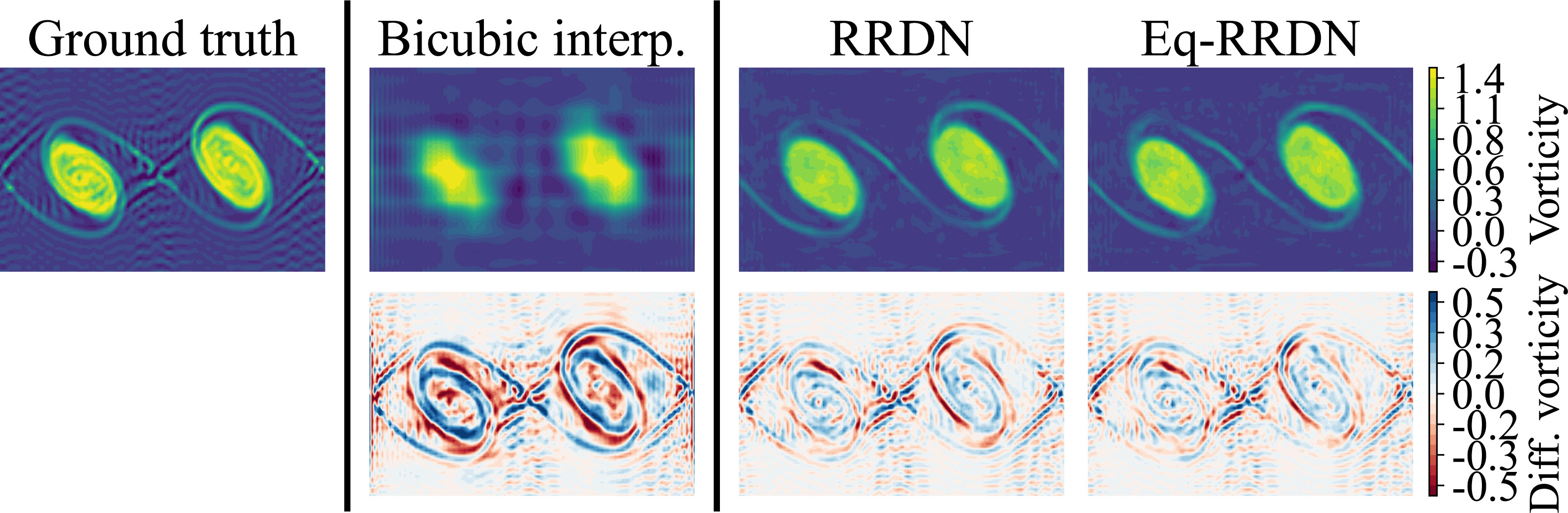}
    \caption{\label{fig:sr-barotropic-instability} An example of the vorticity calculated from the super-resolved velocity in the barotropic instability. The input low-resolution velocity was generated by subsampling with the scale factor $s=9$. ``Bicubic interp.'' stands for bicubic interpolation. The bottom row shows the difference from the ground truth.}
\end{figure*}

\begin{figure*}[htbp]
    \includegraphics[width=8.5cm]{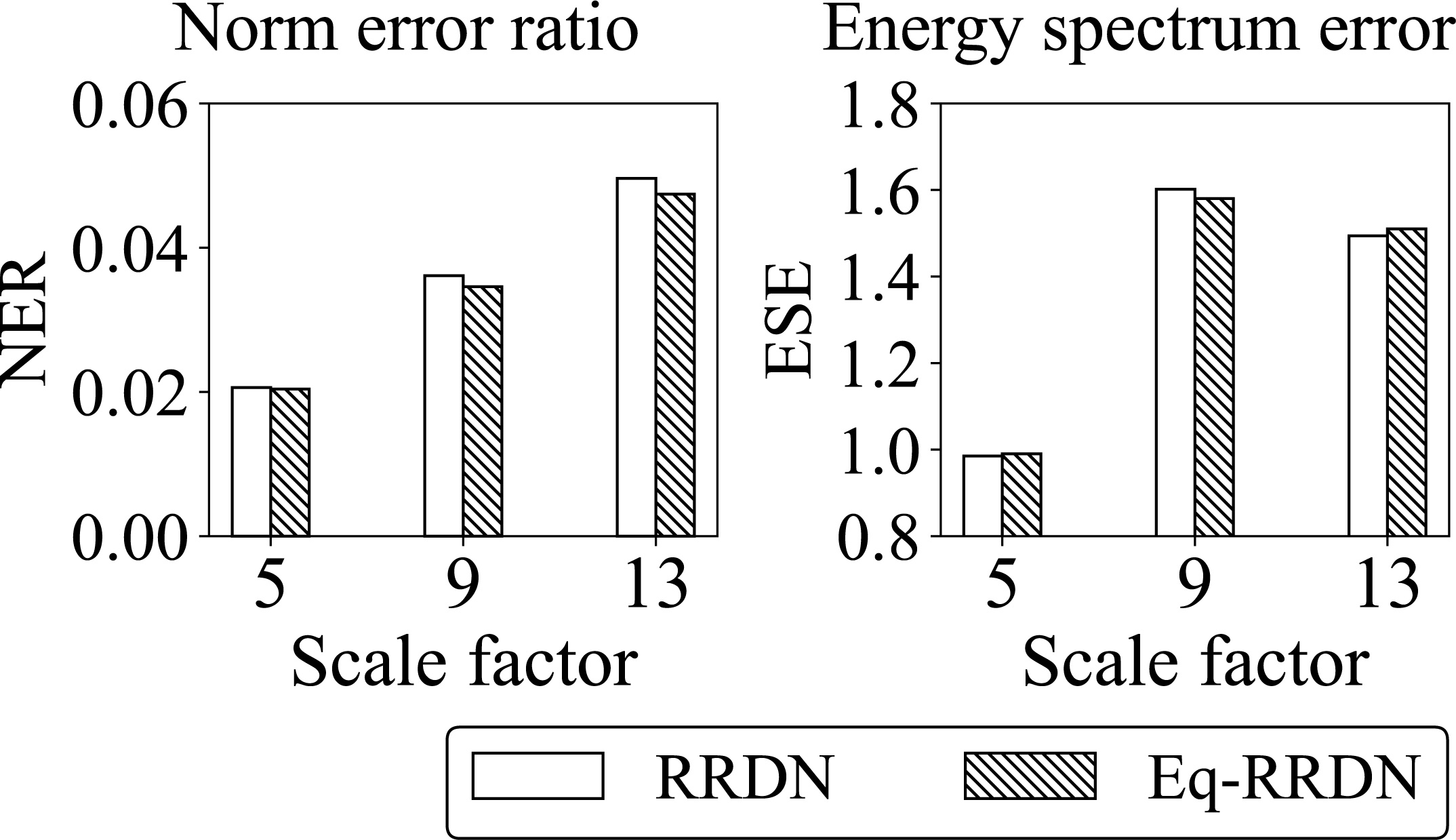}
    \caption{\label{fig:test-errors-barotropic-instability} Test errors of the super-resolved velocity in the barotropic instability. The input low-resolution velocity was generated by subsampling. The norm error ratio (NER) and energy spectrum error (ESE) are defined by (\ref{eq:NER}) and (\ref{eq:ESE}), respectively.}
\end{figure*}

The Eq-RRDN reproduced HR velocity fields similar to those of RRDN. Figure \ref{fig:sr-barotropic-instability} shows an example of the vorticity calculated from the super-resolved velocity when $s=9$. The vortices obtained by bicubic interpolation have indistinct shapes. In contrast, the Eq-RRDN and RRDN accurately reproduce the change in the vortex shape associated with the collapse of the shear flow. Figure \ref{fig:test-errors-barotropic-instability} shows the NER (\ref{eq:NER}) and ESE (\ref{eq:ESE}) at $s = 5$, $9$, and $13$. At each $s$, both types of errors are comparable between the RRDN and Eq-RRDN.

\begin{figure*}[htbp]
    \includegraphics[width=17cm]{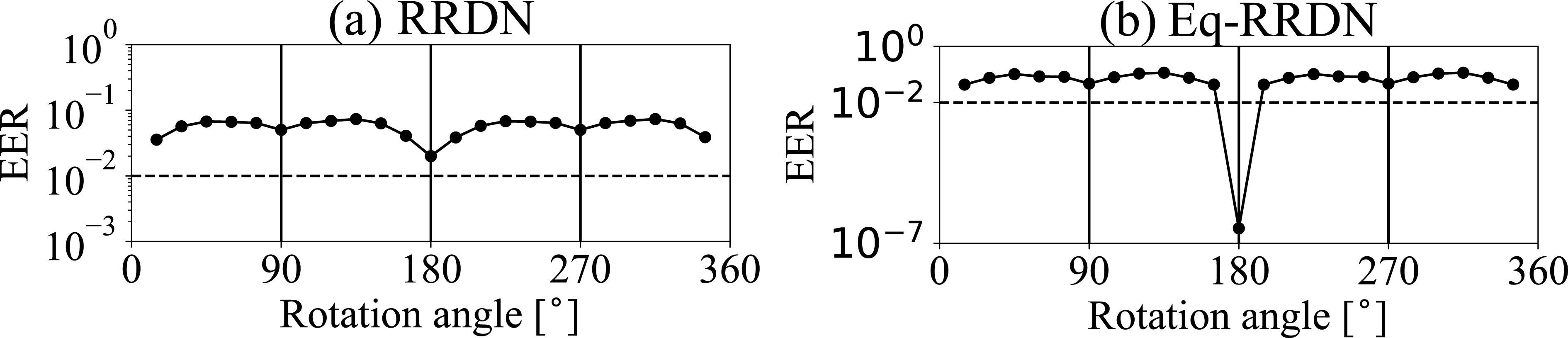}
    \caption{\label{fig:equiv-errors-barotropic-instability} Equivariance error ratio (EER) against the rotation angle in the barotropic instability. The EER is defined by (\ref{eq:EER}). The input low-resolution velocity was generated by subsampling with scale factor $s=9$. The rotation angle intervals are 15${}^{\circ}$, where 0${}^{\circ}$ and 360${}^{\circ}$ are omitted. The horizontal dashed lines show $EER=10^{-2}$.}
\end{figure*}

The RRDN acquired rotational equivariance through training. Figure \ref{fig:equiv-errors-barotropic-instability} shows the EERs (\ref{eq:EER}) against the various rotation angles when $s=9$. The Eq-RRDN shows a relatively small EER ($\sim 10^{-7}$) at 180 degrees due to the imposed $C_2$-equivariance [Fig. \ref{fig:equiv-errors-barotropic-instability}(b)]. The EER of RRDN is the smallest ($\sim 10^{-2}$) at 180 degrees [Fig. \ref{fig:equiv-errors-barotropic-instability}(a)]. The result suggests that the RRDN has learned the $C_2$-equivariance from the rotationally consistent dataset, and this consistency may reflect the 180-degree rotational symmetry of the periodic channel (Sections \ref{subsec:data-consistency} and \ref{subsec:symmetry-data-consistency}).

In addition, Fig. \ref{fig:equiv-errors-barotropic-instability} implies that the Eq-RRDN and RRDN have learned the local equivariance, which may reflect the symmetry for local rotations. Compared with the other angles, the EERs are smaller not only at 180 degrees but also at 90 and 270 degrees. The rotations of 90, 180, and 270 degrees do not require spatial interpolation or extrapolation of velocity fields. This fact means that the EERs at those angles are solely due to the equivariance of an SR model. The rotational equivariance learned by the SR model may reflect some symmetry of the fluid system. However, the small EER at 90 or 270 degrees is not attributable to the global symmetry because the periodic channel is not rotationally symmetric at both angles. A possible explanation is that the local equivariance may reflect the symmetry with respect to local rotation in the channel flow.

\subsection{\label{sec:discussions} Demonstration of breaking of rotational consistency in dataset}

This subsection examines the example proposed in Section \ref{subsec:example-data-inconsistency}: the rotational consistency in a dataset can be broken by subsampling with an even scale factor $s$. In this case, CNNs do not learn rotational equivariance (Section \ref{subsec:data-consistency}). Further, if rotational equivariance is forcefully imposed on an SR model, its accuracy is undermined. To demonstrate these points, the decaying turbulence experiment is investigated by using the RRDN and Eq-RRDN. Similar results were obtained in the barotropic instability experiment. We also confirmed that the results are not strongly dependent on the network architecture, namely the RRDN or DSC/MS.

\begin{figure*}[htbp]
    \includegraphics[width=17cm]{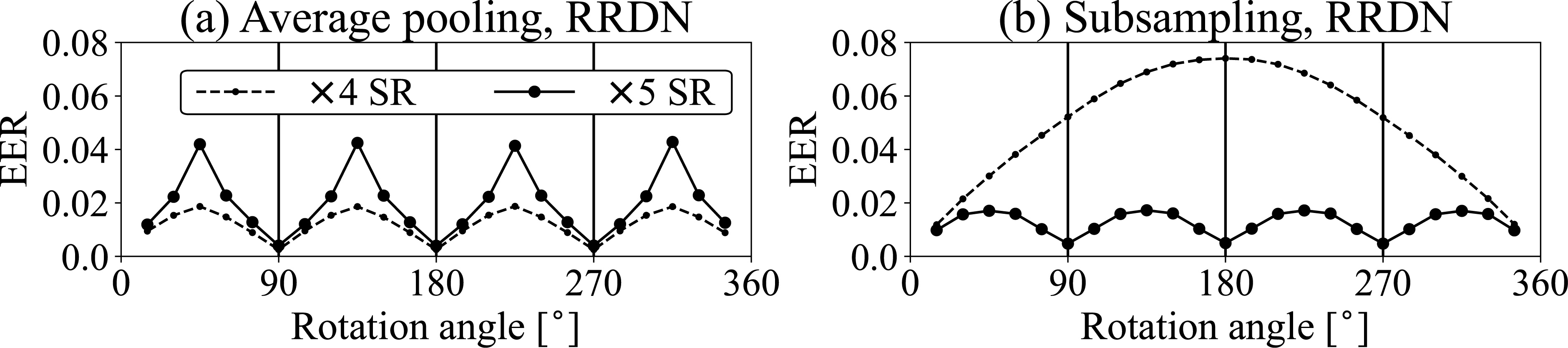}
    \caption{\label{fig:equiv-errors-decaying-turbulence-x8-x9} Equivariance error ratio (EER) against the rotation angle in the decaying turbulence. The EER is defined by (\ref{eq:EER}). The low-resolution velocity was generated with (a) the average pooling and (b) the subsampling, where the scale factor $s=4$ (dashed lines) or $5$ (solid lines). The rotation angle intervals are 15${}^{\circ}$, where 0${}^{\circ}$ and 360${}^{\circ}$ are omitted.}
\end{figure*}

Figure \ref{fig:equiv-errors-decaying-turbulence-x8-x9} shows the curves of EER (\ref{eq:EER}) when $s=4$ or $5$. In the average pooling [Fig. \ref{fig:equiv-errors-decaying-turbulence-x8-x9}(a)], the EER is small at the multiples of 90 degrees regardless of $s$. The average pooling always commutes with the 90-degree rotation and the training dataset is rotationally consistent, which indicates that the RRDN has learned the $C_4$-equivariance. In the subsampling [Fig. \ref{fig:equiv-errors-decaying-turbulence-x8-x9}(b)], when $s=5$ the EER curve is similar to that of average pooling, while it is not when $s=4$; that is, the EER is the largest at 180 degrees. This result demonstrates that the subsampling of an even $s$ breaks the rotational consistency in datasets and CNNs do not learn rotational equivariance.
 
\begin{figure*}[htbp]
    \includegraphics[width=17cm]{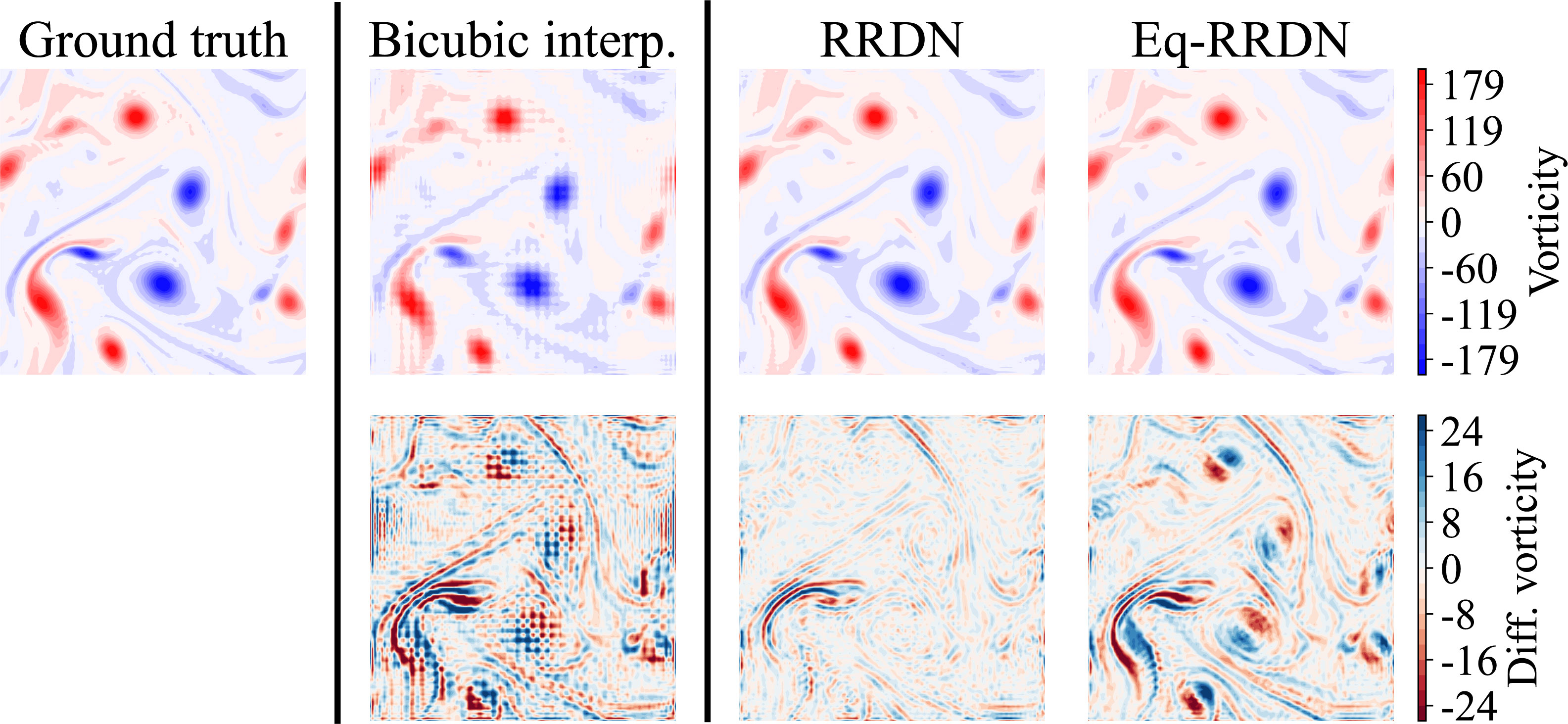}
    \caption{\label{fig:sr-decaying-turbulence-x4} An example of the vorticity calculated from the super-resolved velocity in the decaying turbulence. The input low-resolution velocity was generated by subsampling with the scale factor $s=4$. ``Bicubic interp.'' stands for bicubic interpolation. The bottom row shows the difference from the ground truth.}
\end{figure*}

The accuracy of Eq-RRDN was undermined owing to the imposed $C_4$-equivariance when the dataset was not rotationally consistent. Figure \ref{fig:sr-decaying-turbulence-x4} shows an example of the vorticity calculated from the super-resolved velocity. Here, the input LR velocity was generated by subsampling with $s=4$. The vorticity field of Eq-RRDN shows larger errors than that of RRDN, and the error magnitude is close to that of bicubic interpolation. This result is different from that shown in Fig. \ref{fig:sr-decaying-turbulence}, where the vorticity fields of all CNNs exhibit smaller errors than that of the bicubic interpolation.

\begin{figure}[htbp]
    \includegraphics[width=8.5cm]{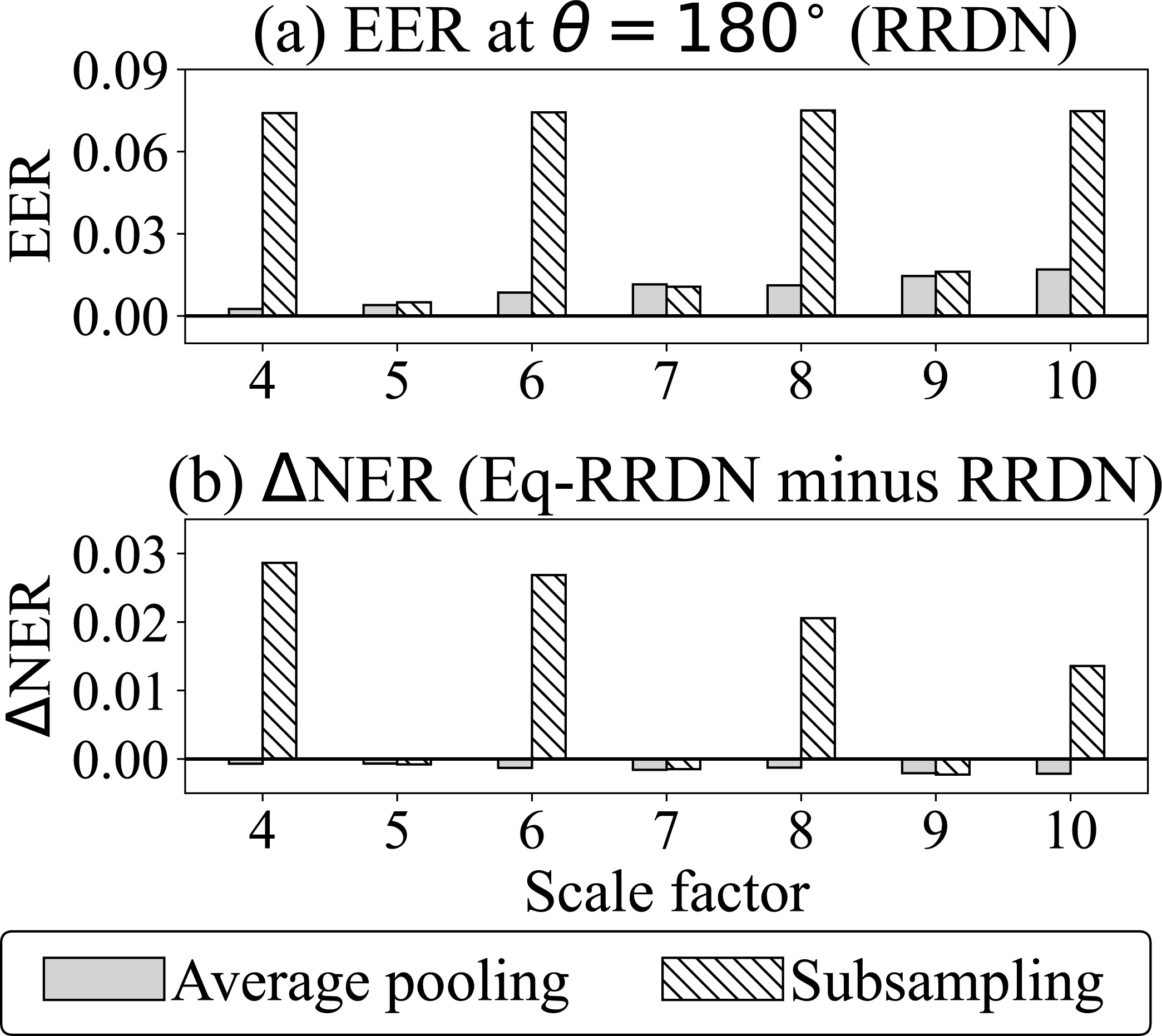}
    \caption{\label{fig:test-errors-decaying-turbulence-x8-x9} Test and equivariance errors for the Eq-RRDN and RRDN at the various scale factors in the decaying turbulence experiment: (a) the difference in NER (norm error ratio) between the Eq-RRDN and RRDN and (b) the EER (equivariance error ration) of RRDN. The NER and EER are defined by (\ref{eq:NER}) and (\ref{eq:EER}), respectively.}
\end{figure}

Similar results were observed at other scale factors $s$. Figure \ref{fig:test-errors-decaying-turbulence-x8-x9} shows the EER and the difference in NER, $\Delta NER$, for $s$ between 4 and 10. Here, $\Delta NER$ is defined as the NER of Eq-RRDN minus that of RRDN. In the average pooling (gray), the EERs of RRDN are on the order of $10^{-2}$ [Fig. \ref{fig:test-errors-decaying-turbulence-x8-x9}(a)], suggesting that the RRDN has learned the $C_4$-equivariance regardless of the parity of $s$. Further, $\Delta NER$ is always quite small [Fig. \ref{fig:test-errors-decaying-turbulence-x8-x9}(b)], which is likely due to the RRDN being effectively $C_4$-equivariant. In the subsampling (hatched), the EER of RRDN is small when $s$ is odd as in the average pooling, while it is large when $s$ is even. This result suggests that in the latter case, the RRDN did not learn the $C_4$-equivariance. At the even $s$, the large $\Delta NER$ is found, where the error of Eq-RRDN was increased by forcefully imposing the $C_4$-equivariance.

\section{\label{sec:conclusions}Conclusions}

This study investigated the super-resolution (SR) of velocity fields in two-dimensional fluids from the viewpoint of rotational equivariance. The equivariance\cite{Weiler2019, Bronstein2021} is generally defined as the property in which a function (e.g., an SR model) commutes with certain transformations (e.g., rotation).

We have theoretically discussed relationships among equivariance, covariance, and symmetry. The covariance in geometry\cite{Schutz1980, Nakahara2003} generally describes the transformation laws of tensors under certain coordinate transformations. A vector is distinguished from a tuple of scalars through the difference in covariance. The equivariance of SR models is a sufficient condition for the covariance of super-resolved velocity vector fields. The rotational consistency in datasets is newly introduced as the invariance of pairs of velocity in high and low resolution (HR and LR). The rotational consistency in datasets is necessary and sufficient for SR models to acquire rotational equivariance through training when the datasets are sufficiently large. Furthermore, it is pointed out that under the following two conditions, the rotational symmetry of fluid systems leads to the rotational consistency of datasets: (i) HR velocity fields are directly generated from fluid experiments or simulations, and (ii) LR velocity fields are generated from the HR ones by an operation $p$ that commutes with rotation.

To demonstrate the above theoretical suggestions, this study performed SR with two existing convolutional neural networks (CNNs): the hybrid downsampled skip-connection/multi-scale model (DSC/MS)\cite{Fukami2019} and the residual in residual dense network (RRDN).\cite{Bode2021} By using two types of CNNs, we confirmed that the present results are not strongly dependent on the network architecture. The rotationally equivariant DSC/MS and RRDN were obtained by replacing each layer with its equivariant counterpart.\cite{Weiler2019}  All CNNs were trained with pairs of HR and LR velocities. The HR velocity data were generated from the two-dimensional fluid dynamics simulations, whereas the LR velocity data were obtained as the subsampling or the average pooling of the HR velocity.

The rotationally equivariant CNNs exhibited higher accuracy in SR than the original CNNs (i.e., DSC/MS and RRDN) when the training data size was small. For the large dataset, the accuracy of the original CNNs was comparable to that of the rotationally equivariant CNNs. This is explained by the fact that the original CNNs acquired rotational equivariance through training. The learning of equivariance may have reduced the effective degrees of freedom of CNNs. A deeper CNN may need more training data to acquire rotational equivariance. The need for such a large dataset can be eliminated by imposing the rotational equivariance as prior knowledge.

Even if a fluid system has rotational symmetry, this symmetry may not carry over to pairs of LR and HR velocities, and the dataset is not rotationally consistent. We have confirmed this hypothesis by comparing the methods that generate LR velocity, namely subsampling and average pooling. In subsampling at even intervals, the patterns of LR velocity depend on the orientation of HR velocity, breaking the rotational consistency in the dataset. In this case, the original CNNs did not learn rotational equivariance. Further, the accuracy in SR was reduced by forcefully imposing rotational equivariance on the CNNs. Therefore, when a rotationally equivariant CNN is used in SR, it is necessary to verify the rotational symmetry of the fluid system and the rotational consistency in the dataset.

There are at least three directions for future work. The first is the investigation of other equivariance in fluid SR, such as the Galilean invariance\cite{Ling2016, Wang2021} or scale equivariance.\cite{Sosnovik2021, Wang2021} The second theme is the condition that characterizes relationships between fluid systems and datasets in terms of symmetry. The above two conditions (i) and (ii) may be too strict. For instance, the proposed conditions cannot be applied to datasets where the LR and HR flows are generated from separate fluid simulations. It will be necessary to relax the conditions for more realistic applications. The third involves SR on curved manifolds such as a sphere. The present paper utilizes the Cartesian coordinates on the flat plane, where the basis is identical at all locations. Generally, basis vectors depend on positions on manifolds, and the convolution must incorporate the difference in those bases.\cite{Cohen2019IcosahedralGauge, Haan2021gauge} Flows on a sphere such as the atmosphere may be suitable for examining the fluid SR on curved manifolds.

\begin{acknowledgments}
This work was supported by the JSPS KAKENHI (Grant Number 20H05751). This work used computational resources TSUBAME3.0 supercomputer provided by Tokyo Institute of Technology through the HPCI System Research Project (Project ID: hp220102).
\end{acknowledgments}

\section*{Data Availability Statement}

The data that support the findings of this study are available within the article. The source code is available on the GitHub repository.\cite{Yasuda2022GH}

\appendix

\section{\label{sec-app:theory-CNN} Review on a theory of equivariant convolution}

\subsection{\label{subsec:se2} Special Euclidean group $SE(2)$}

We briefly review a theory on the convolution equivariant to translation and rotation.\cite{Weiler2018, Weiler2019, Cohen2019} This subsection introduces the combination of translation and rotation, namely the special Euclidean group $SE(2)$.

The translation is first formulated for a position vector $\bm{x} = (x, y)^{\mathrm{T}}$:
\begin{eqnarray}
\mathscr{T}_{\bm{t}} \bullet \bm{x} &=& \bm{x} + \bm{t}\;, \label{eq:def-translation} \\
\end{eqnarray}
where $\bm{t}$ is a translation vector. An element of $SE(2)$, $\mathscr{G}_{\theta,\bm{t}}$ acts on $\bm{x}$ as follows:
\begin{equation}
   \mathscr{G}_{\theta,\bm{t}} \bullet\bm{x} = \left(\mathscr{T}_{\bm{t}} \mathscr{R}_{\theta}\right) \bullet \bm{x} = R_{\theta} \bm{x} + \bm{t}\;. \label{eq:def-translation-rotation}
\end{equation}

The covariance\cite{Schutz1980, Nakahara2003} with respect to $SE(2)$ is described for a scalar field $\omega(\bm{x})$ and a vector field $\bm{v}(\bm{x})$ in the following:
\begin{subequations}
\label{eq:se2-covariance-whole}
\begin{eqnarray}
    \mathscr{G}_{\theta,\bm{t}} \bullet \omega(\bm{x}) &=& \omega(R^{-1}_{\theta} (\bm{x} - \bm{t}))\;, \label{eq:se2-covariance-scalar} \\
    \mathscr{G}_{\theta,\bm{t}} \bullet \bm{v}(\bm{x}) &=& R_{\theta} \bm{v}(R^{-1}_{\theta} (\bm{x} - \bm{t}))\;, \label{eq:se2-covariance-vector} \\
    \mathscr{G}_{\theta,\bm{t}} \bullet F(\bm{x}) &=& Q_\theta F(R^{-1}_{\theta} (\bm{x} - \bm{t}))\;. \label{eq:se2-covariance-general-tensor}
\end{eqnarray}
\end{subequations}
As in (\ref{eq:covariance-whole}), only the referred position is changed for $\omega(\bm{x})$ in (\ref{eq:se2-covariance-scalar}), whereas both the referred position and vector components are changed for $\bm{v}(\bm{x})$ in (\ref{eq:se2-covariance-vector}). Equation (\ref{eq:se2-covariance-general-tensor}) is used in general cases, where the form of the matrix $Q_\theta$ is determined by the type of $F$. For instance, when $F$ is a stack of vectors, $Q_{\theta}$ is a block diagonal matrix consisting of $R_\theta$. When $F$ is a tuple of scalars such as colors, $Q_{\theta}$ becomes unity.

\subsection{\label{subsec:linear-constraint-equivariance}Linear constraint on equivariant kernels of convolution}

The equivariance is achieved by sharing the weights in convolution kernels.\cite{Weiler2018, Weiler2019, Cohen2019} The pattern of weight sharing is obtained by solving the linear constraint described below.

The convolution is defined as
\begin{equation}
    F^{(\rm out)}(\bm{y}) = \kappa \ast F^{(\rm in)} = \int \kappa(\bm{y} - \bm{x}) F^{(\rm in)}(\bm{x}) \;{\rm d}\bm{x}\;, \label{eq:def-convolution}
\end{equation}
where $\ast$ stands for the convolution, input $F^{(\rm in)}$ and output $F^{(\rm out)}$ are tensors generally having different ranks, $\kappa$ is a kernel, and the integral is performed over the continuous Euclidean plane without boundary. The multiplication between $\kappa$ and $F^{(\rm in)}$ is generally a matrix operation. The tensors $F^{(\rm in)}$ and $F^{(\rm out)}$ are transformed by an element of $SE(2)$ as in Eq. (\ref{eq:se2-covariance-general-tensor}):
\begin{subequations}
\label{eq:transformation-Fin-Fout}
\begin{eqnarray}
    \mathscr{G}_{\theta,\bm{t}} \bullet F^{(\rm in)}(\bm{x}) &=& Q_{\theta}^{(\rm in)} F^{(\rm in)}(R^{-1}_{\theta}(\bm{x} - \bm{t}))\;,\\
    \mathscr{G}_{\theta,\bm{t}} \bullet F^{(\rm out)}(\bm{x}) &=& Q_{\theta}^{(\rm out)} F^{(\rm out)}(R^{-1}_{\theta}(\bm{x} - \bm{t}))\;.
\end{eqnarray}
\end{subequations}

The covariance (\ref{eq:transformation-Fin-Fout}) implies the equivariance of the convolution (\ref{eq:def-convolution}). The sufficient condition for the equivariance is the following linear constraint on $\kappa$:\cite{Weiler2018, Weiler2019, Cohen2019}
\begin{equation}
    Q_{\theta}^{(\rm out)} \kappa(\bm{x}) = \kappa(R_\theta \bm{x} ) Q_{\theta}^{(\rm in)}\;. \label{eq:linear-constraint-equivariance}
\end{equation}
Following Ref.~\onlinecite{Weiler2019}, the equivariance is confirmed:
\begin{eqnarray}
    \kappa \ast \left( \mathscr{G}_{\theta,\bm{t}} \bullet F^{(\rm in)} \right) &=& \int \kappa(\bm{y} - \bm{x}) Q_{\theta}^{(\rm in)} F^{(\rm in)}(\underbrace{R^{-1}_{\theta}(\bm{x}-\bm{t})}_{\bm{x^{\prime}}}) \;{\rm d}\bm{x}\;, \nonumber \\
    &=&  \int \kappa(\bm{y} - \bm{t} - R_\theta\bm{x^{\prime}}) Q_\theta^{(\rm in)} F^{(\rm in)}(\bm{x^{\prime}}) \;{\rm d}\bm{x^{\prime}}\;, \nonumber \\
    &=&  \int \underbrace{\kappa(R_{\theta}(R^{-1}_{\theta}(\bm{y} - \bm{t})- \bm{x^{\prime}})) Q^{(\rm in)}}_{Q_\theta^{(\rm out)}\kappa(R^{-1}_{\theta}(\bm{y} - \bm{t})-\bm{x^{\prime}})} F^{(\rm in)}(\bm{x^{\prime}}) \;{\rm d}\bm{x^{\prime}}\;, \nonumber \\
    &=& Q_\theta^{(\rm out)}F^{(\rm out)}(R^{-1}_{\theta}(\bm{y} - \bm{t}))\;, \nonumber \\
    &=& \mathscr{G}_{\theta,\bm{t}} \bullet F^{(\rm out)}\;,\label{eq:confirmation-equiv}
\end{eqnarray}
where the fact that the Jacobian is unity under rotation is exploited. Equation (\ref{eq:confirmation-equiv}) indicates that the output $F^{(\rm out)}$ is transformed according to a transformation of the input $F^{(\rm in)}$, which is the definition of the equivariance (\ref{eq:def-equivariance}). In application, a convolution kernel is given by a linear combination of the solution of (\ref{eq:linear-constraint-equivariance}).

A few important mathematical results are mentioned here. The constraint (\ref{eq:linear-constraint-equivariance}) is not only sufficient but also necessary for the equivariance of convolution.\cite{Weiler2018} Moreover, a linear map is equivariant if and only if the map is the equivariant convolution.\cite{Weiler2018} In the derivation of (\ref{eq:linear-constraint-equivariance}), the two-dimensionality is not exploited; hence, the same constraint can be derived in the three-dimensional Euclidean space.\cite{Weiler2018} The convolution satisfying (\ref{eq:linear-constraint-equivariance}) is equivariant not only to the global rotation but also to local rotation;\cite{Weiler2019} this is known as the gauge transformation.\cite{Cohen2019IcosahedralGauge}

\subsection{\label{subsec:ex1-equivariance}Example 1: Trivial representation with spatially varying kernel}

The linear constraint (\ref{eq:linear-constraint-equivariance}) is interpreted using two simple examples. We discuss rotation in these examples, while the kernel satisfying (\ref{eq:linear-constraint-equivariance}) is equivariant to translation and rotation as in (\ref{eq:confirmation-equiv}).

The effect of changes in position is discussed first. Consider that the input and output are scalars. Scalar has a trivial representation;\cite{Gilmore2008, Zee2016} i.e., all $Q_{\theta}$ such as in (\ref{eq:transformation-Fin-Fout}) become unity. Consequently, the constraint (\ref{eq:linear-constraint-equivariance}) becomes
\begin{equation}
    \kappa(\bm{x}) = \kappa(R_\theta \bm{x} )\;, \label{eq:linear-constraint-equivariance-simple2}
\end{equation}
where $\kappa(\bm{x})$ is a real number depending on $\bm{x}$ and acts on a scalar field.

The rotation in multiples of 120 degrees is discussed as an example. A pattern having the wavenumber-3 structure such as $\cos(3\phi)$ is a solution to (\ref{eq:linear-constraint-equivariance-simple2}), where $\phi$ is an azimuthal angle. Such a kernel is invariant to the rotation, implying that the output is rotated according to a rotation of the input. The radial structure is not determined by (\ref{eq:linear-constraint-equivariance-simple2}); usually, a certain form such as Gaussian is assumed, and its amplitude is optimized by training.\cite{Worrall2017, Weiler2019}

There is a limitation to expressive power due to the rotational equivariance. For the conventional convolution, a kernel can learn an azimuthal pattern from data, whereas it is not necessarily equivariant to rotation. This is in contrast to the equivariant kernels: the azimuthal pattern is determined by the constraint (\ref{eq:linear-constraint-equivariance-simple2}), which guarantees the rotational equivariance. This example indicates that the weights of equivariant kernels are shared along the azimuthal direction, and this weight sharing reduces the degrees of freedom of kernels.

\subsection{\label{subsec-app:example-regular-repr}Example 2: regular representation with point-wise kernel}

The effect of $Q_{\theta}$ in (\ref{eq:linear-constraint-equivariance}) is discussed next. Consider that the convolution is point-wise; that is, the kernel size is $1\times1$ pixel if the space is discrete. The input and output are assumed to be the same feature type. Equation (\ref{eq:linear-constraint-equivariance}) is then simplified as follows:
\begin{equation}
    Q_{\theta} \kappa = \kappa Q_{\theta}\;,\label{eq:linear-constraint-equivariance-simple1}
\end{equation}
where the superscripts of $(\rm out)$ and $(\rm in)$ are omitted and the matrix $\kappa$ spans the channel dimension.

\begin{figure}[htbp]
    \includegraphics[width=8.5cm]{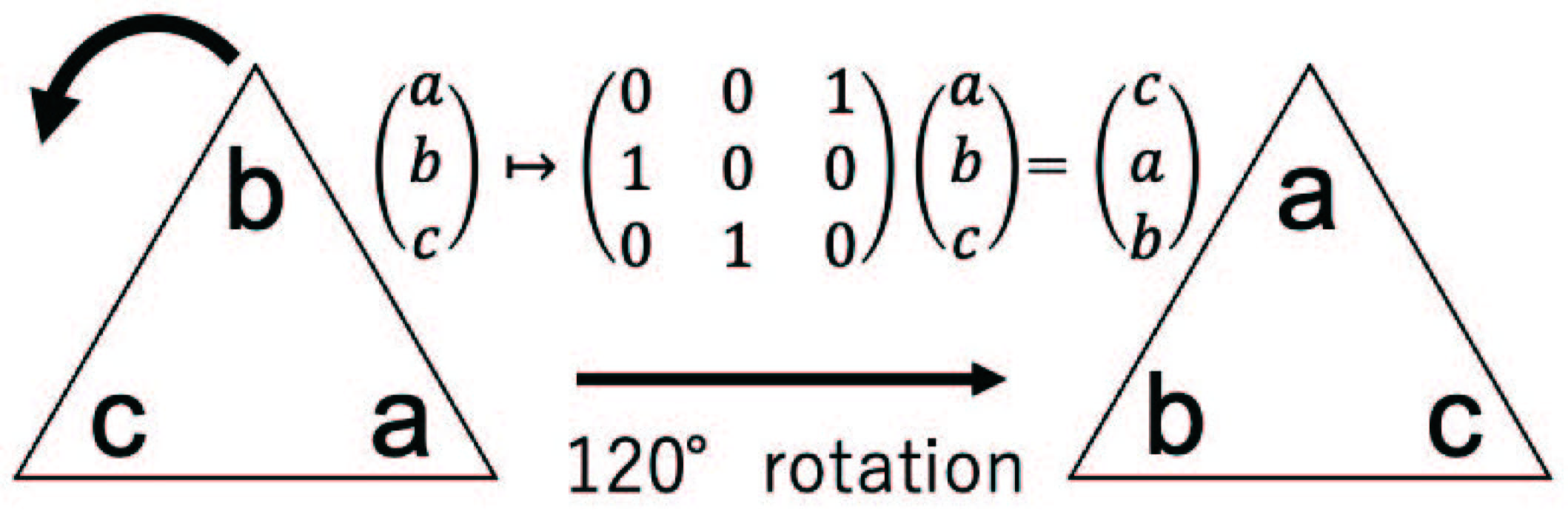}
    \caption{\label{fig:schematic-regular-representation} Schematic describing the action of the $120$-degree rotation on a feature set $(a,b,c)^{\rm T}$ with the regular representation. The matrix of $Q_{120^{\circ}}$ in (\ref{eq:regular-representation-120-rot}) is used in the figure.}
\end{figure}

The rotation in multiples of 120 degrees is discussed again. The regular representation\cite{Gilmore2008, Zee2016} is employed here, where the rotation is formulated as a permutation. Figure \ref{fig:schematic-regular-representation} is a schematic describing the action of the 120-degree rotation on a feature set. Each feature is identified as a vertex of an equilateral triangle. The rotation is regarded as the permutation of these vertices. Mathematically, the feature set in Fig. \ref{fig:schematic-regular-representation}, $(a,b,c)^{\rm T}$, is not a geometric vector; rather, it is regarded as a linearly transformed vector. The set $(a,b,c)^{\rm T}$ can contain complex numbers, also implying that it is not a geometric vector. Ref.~\onlinecite{Weiler2019} discusses the difference between the real and complex representations. In the regular representation, all rotations are described with the following three matrices and combinations thereof:
\begin{subequations}
\label{eq:regular-representation-c3}
\begin{eqnarray}
    Q_{0^{\circ}} &=& \begin{pmatrix}
                            1 & 0 & 0 \\
                            0 & 1 & 0 \\
                            0 & 0 & 1 \\
                        \end{pmatrix}\;, \\
    Q_{120^{\circ}} &=& \begin{pmatrix}
                            0 & 0 & 1 \\
                            1 & 0 & 0 \\
                            0 & 1 & 0 \\
                        \end{pmatrix}\;, \label{eq:regular-representation-120-rot} \\
    Q_{240^{\circ}} &=& \begin{pmatrix}
                            0 & 1 & 0 \\
                            0 & 0 & 1 \\
                            1 & 0 & 0 \\
                        \end{pmatrix}\;.
\end{eqnarray}
\end{subequations}

The constraint (\ref{eq:linear-constraint-equivariance-simple1}) must hold for all three matrices in (\ref{eq:regular-representation-c3}). The solution is
\begin{equation}
    \kappa = \begin{pmatrix}
        \lambda_1 & \lambda_2 & \lambda_3 \\
        \lambda_3 & \lambda_1 & \lambda_2 \\
        \lambda_2 & \lambda_3 & \lambda_1 \\
    \end{pmatrix}\;, \label{eq:equivariant-kernel-solution1}\\
\end{equation}
where $\lambda_1$, $\lambda_2$, and $\lambda_3$ are learnable parameters that are determined by training. The kernel (\ref{eq:equivariant-kernel-solution1}) has a regular order of $\lambda_1$, $\lambda_2$, and $\lambda_3$, which makes $\kappa$ commutative with a permutation of rotation.

A limitation to expressive power is found again. The equivariant kernel (\ref{eq:equivariant-kernel-solution1}) has only 3 learnable parameters, although it is a $3\times3$ matrix. In contrast, all 9 parameters are learnable in the conventional convolution, which is not necessarily equivariant to rotation. This example indicates that the weights of equivariant kernels are shared along the channel direction, and this weight-sharing reduces the degrees of freedom of kernels.

\section{\label{subsec-app:dependency-cn} Dependency on the order of cyclic groups}

This section examines the dependency that the rotationally equivariant CNNs have on the order $N$ of the cyclic group $C_N$. The LR velocity generated by subsampling was super-resolved with the scale factor $s=9$ in the decaying turbulence experiment. The test errors did not largely decrease as $N$ was increased, which is in contrast to the result in image classification.\cite{Marcos2017, Bekkers2018, Weiler2019}  The results here suggest that $C_4$ is a good selection for the hyper-parameter $C_N$.

Eq-DSC/MS is first examined. This model has various spatial sizes of kernels in the convolution layers (Fig. \ref{fig:DSC/MS}), all of which were fixed here. In each layer, the number of features was also fixed. For instance, if both input and output of a convolution layer are described with two sets of the regular representation, the number of channels is $16$ for $C_8$ and $24$ for $C_{12}$. Thus, as $N$ is increased, the number of trainable parameters increases.

\begin{figure}[htbp]
    \includegraphics[width=8.5cm]{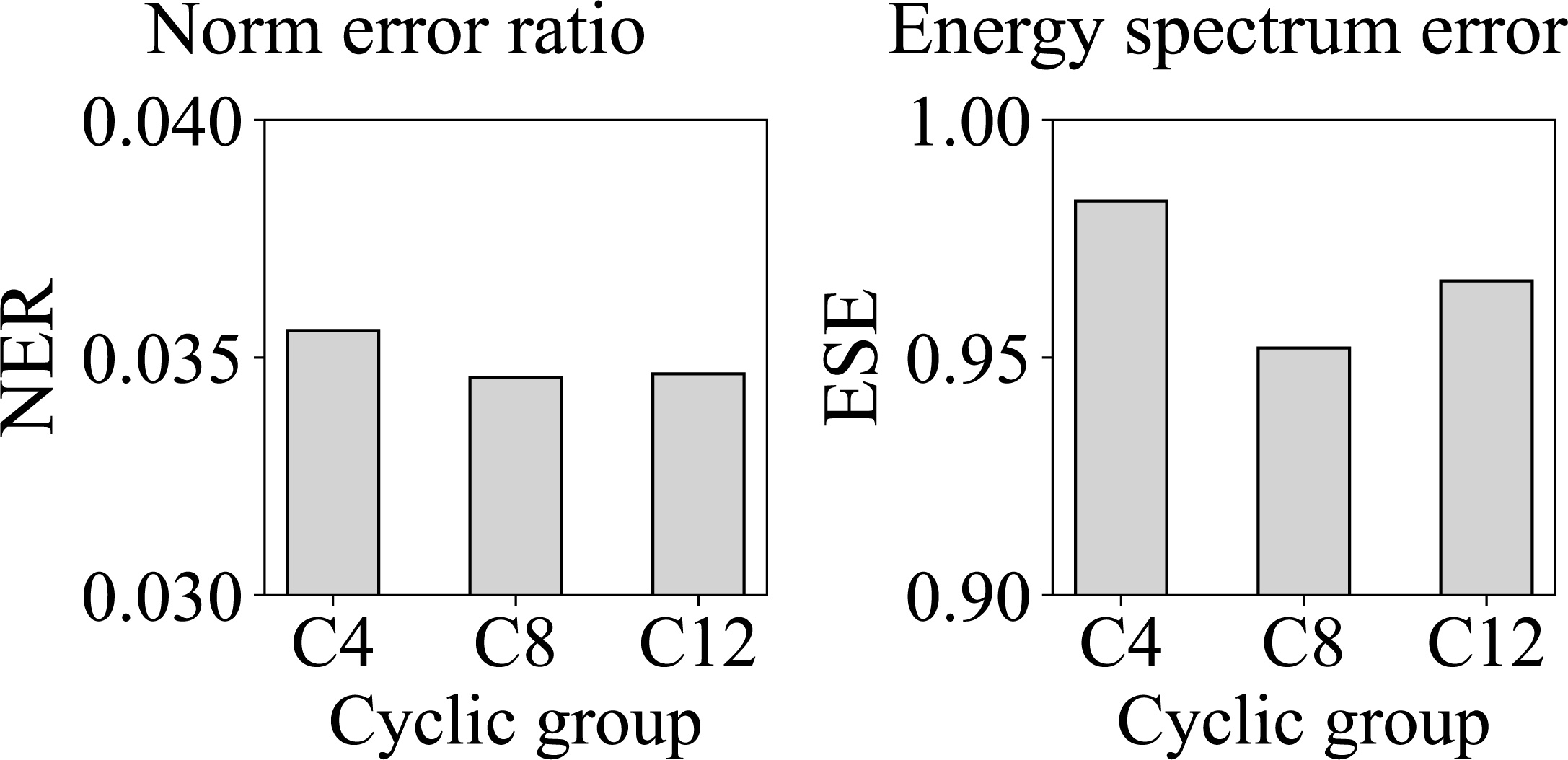}
    \caption{\label{fig:dependency-cn-dscms} Test errors against the orders of cyclic group $C_N$ used in the Eq-DSC/MS. The low-resolution velocity was generated by subsampling with the scale factor $s=9$ from the decaying turbulence experiment. The NER (norm error ratio) and ESE (energy spectrum error) are defined by (\ref{eq:NER}) and (\ref{eq:ESE}), respectively.}
\end{figure}

The test errors did not significantly decrease with an increase in $N$. Figure \ref{fig:dependency-cn-dscms} shows the NER (\ref{eq:NER}) and ESE (\ref{eq:ESE}) of Eq-DSC/MS for $C_4$, $C_8$, and $C_{12}$. Note that $C_4$ is a subgroup of $C_8$ and $C_{12}$. Both NER and ESE are the minima at $C_8$. The reductions in NER and ESE are only approximately 3\% compared with those at $C_4$, even though the number of trainable parameters is approximately doubled from $C_4$ to $C_8$.

The Eq-RRDN is examined next. Generally, the spatial size of kernels needs to be larger when the order of $C_N$ is increased because spatial interpolation is used to determine the equivariant kernel weights.\cite{Worrall2017, Weiler2019} Eq-RRDN is appropriate for an experiment varying both $C_N$ and kernel size because all kernels have the same spatial size (Fig \ref{fig:RRDN}). The Eq-RRDN consisting of one dense block (Fig. \ref{fig:RRDN}) was employed here to reduce the training cost. Each convolution layer utilized eight sets of the regular representation. For instance, the number of channels is 32 for $C_4$ and 96 for $C_{12}$. Thus, as $N$ is increased, the number of trainable parameters increases.

\begin{figure}[htbp]
    \includegraphics[width=8.5cm]{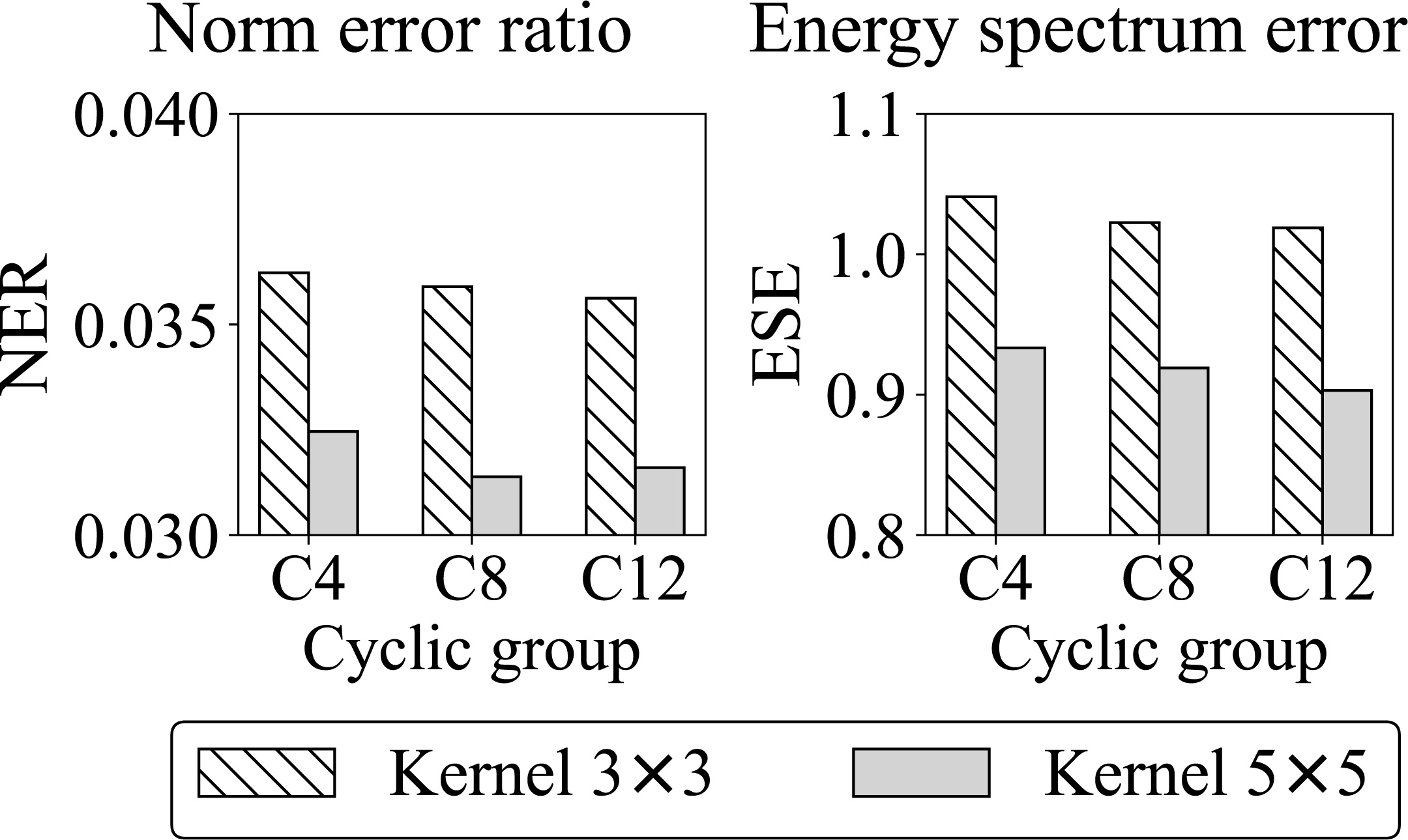}
    \caption{\label{fig:dependency-cn-rrdn} Test errors against the various $C_N$ and kernel sizes used in the Eq-RRDN. The low-resolution velocity was generated by subsampling with the scale factor $s=9$ from the decaying turbulence experiment. The NER (norm error ratio) and ESE (energy spectrum error) are defined by (\ref{eq:NER}) and (\ref{eq:ESE}), respectively.}
\end{figure}

Again, the test errors did not significantly decrease with an increase in $N$. Figure \ref{fig:dependency-cn-rrdn} shows the NERs (\ref{eq:NER}) and ESEs (\ref{eq:ESE}) of Eq-RRDN against the various $C_N$ and spatial sizes of kernels. The error reduction achieved by increasing $N$ is not as strong as that achieved by increasing the kernel size. A similar error reduction was observed for the RRDN, suggesting that the error reduction realized by increasing the kernel size was not due to the imposed equivariance. A comparison between $C_4$ and $C_{8}$ in the $3\times3$ kernel (Fig. \ref{fig:dependency-cn-rrdn}) shows that the NER and ESE decrease by only approximately 2\%, even though the number of parameters is doubled from $C_4$ to $C_{8}$. The $C_4$-equivariance may be optimal in the decaying turbulence experiment from the viewpoint of the balance between the number of trainable parameters and the magnitude of test errors.

\bibliography{references}

\begin{thebibliography}{96}%
\makeatletter
\providecommand \@ifxundefined [1]{%
 \@ifx{#1\undefined}
}%
\providecommand \@ifnum [1]{%
 \ifnum #1\expandafter \@firstoftwo
 \else \expandafter \@secondoftwo
 \fi
}%
\providecommand \@ifx [1]{%
 \ifx #1\expandafter \@firstoftwo
 \else \expandafter \@secondoftwo
 \fi
}%
\providecommand \natexlab [1]{#1}%
\providecommand \enquote  [1]{``#1''}%
\providecommand \bibnamefont  [1]{#1}%
\providecommand \bibfnamefont [1]{#1}%
\providecommand \citenamefont [1]{#1}%
\providecommand \href@noop [0]{\@secondoftwo}%
\providecommand \href [0]{\begingroup \@sanitize@url \@href}%
\providecommand \@href[1]{\@@startlink{#1}\@@href}%
\providecommand \@@href[1]{\endgroup#1\@@endlink}%
\providecommand \@sanitize@url [0]{\catcode `\\12\catcode `\$12\catcode
  `\&12\catcode `\#12\catcode `\^12\catcode `\_12\catcode `\%12\relax}%
\providecommand \@@startlink[1]{}%
\providecommand \@@endlink[0]{}%
\providecommand \url  [0]{\begingroup\@sanitize@url \@url }%
\providecommand \@url [1]{\endgroup\@href {#1}{\urlprefix }}%
\providecommand \urlprefix  [0]{URL }%
\providecommand \Eprint [0]{\href }%
\providecommand \doibase [0]{http://dx.doi.org/}%
\providecommand \selectlanguage [0]{\@gobble}%
\providecommand \bibinfo  [0]{\@secondoftwo}%
\providecommand \bibfield  [0]{\@secondoftwo}%
\providecommand \translation [1]{[#1]}%
\providecommand \BibitemOpen [0]{}%
\providecommand \bibitemStop [0]{}%
\providecommand \bibitemNoStop [0]{.\EOS\space}%
\providecommand \EOS [0]{\spacefactor3000\relax}%
\providecommand \BibitemShut  [1]{\csname bibitem#1\endcsname}%
\let\auto@bib@innerbib\@empty
\bibitem [{\citenamefont {Brunton}, \citenamefont {Noack},\ and\ \citenamefont
  {Koumoutsakos}(2020)}]{Brunton2020}%
  \BibitemOpen
  \bibfield  {author} {\bibinfo {author} {\bibfnamefont {S.~L.}\ \bibnamefont
  {Brunton}}, \bibinfo {author} {\bibfnamefont {B.~R.}\ \bibnamefont {Noack}},
  \ and\ \bibinfo {author} {\bibfnamefont {P.}~\bibnamefont {Koumoutsakos}},\
  }\bibfield  {title} {\enquote {\bibinfo {title} {Machine learning for fluid
  mechanics},}\ }\href {\doibase 10.1146/annurev-fluid-010719-060214}
  {\bibfield  {journal} {\bibinfo  {journal} {Annual Review of Fluid
  Mechanics}\ }\textbf {\bibinfo {volume} {52}},\ \bibinfo {pages} {477--508}
  (\bibinfo {year} {2020})},\ \Eprint
  {http://arxiv.org/abs/https://doi.org/10.1146/annurev-fluid-010719-060214}
  {https://doi.org/10.1146/annurev-fluid-010719-060214} \BibitemShut {NoStop}%
\bibitem [{\citenamefont {Duraisamy}(2021)}]{Duraisamy2021}%
  \BibitemOpen
  \bibfield  {author} {\bibinfo {author} {\bibfnamefont {K.}~\bibnamefont
  {Duraisamy}},\ }\bibfield  {title} {\enquote {\bibinfo {title} {Perspectives
  on machine learning-augmented reynolds-averaged and large eddy simulation
  models of turbulence},}\ }\href {\doibase 10.1103/PhysRevFluids.6.050504}
  {\bibfield  {journal} {\bibinfo  {journal} {Phys. Rev. Fluids}\ }\textbf
  {\bibinfo {volume} {6}},\ \bibinfo {pages} {050504} (\bibinfo {year}
  {2021})}\BibitemShut {NoStop}%
\bibitem [{\citenamefont {Vinuesa}\ and\ \citenamefont
  {Brunton}(2021)}]{Vinuesa2021}%
  \BibitemOpen
  \bibfield  {author} {\bibinfo {author} {\bibfnamefont {R.}~\bibnamefont
  {Vinuesa}}\ and\ \bibinfo {author} {\bibfnamefont {S.~L.}\ \bibnamefont
  {Brunton}},\ }\href@noop {} {\enquote {\bibinfo {title} {The potential of
  machine learning to enhance computational fluid dynamics},}\ } (\bibinfo
  {year} {2021}),\ \Eprint {http://arxiv.org/abs/2110.02085} {arXiv:2110.02085
  [physics.flu-dyn]} \BibitemShut {NoStop}%
\bibitem [{\citenamefont {Brunton}(2022)}]{Brunton2022}%
  \BibitemOpen
  \bibfield  {author} {\bibinfo {author} {\bibfnamefont {S.~L.}\ \bibnamefont
  {Brunton}},\ }\bibfield  {title} {\enquote {\bibinfo {title} {Applying
  machine learning to study fluid mechanics},}\ }\href {\doibase
  10.1007/s10409-021-01143-6} {\bibfield  {journal} {\bibinfo  {journal} {Acta
  Mechanica Sinica}\ } (\bibinfo {year} {2022}),\
  10.1007/s10409-021-01143-6}\BibitemShut {NoStop}%
\bibitem [{\citenamefont {Raissi}, \citenamefont {Perdikaris},\ and\
  \citenamefont {Karniadakis}(2019)}]{Raissi2019}%
  \BibitemOpen
  \bibfield  {author} {\bibinfo {author} {\bibfnamefont {M.}~\bibnamefont
  {Raissi}}, \bibinfo {author} {\bibfnamefont {P.}~\bibnamefont {Perdikaris}},
  \ and\ \bibinfo {author} {\bibfnamefont {G.}~\bibnamefont {Karniadakis}},\
  }\bibfield  {title} {\enquote {\bibinfo {title} {Physics-informed neural
  networks: A deep learning framework for solving forward and inverse problems
  involving nonlinear partial differential equations},}\ }\href {\doibase
  https://doi.org/10.1016/j.jcp.2018.10.045} {\bibfield  {journal} {\bibinfo
  {journal} {Journal of Computational Physics}\ }\textbf {\bibinfo {volume}
  {378}},\ \bibinfo {pages} {686--707} (\bibinfo {year} {2019})}\BibitemShut
  {NoStop}%
\bibitem [{\citenamefont {Kashinath}\ \emph {et~al.}(2021)\citenamefont
  {Kashinath}, \citenamefont {Mustafa}, \citenamefont {Albert}, \citenamefont
  {Wu}, \citenamefont {Jiang}, \citenamefont {Esmaeilzadeh}, \citenamefont
  {Azizzadenesheli}, \citenamefont {Wang}, \citenamefont {Chattopadhyay},
  \citenamefont {Singh}, \citenamefont {Manepalli}, \citenamefont {Chirila},
  \citenamefont {Yu}, \citenamefont {Walters}, \citenamefont {White},
  \citenamefont {Xiao}, \citenamefont {Tchelepi}, \citenamefont {Marcus},
  \citenamefont {Anandkumar}, \citenamefont {Hassanzadeh},\ and\ \citenamefont
  {Prabhat}}]{Kashinath2021}%
  \BibitemOpen
  \bibfield  {author} {\bibinfo {author} {\bibfnamefont {K.}~\bibnamefont
  {Kashinath}}, \bibinfo {author} {\bibfnamefont {M.}~\bibnamefont {Mustafa}},
  \bibinfo {author} {\bibfnamefont {A.}~\bibnamefont {Albert}}, \bibinfo
  {author} {\bibfnamefont {J.-L.}\ \bibnamefont {Wu}}, \bibinfo {author}
  {\bibfnamefont {C.}~\bibnamefont {Jiang}}, \bibinfo {author} {\bibfnamefont
  {S.}~\bibnamefont {Esmaeilzadeh}}, \bibinfo {author} {\bibfnamefont
  {K.}~\bibnamefont {Azizzadenesheli}}, \bibinfo {author} {\bibfnamefont
  {R.}~\bibnamefont {Wang}}, \bibinfo {author} {\bibfnamefont {A.}~\bibnamefont
  {Chattopadhyay}}, \bibinfo {author} {\bibfnamefont {A.}~\bibnamefont
  {Singh}}, \bibinfo {author} {\bibfnamefont {A.}~\bibnamefont {Manepalli}},
  \bibinfo {author} {\bibfnamefont {D.}~\bibnamefont {Chirila}}, \bibinfo
  {author} {\bibfnamefont {R.}~\bibnamefont {Yu}}, \bibinfo {author}
  {\bibfnamefont {R.}~\bibnamefont {Walters}}, \bibinfo {author} {\bibfnamefont
  {B.}~\bibnamefont {White}}, \bibinfo {author} {\bibfnamefont
  {H.}~\bibnamefont {Xiao}}, \bibinfo {author} {\bibfnamefont {H.~A.}\
  \bibnamefont {Tchelepi}}, \bibinfo {author} {\bibfnamefont {P.}~\bibnamefont
  {Marcus}}, \bibinfo {author} {\bibfnamefont {A.}~\bibnamefont {Anandkumar}},
  \bibinfo {author} {\bibfnamefont {P.}~\bibnamefont {Hassanzadeh}}, \ and\
  \bibinfo {author} {\bibfnamefont {n.}~\bibnamefont {Prabhat}},\ }\bibfield
  {title} {\enquote {\bibinfo {title} {Physics-informed machine learning: case
  studies for weather and climate modelling},}\ }\href {\doibase
  10.1098/rsta.2020.0093} {\bibfield  {journal} {\bibinfo  {journal}
  {Philosophical Transactions of the Royal Society A: Mathematical, Physical
  and Engineering Sciences}\ }\textbf {\bibinfo {volume} {379}},\ \bibinfo
  {pages} {20200093} (\bibinfo {year} {2021})},\ \Eprint
  {http://arxiv.org/abs/https://royalsocietypublishing.org/doi/pdf/10.1098/rsta.2020.0093}
  {https://royalsocietypublishing.org/doi/pdf/10.1098/rsta.2020.0093}
  \BibitemShut {NoStop}%
\bibitem [{\citenamefont {Cai}\ \emph {et~al.}(2022)\citenamefont {Cai},
  \citenamefont {Mao}, \citenamefont {Wang}, \citenamefont {Yin}, \citenamefont
  {George},\ and\ \citenamefont {Karniadakis}}]{Cai2022}%
  \BibitemOpen
  \bibfield  {author} {\bibinfo {author} {\bibfnamefont {S.}~\bibnamefont
  {Cai}}, \bibinfo {author} {\bibfnamefont {Z.}~\bibnamefont {Mao}}, \bibinfo
  {author} {\bibfnamefont {Z.}~\bibnamefont {Wang}}, \bibinfo {author}
  {\bibfnamefont {M.}~\bibnamefont {Yin}}, \bibinfo {author} {\bibnamefont
  {George}}, \ and\ \bibinfo {author} {\bibfnamefont {E.}~\bibnamefont
  {Karniadakis}},\ }\bibfield  {title} {\enquote {\bibinfo {title}
  {Physics-informed neural networks (pinns) for fluid mechanics: a review},}\
  }\href {\doibase 10.1007/s10409-021-01148-1} {\bibfield  {journal} {\bibinfo
  {journal} {Acta Mechanica Sinica}\ } (\bibinfo {year} {2022}),\
  10.1007/s10409-021-01148-1}\BibitemShut {NoStop}%
\bibitem [{\citenamefont {Schutz}(1980)}]{Schutz1980}%
  \BibitemOpen
  \bibfield  {author} {\bibinfo {author} {\bibfnamefont {B.~F.}\ \bibnamefont
  {Schutz}},\ }\href@noop {} {\emph {\bibinfo {title} {Geometrical Methods of
  Mathematical Physics}}},\ \bibinfo {edition} {first edition}\ ed.\ (\bibinfo
  {publisher} {Cambridge University Press},\ \bibinfo {year}
  {1980})\BibitemShut {NoStop}%
\bibitem [{\citenamefont {Nakahara}(2003)}]{Nakahara2003}%
  \BibitemOpen
  \bibfield  {author} {\bibinfo {author} {\bibfnamefont {M.}~\bibnamefont
  {Nakahara}},\ }\href@noop {} {\emph {\bibinfo {title} {Geometry, Topology and
  Physics}}},\ \bibinfo {edition} {second edition}\ ed.\ (\bibinfo  {publisher}
  {CRC Press},\ \bibinfo {year} {2003})\BibitemShut {NoStop}%
\bibitem [{\citenamefont {Dirac}(1996)}]{Dirac1996}%
  \BibitemOpen
  \bibfield  {author} {\bibinfo {author} {\bibfnamefont {P.~A.~M.}\
  \bibnamefont {Dirac}},\ }\href@noop {} {\emph {\bibinfo {title} {General
  Theory of Relativity}}}\ (\bibinfo  {publisher} {Princeton University
  Press},\ \bibinfo {year} {1996})\BibitemShut {NoStop}%
\bibitem [{\citenamefont {Dong}\ \emph {et~al.}(2014)\citenamefont {Dong},
  \citenamefont {Loy}, \citenamefont {He},\ and\ \citenamefont
  {Tang}}]{Dong2014}%
  \BibitemOpen
  \bibfield  {author} {\bibinfo {author} {\bibfnamefont {C.}~\bibnamefont
  {Dong}}, \bibinfo {author} {\bibfnamefont {C.~C.}\ \bibnamefont {Loy}},
  \bibinfo {author} {\bibfnamefont {K.}~\bibnamefont {He}}, \ and\ \bibinfo
  {author} {\bibfnamefont {X.}~\bibnamefont {Tang}},\ }\bibfield  {title}
  {\enquote {\bibinfo {title} {Learning a deep convolutional network for image
  super-resolution},}\ }in\ \href@noop {} {\emph {\bibinfo {booktitle}
  {Computer Vision -- ECCV 2014}}},\ \bibinfo {editor} {edited by\ \bibinfo
  {editor} {\bibfnamefont {D.}~\bibnamefont {Fleet}}, \bibinfo {editor}
  {\bibfnamefont {T.}~\bibnamefont {Pajdla}}, \bibinfo {editor} {\bibfnamefont
  {B.}~\bibnamefont {Schiele}}, \ and\ \bibinfo {editor} {\bibfnamefont
  {T.}~\bibnamefont {Tuytelaars}}}\ (\bibinfo  {publisher} {Springer
  International Publishing},\ \bibinfo {address} {Cham},\ \bibinfo {year}
  {2014})\ pp.\ \bibinfo {pages} {184--199}\BibitemShut {NoStop}%
\bibitem [{\citenamefont {Ledig}\ \emph {et~al.}(2017)\citenamefont {Ledig},
  \citenamefont {Theis}, \citenamefont {Huszar}, \citenamefont {Caballero},
  \citenamefont {Cunningham}, \citenamefont {Acosta}, \citenamefont {Aitken},
  \citenamefont {Tejani}, \citenamefont {Totz}, \citenamefont {Wang},\ and\
  \citenamefont {Shi}}]{Ledig2017SRGAN}%
  \BibitemOpen
  \bibfield  {author} {\bibinfo {author} {\bibfnamefont {C.}~\bibnamefont
  {Ledig}}, \bibinfo {author} {\bibfnamefont {L.}~\bibnamefont {Theis}},
  \bibinfo {author} {\bibfnamefont {F.}~\bibnamefont {Huszar}}, \bibinfo
  {author} {\bibfnamefont {J.}~\bibnamefont {Caballero}}, \bibinfo {author}
  {\bibfnamefont {A.}~\bibnamefont {Cunningham}}, \bibinfo {author}
  {\bibfnamefont {A.}~\bibnamefont {Acosta}}, \bibinfo {author} {\bibfnamefont
  {A.}~\bibnamefont {Aitken}}, \bibinfo {author} {\bibfnamefont
  {A.}~\bibnamefont {Tejani}}, \bibinfo {author} {\bibfnamefont
  {J.}~\bibnamefont {Totz}}, \bibinfo {author} {\bibfnamefont {Z.}~\bibnamefont
  {Wang}}, \ and\ \bibinfo {author} {\bibfnamefont {W.}~\bibnamefont {Shi}},\
  }\bibfield  {title} {\enquote {\bibinfo {title} {Photo-realistic single image
  super-resolution using a generative adversarial network},}\ }in\ \href@noop
  {} {\emph {\bibinfo {booktitle} {Proceedings of the IEEE Conference on
  Computer Vision and Pattern Recognition (CVPR)}}}\ (\bibinfo {year}
  {2017})\BibitemShut {NoStop}%
\bibitem [{\citenamefont {Wang}\ \emph {et~al.}(2018)\citenamefont {Wang},
  \citenamefont {Yu}, \citenamefont {Wu}, \citenamefont {Gu}, \citenamefont
  {Liu}, \citenamefont {Dong}, \citenamefont {Qiao},\ and\ \citenamefont
  {Change~Loy}}]{Wang2018ESRGAN}%
  \BibitemOpen
  \bibfield  {author} {\bibinfo {author} {\bibfnamefont {X.}~\bibnamefont
  {Wang}}, \bibinfo {author} {\bibfnamefont {K.}~\bibnamefont {Yu}}, \bibinfo
  {author} {\bibfnamefont {S.}~\bibnamefont {Wu}}, \bibinfo {author}
  {\bibfnamefont {J.}~\bibnamefont {Gu}}, \bibinfo {author} {\bibfnamefont
  {Y.}~\bibnamefont {Liu}}, \bibinfo {author} {\bibfnamefont {C.}~\bibnamefont
  {Dong}}, \bibinfo {author} {\bibfnamefont {Y.}~\bibnamefont {Qiao}}, \ and\
  \bibinfo {author} {\bibfnamefont {C.}~\bibnamefont {Change~Loy}},\ }\bibfield
   {title} {\enquote {\bibinfo {title} {Esrgan: Enhanced super-resolution
  generative adversarial networks},}\ }in\ \href@noop {} {\emph {\bibinfo
  {booktitle} {Proceedings of the European Conference on Computer Vision (ECCV)
  Workshops}}}\ (\bibinfo {year} {2018})\BibitemShut {NoStop}%
\bibitem [{\citenamefont {Ha}\ \emph {et~al.}(2019)\citenamefont {Ha},
  \citenamefont {Ren}, \citenamefont {Xu}, \citenamefont {Zhao}, \citenamefont
  {Xie}, \citenamefont {Masero},\ and\ \citenamefont {Hussain}}]{Ha2019}%
  \BibitemOpen
  \bibfield  {author} {\bibinfo {author} {\bibfnamefont {V.~K.}\ \bibnamefont
  {Ha}}, \bibinfo {author} {\bibfnamefont {J.-C.}\ \bibnamefont {Ren}},
  \bibinfo {author} {\bibfnamefont {X.-Y.}\ \bibnamefont {Xu}}, \bibinfo
  {author} {\bibfnamefont {S.}~\bibnamefont {Zhao}}, \bibinfo {author}
  {\bibfnamefont {G.}~\bibnamefont {Xie}}, \bibinfo {author} {\bibfnamefont
  {V.}~\bibnamefont {Masero}}, \ and\ \bibinfo {author} {\bibfnamefont
  {A.}~\bibnamefont {Hussain}},\ }\bibfield  {title} {\enquote {\bibinfo
  {title} {Deep learning based single image super-resolution: A survey},}\
  }\href {\doibase 10.1007/s11633-019-1183-x} {\bibfield  {journal} {\bibinfo
  {journal} {International Journal of Automation and Computing}\ } (\bibinfo
  {year} {2019}),\ 10.1007/s11633-019-1183-x}\BibitemShut {NoStop}%
\bibitem [{\citenamefont {Anwar}, \citenamefont {Khan},\ and\ \citenamefont
  {Barnes}(2020)}]{Anwar2020}%
  \BibitemOpen
  \bibfield  {author} {\bibinfo {author} {\bibfnamefont {S.}~\bibnamefont
  {Anwar}}, \bibinfo {author} {\bibfnamefont {S.}~\bibnamefont {Khan}}, \ and\
  \bibinfo {author} {\bibfnamefont {N.}~\bibnamefont {Barnes}},\ }\bibfield
  {title} {\enquote {\bibinfo {title} {A deep journey into super-resolution: A
  survey},}\ }\href {\doibase 10.1145/3390462} {\bibfield  {journal} {\bibinfo
  {journal} {ACM Comput. Surv.}\ }\textbf {\bibinfo {volume} {53}} (\bibinfo
  {year} {2020}),\ 10.1145/3390462}\BibitemShut {NoStop}%
\bibitem [{\citenamefont {Deng}\ \emph {et~al.}(2019)\citenamefont {Deng},
  \citenamefont {He}, \citenamefont {Liu},\ and\ \citenamefont
  {Kim}}]{Deng2019}%
  \BibitemOpen
  \bibfield  {author} {\bibinfo {author} {\bibfnamefont {Z.}~\bibnamefont
  {Deng}}, \bibinfo {author} {\bibfnamefont {C.}~\bibnamefont {He}}, \bibinfo
  {author} {\bibfnamefont {Y.}~\bibnamefont {Liu}}, \ and\ \bibinfo {author}
  {\bibfnamefont {K.~C.}\ \bibnamefont {Kim}},\ }\bibfield  {title} {\enquote
  {\bibinfo {title} {Super-resolution reconstruction of turbulent velocity
  fields using a generative adversarial network-based artificial intelligence
  framework},}\ }\href {\doibase 10.1063/1.5127031} {\bibfield  {journal}
  {\bibinfo  {journal} {Physics of Fluids}\ }\textbf {\bibinfo {volume} {31}}
  (\bibinfo {year} {2019}),\ 10.1063/1.5127031}\BibitemShut {NoStop}%
\bibitem [{\citenamefont {Fukami}, \citenamefont {Fukagata},\ and\
  \citenamefont {Taira}(2019)}]{Fukami2019}%
  \BibitemOpen
  \bibfield  {author} {\bibinfo {author} {\bibfnamefont {K.}~\bibnamefont
  {Fukami}}, \bibinfo {author} {\bibfnamefont {K.}~\bibnamefont {Fukagata}}, \
  and\ \bibinfo {author} {\bibfnamefont {K.}~\bibnamefont {Taira}},\ }\bibfield
   {title} {\enquote {\bibinfo {title} {Super-resolution reconstruction of
  turbulent flows with machine learning},}\ }\href {\doibase
  10.1017/jfm.2019.238} {\bibfield  {journal} {\bibinfo  {journal} {Journal of
  Fluid Mechanics}\ }\textbf {\bibinfo {volume} {870}},\ \bibinfo {pages}
  {106--120} (\bibinfo {year} {2019})}\BibitemShut {NoStop}%
\bibitem [{\citenamefont {Maulik}\ \emph {et~al.}(2020)\citenamefont {Maulik},
  \citenamefont {Fukami}, \citenamefont {Ramachandra}, \citenamefont
  {Fukagata},\ and\ \citenamefont {Taira}}]{Maulik2020}%
  \BibitemOpen
  \bibfield  {author} {\bibinfo {author} {\bibfnamefont {R.}~\bibnamefont
  {Maulik}}, \bibinfo {author} {\bibfnamefont {K.}~\bibnamefont {Fukami}},
  \bibinfo {author} {\bibfnamefont {N.}~\bibnamefont {Ramachandra}}, \bibinfo
  {author} {\bibfnamefont {K.}~\bibnamefont {Fukagata}}, \ and\ \bibinfo
  {author} {\bibfnamefont {K.}~\bibnamefont {Taira}},\ }\bibfield  {title}
  {\enquote {\bibinfo {title} {Probabilistic neural networks for fluid flow
  surrogate modeling and data recovery},}\ }\href {\doibase
  10.1103/PhysRevFluids.5.104401} {\bibfield  {journal} {\bibinfo  {journal}
  {Phys. Rev. Fluids}\ }\textbf {\bibinfo {volume} {5}},\ \bibinfo {pages}
  {104401} (\bibinfo {year} {2020})}\BibitemShut {NoStop}%
\bibitem [{\citenamefont {Wang}\ \emph {et~al.}(2020)\citenamefont {Wang},
  \citenamefont {Bentivegna}, \citenamefont {Zhou}, \citenamefont {Klein},\
  and\ \citenamefont {Elmegreen}}]{Wang2020}%
  \BibitemOpen
  \bibfield  {author} {\bibinfo {author} {\bibfnamefont {C.}~\bibnamefont
  {Wang}}, \bibinfo {author} {\bibfnamefont {E.}~\bibnamefont {Bentivegna}},
  \bibinfo {author} {\bibfnamefont {W.}~\bibnamefont {Zhou}}, \bibinfo {author}
  {\bibfnamefont {L.}~\bibnamefont {Klein}}, \ and\ \bibinfo {author}
  {\bibfnamefont {B.}~\bibnamefont {Elmegreen}},\ }\bibfield  {title} {\enquote
  {\bibinfo {title} {Physics-informed neural network super resolution for
  advection-diffusion models},}\ }in\ \href
  {https://ml4physicalsciences.github.io/2020/} {\emph {\bibinfo {booktitle}
  {Third Workshop on Machine Learning and the Physical Sciences (NeurIPS
  2020)}}}\ (\bibinfo {year} {2020})\BibitemShut {NoStop}%
\bibitem [{\citenamefont {Fukami}, \citenamefont {Fukagata},\ and\
  \citenamefont {Taira}(2020)}]{Fukami2020}%
  \BibitemOpen
  \bibfield  {author} {\bibinfo {author} {\bibfnamefont {K.}~\bibnamefont
  {Fukami}}, \bibinfo {author} {\bibfnamefont {K.}~\bibnamefont {Fukagata}}, \
  and\ \bibinfo {author} {\bibfnamefont {K.}~\bibnamefont {Taira}},\ }\bibfield
   {title} {\enquote {\bibinfo {title} {Assessment of supervised machine
  learning methods for fluid flows},}\ }\href {\doibase
  10.1007/s00162-020-00518-y} {\bibfield  {journal} {\bibinfo  {journal}
  {Theoretical and Computational Fluid Dynamics}\ }\textbf {\bibinfo {volume}
  {34}},\ \bibinfo {pages} {497--519} (\bibinfo {year} {2020})}\BibitemShut
  {NoStop}%
\bibitem [{\citenamefont {Fukami}, \citenamefont {Fukagata},\ and\
  \citenamefont {Taira}(2021)}]{Fukami2021}%
  \BibitemOpen
  \bibfield  {author} {\bibinfo {author} {\bibfnamefont {K.}~\bibnamefont
  {Fukami}}, \bibinfo {author} {\bibfnamefont {K.}~\bibnamefont {Fukagata}}, \
  and\ \bibinfo {author} {\bibfnamefont {K.}~\bibnamefont {Taira}},\ }\bibfield
   {title} {\enquote {\bibinfo {title} {Machine-learning-based spatio-temporal
  super resolution reconstruction of turbulent flows},}\ }\href {\doibase
  10.1017/jfm.2020.948} {\bibfield  {journal} {\bibinfo  {journal} {Journal of
  Fluid Mechanics}\ }\textbf {\bibinfo {volume} {909}},\ \bibinfo {pages} {A9}
  (\bibinfo {year} {2021})}\BibitemShut {NoStop}%
\bibitem [{\citenamefont {Wang}\ \emph
  {et~al.}(2021{\natexlab{a}})\citenamefont {Wang}, \citenamefont {Luo},
  \citenamefont {Xu}, \citenamefont {Luo},\ and\ \citenamefont
  {Yuan}}]{Wang2021DSCMS}%
  \BibitemOpen
  \bibfield  {author} {\bibinfo {author} {\bibfnamefont {L.}~\bibnamefont
  {Wang}}, \bibinfo {author} {\bibfnamefont {Z.}~\bibnamefont {Luo}}, \bibinfo
  {author} {\bibfnamefont {J.}~\bibnamefont {Xu}}, \bibinfo {author}
  {\bibfnamefont {W.}~\bibnamefont {Luo}}, \ and\ \bibinfo {author}
  {\bibfnamefont {J.}~\bibnamefont {Yuan}},\ }\bibfield  {title} {\enquote
  {\bibinfo {title} {A novel framework for cost-effectively reconstructing the
  global flow field by super-resolution},}\ }\href {\doibase 10.1063/5.0062775}
  {\bibfield  {journal} {\bibinfo  {journal} {Physics of Fluids}\ }\textbf
  {\bibinfo {volume} {33}},\ \bibinfo {pages} {095105} (\bibinfo {year}
  {2021}{\natexlab{a}})},\ \Eprint
  {http://arxiv.org/abs/https://doi.org/10.1063/5.0062775}
  {https://doi.org/10.1063/5.0062775} \BibitemShut {NoStop}%
\bibitem [{\citenamefont {Liu}\ \emph {et~al.}(2020)\citenamefont {Liu},
  \citenamefont {Tang}, \citenamefont {Huang},\ and\ \citenamefont
  {Lu}}]{Liu2020}%
  \BibitemOpen
  \bibfield  {author} {\bibinfo {author} {\bibfnamefont {B.}~\bibnamefont
  {Liu}}, \bibinfo {author} {\bibfnamefont {J.}~\bibnamefont {Tang}}, \bibinfo
  {author} {\bibfnamefont {H.}~\bibnamefont {Huang}}, \ and\ \bibinfo {author}
  {\bibfnamefont {X.-Y.}\ \bibnamefont {Lu}},\ }\bibfield  {title} {\enquote
  {\bibinfo {title} {Deep learning methods for super-resolution reconstruction
  of turbulent flows},}\ }\href {\doibase 10.1063/1.5140772} {\bibfield
  {journal} {\bibinfo  {journal} {Physics of Fluids}\ }\textbf {\bibinfo
  {volume} {32}},\ \bibinfo {pages} {025105} (\bibinfo {year}
  {2020})}\BibitemShut {NoStop}%
\bibitem [{\citenamefont {Bode}\ \emph {et~al.}(2021)\citenamefont {Bode},
  \citenamefont {Gauding}, \citenamefont {Lian}, \citenamefont {Denker},
  \citenamefont {Davidovic}, \citenamefont {Kleinheinz}, \citenamefont
  {Jitsev},\ and\ \citenamefont {Pitsch}}]{Bode2021}%
  \BibitemOpen
  \bibfield  {author} {\bibinfo {author} {\bibfnamefont {M.}~\bibnamefont
  {Bode}}, \bibinfo {author} {\bibfnamefont {M.}~\bibnamefont {Gauding}},
  \bibinfo {author} {\bibfnamefont {Z.}~\bibnamefont {Lian}}, \bibinfo {author}
  {\bibfnamefont {D.}~\bibnamefont {Denker}}, \bibinfo {author} {\bibfnamefont
  {M.}~\bibnamefont {Davidovic}}, \bibinfo {author} {\bibfnamefont
  {K.}~\bibnamefont {Kleinheinz}}, \bibinfo {author} {\bibfnamefont
  {J.}~\bibnamefont {Jitsev}}, \ and\ \bibinfo {author} {\bibfnamefont
  {H.}~\bibnamefont {Pitsch}},\ }\bibfield  {title} {\enquote {\bibinfo {title}
  {Using physics-informed enhanced super-resolution generative adversarial
  networks for subfilter modeling in turbulent reactive flows},}\ }\href
  {\doibase 10.1016/J.PROCI.2020.06.022} {\bibfield  {journal} {\bibinfo
  {journal} {Proceedings of the Combustion Institute}\ }\textbf {\bibinfo
  {volume} {38}},\ \bibinfo {pages} {2617--2625} (\bibinfo {year}
  {2021})}\BibitemShut {NoStop}%
\bibitem [{\citenamefont {Kim}\ \emph {et~al.}(2021)\citenamefont {Kim},
  \citenamefont {Kim}, \citenamefont {Won},\ and\ \citenamefont
  {Lee}}]{Kim2021}%
  \BibitemOpen
  \bibfield  {author} {\bibinfo {author} {\bibfnamefont {H.}~\bibnamefont
  {Kim}}, \bibinfo {author} {\bibfnamefont {J.}~\bibnamefont {Kim}}, \bibinfo
  {author} {\bibfnamefont {S.}~\bibnamefont {Won}}, \ and\ \bibinfo {author}
  {\bibfnamefont {C.}~\bibnamefont {Lee}},\ }\bibfield  {title} {\enquote
  {\bibinfo {title} {Unsupervised deep learning for super-resolution
  reconstruction of turbulence},}\ }\href {\doibase 10.1017/jfm.2020.1028}
  {\bibfield  {journal} {\bibinfo  {journal} {Journal of Fluid Mechanics}\
  }\textbf {\bibinfo {volume} {910}},\ \bibinfo {pages} {A29} (\bibinfo {year}
  {2021})}\BibitemShut {NoStop}%
\bibitem [{\citenamefont {Bao}\ \emph {et~al.}(2022)\citenamefont {Bao},
  \citenamefont {Chen}, \citenamefont {Johnson}, \citenamefont {Givi},
  \citenamefont {Sammak},\ and\ \citenamefont {Jia}}]{Bao2022physics}%
  \BibitemOpen
  \bibfield  {author} {\bibinfo {author} {\bibfnamefont {T.}~\bibnamefont
  {Bao}}, \bibinfo {author} {\bibfnamefont {S.}~\bibnamefont {Chen}}, \bibinfo
  {author} {\bibfnamefont {T.~T.}\ \bibnamefont {Johnson}}, \bibinfo {author}
  {\bibfnamefont {P.}~\bibnamefont {Givi}}, \bibinfo {author} {\bibfnamefont
  {S.}~\bibnamefont {Sammak}}, \ and\ \bibinfo {author} {\bibfnamefont
  {X.}~\bibnamefont {Jia}},\ }\bibfield  {title} {\enquote {\bibinfo {title}
  {Physics guided neural networks for spatio-temporal super-resolution of
  turbulent flows},}\ }in\ \href {https://openreview.net/forum?id=S98VJL8jcxq}
  {\emph {\bibinfo {booktitle} {The 38th Conference on Uncertainty in
  Artificial Intelligence}}}\ (\bibinfo {year} {2022})\BibitemShut {NoStop}%
\bibitem [{\citenamefont {Jiang}\ \emph {et~al.}(2020)\citenamefont {Jiang},
  \citenamefont {Esmaeilzadeh}, \citenamefont {Azizzadenesheli}, \citenamefont
  {Kashinath}, \citenamefont {Mustafa}, \citenamefont {Tchelepi}, \citenamefont
  {Marcus}, \citenamefont {Prabhat},\ and\ \citenamefont
  {Anandkumar}}]{Jiang2020}%
  \BibitemOpen
  \bibfield  {author} {\bibinfo {author} {\bibfnamefont {C.}~\bibnamefont
  {Jiang}}, \bibinfo {author} {\bibfnamefont {S.}~\bibnamefont {Esmaeilzadeh}},
  \bibinfo {author} {\bibfnamefont {K.}~\bibnamefont {Azizzadenesheli}},
  \bibinfo {author} {\bibfnamefont {K.}~\bibnamefont {Kashinath}}, \bibinfo
  {author} {\bibfnamefont {M.}~\bibnamefont {Mustafa}}, \bibinfo {author}
  {\bibfnamefont {H.~A.}\ \bibnamefont {Tchelepi}}, \bibinfo {author}
  {\bibfnamefont {P.}~\bibnamefont {Marcus}}, \bibinfo {author} {\bibfnamefont
  {M.}~\bibnamefont {Prabhat}}, \ and\ \bibinfo {author} {\bibfnamefont
  {A.}~\bibnamefont {Anandkumar}},\ }\bibfield  {title} {\enquote {\bibinfo
  {title} {Meshfreeflownet: A physics-constrained deep continuous space-time
  super-resolution framework},}\ }in\ \href {\doibase
  10.1109/SC41405.2020.00013} {\emph {\bibinfo {booktitle} {2020 SC20:
  International Conference for High Performance Computing, Networking, Storage
  and Analysis (SC)}}}\ (\bibinfo  {publisher} {IEEE Computer Society},\
  \bibinfo {address} {Los Alamitos, CA, USA},\ \bibinfo {year} {2020})\ pp.\
  \bibinfo {pages} {1--15}\BibitemShut {NoStop}%
\bibitem [{\citenamefont {Xie}\ \emph {et~al.}(2018)\citenamefont {Xie},
  \citenamefont {Franz}, \citenamefont {Chu},\ and\ \citenamefont
  {Thuerey}}]{Xie2018}%
  \BibitemOpen
  \bibfield  {author} {\bibinfo {author} {\bibfnamefont {Y.}~\bibnamefont
  {Xie}}, \bibinfo {author} {\bibfnamefont {E.}~\bibnamefont {Franz}}, \bibinfo
  {author} {\bibfnamefont {M.}~\bibnamefont {Chu}}, \ and\ \bibinfo {author}
  {\bibfnamefont {N.}~\bibnamefont {Thuerey}},\ }\bibfield  {title} {\enquote
  {\bibinfo {title} {Tempogan: A temporally coherent, volumetric gan for
  super-resolution fluid flow},}\ }\href {\doibase 10.1145/3197517.3201304}
  {\bibfield  {journal} {\bibinfo  {journal} {ACM Trans. Graph.}\ }\textbf
  {\bibinfo {volume} {37}} (\bibinfo {year} {2018}),\
  10.1145/3197517.3201304}\BibitemShut {NoStop}%
\bibitem [{\citenamefont {Werhahn}\ \emph {et~al.}(2019)\citenamefont
  {Werhahn}, \citenamefont {Xie}, \citenamefont {Chu},\ and\ \citenamefont
  {Thuerey}}]{Werhahn2019}%
  \BibitemOpen
  \bibfield  {author} {\bibinfo {author} {\bibfnamefont {M.}~\bibnamefont
  {Werhahn}}, \bibinfo {author} {\bibfnamefont {Y.}~\bibnamefont {Xie}},
  \bibinfo {author} {\bibfnamefont {M.}~\bibnamefont {Chu}}, \ and\ \bibinfo
  {author} {\bibfnamefont {N.}~\bibnamefont {Thuerey}},\ }\bibfield  {title}
  {\enquote {\bibinfo {title} {A multi-pass gan for fluid flow
  super-resolution},}\ }\href {\doibase 10.1145/3340251} {\bibfield  {journal}
  {\bibinfo  {journal} {Proc. ACM Comput. Graph. Interact. Tech.}\ }\textbf
  {\bibinfo {volume} {2}} (\bibinfo {year} {2019}),\
  10.1145/3340251}\BibitemShut {NoStop}%
\bibitem [{\citenamefont {Bai}\ \emph {et~al.}(2020)\citenamefont {Bai},
  \citenamefont {Li}, \citenamefont {Desbrun},\ and\ \citenamefont
  {Liu}}]{Bai2020}%
  \BibitemOpen
  \bibfield  {author} {\bibinfo {author} {\bibfnamefont {K.}~\bibnamefont
  {Bai}}, \bibinfo {author} {\bibfnamefont {W.}~\bibnamefont {Li}}, \bibinfo
  {author} {\bibfnamefont {M.}~\bibnamefont {Desbrun}}, \ and\ \bibinfo
  {author} {\bibfnamefont {X.}~\bibnamefont {Liu}},\ }\bibfield  {title}
  {\enquote {\bibinfo {title} {Dynamic upsampling of smoke through
  dictionary-based learning},}\ }\href {\doibase 10.1145/3412360} {\bibfield
  {journal} {\bibinfo  {journal} {ACM Trans. Graph.}\ }\textbf {\bibinfo
  {volume} {40}} (\bibinfo {year} {2020}),\ 10.1145/3412360}\BibitemShut
  {NoStop}%
\bibitem [{\citenamefont {Ferdian}\ \emph {et~al.}(2020)\citenamefont
  {Ferdian}, \citenamefont {Suinesiaputra}, \citenamefont {Dubowitz},
  \citenamefont {Zhao}, \citenamefont {Wang}, \citenamefont {Cowan},\ and\
  \citenamefont {Young}}]{Ferdian2020}%
  \BibitemOpen
  \bibfield  {author} {\bibinfo {author} {\bibfnamefont {E.}~\bibnamefont
  {Ferdian}}, \bibinfo {author} {\bibfnamefont {A.}~\bibnamefont
  {Suinesiaputra}}, \bibinfo {author} {\bibfnamefont {D.~J.}\ \bibnamefont
  {Dubowitz}}, \bibinfo {author} {\bibfnamefont {D.}~\bibnamefont {Zhao}},
  \bibinfo {author} {\bibfnamefont {A.}~\bibnamefont {Wang}}, \bibinfo {author}
  {\bibfnamefont {B.}~\bibnamefont {Cowan}}, \ and\ \bibinfo {author}
  {\bibfnamefont {A.~A.}\ \bibnamefont {Young}},\ }\bibfield  {title} {\enquote
  {\bibinfo {title} {4dflownet: Super-resolution 4d flow mri using deep
  learning and computational fluid dynamics},}\ }\href {\doibase
  10.3389/fphy.2020.00138} {\bibfield  {journal} {\bibinfo  {journal}
  {Frontiers in Physics}\ }\textbf {\bibinfo {volume} {8}},\ \bibinfo {pages}
  {138} (\bibinfo {year} {2020})}\BibitemShut {NoStop}%
\bibitem [{\citenamefont {Sun}\ and\ \citenamefont {Wang}(2020)}]{Sun2020}%
  \BibitemOpen
  \bibfield  {author} {\bibinfo {author} {\bibfnamefont {L.}~\bibnamefont
  {Sun}}\ and\ \bibinfo {author} {\bibfnamefont {J.-X.}\ \bibnamefont {Wang}},\
  }\bibfield  {title} {\enquote {\bibinfo {title} {Physics-constrained bayesian
  neural network for fluid flow reconstruction with sparse and noisy data},}\
  }\href {\doibase https://doi.org/10.1016/j.taml.2020.01.031} {\bibfield
  {journal} {\bibinfo  {journal} {Theoretical and Applied Mechanics Letters}\
  }\textbf {\bibinfo {volume} {10}},\ \bibinfo {pages} {161--169} (\bibinfo
  {year} {2020})}\BibitemShut {NoStop}%
\bibitem [{\citenamefont {Gao}, \citenamefont {Sun},\ and\ \citenamefont
  {Wang}(2021)}]{Gao2021}%
  \BibitemOpen
  \bibfield  {author} {\bibinfo {author} {\bibfnamefont {H.}~\bibnamefont
  {Gao}}, \bibinfo {author} {\bibfnamefont {L.}~\bibnamefont {Sun}}, \ and\
  \bibinfo {author} {\bibfnamefont {J.-X.}\ \bibnamefont {Wang}},\ }\bibfield
  {title} {\enquote {\bibinfo {title} {Super-resolution and denoising of fluid
  flow using physics-informed convolutional neural networks without
  high-resolution labels},}\ }\href {\doibase 10.1063/5.0054312} {\bibfield
  {journal} {\bibinfo  {journal} {Physics of Fluids}\ }\textbf {\bibinfo
  {volume} {33}},\ \bibinfo {pages} {073603} (\bibinfo {year}
  {2021})}\BibitemShut {NoStop}%
\bibitem [{\citenamefont {Ducournau}\ and\ \citenamefont
  {Fablet}(2016)}]{Ducournau2016}%
  \BibitemOpen
  \bibfield  {author} {\bibinfo {author} {\bibfnamefont {A.}~\bibnamefont
  {Ducournau}}\ and\ \bibinfo {author} {\bibfnamefont {R.}~\bibnamefont
  {Fablet}},\ }\bibfield  {title} {\enquote {\bibinfo {title} {Deep learning
  for ocean remote sensing: an application of convolutional neural networks for
  super-resolution on satellite-derived sst data},}\ \ }(\bibinfo {year}
  {2016})\ pp.\ \bibinfo {pages} {1--6}\BibitemShut {NoStop}%
\bibitem [{\citenamefont {Cannon}(2011)}]{Cannon2011}%
  \BibitemOpen
  \bibfield  {author} {\bibinfo {author} {\bibfnamefont {A.~J.}\ \bibnamefont
  {Cannon}},\ }\bibfield  {title} {\enquote {\bibinfo {title} {Quantile
  regression neural networks: Implementation in r and application to
  precipitation downscaling},}\ }\href {\doibase 10.1016/j.cageo.2010.07.005}
  {\bibfield  {journal} {\bibinfo  {journal} {Computers and Geosciences}\
  }\textbf {\bibinfo {volume} {37}},\ \bibinfo {pages} {1277--1284} (\bibinfo
  {year} {2011})}\BibitemShut {NoStop}%
\bibitem [{\citenamefont {Vandal}\ \emph {et~al.}(2017)\citenamefont {Vandal},
  \citenamefont {Kodra}, \citenamefont {Ganguly}, \citenamefont {Michaelis},
  \citenamefont {Nemani},\ and\ \citenamefont {Ganguly}}]{Vandal2017}%
  \BibitemOpen
  \bibfield  {author} {\bibinfo {author} {\bibfnamefont {T.}~\bibnamefont
  {Vandal}}, \bibinfo {author} {\bibfnamefont {E.}~\bibnamefont {Kodra}},
  \bibinfo {author} {\bibfnamefont {S.}~\bibnamefont {Ganguly}}, \bibinfo
  {author} {\bibfnamefont {A.}~\bibnamefont {Michaelis}}, \bibinfo {author}
  {\bibfnamefont {R.}~\bibnamefont {Nemani}}, \ and\ \bibinfo {author}
  {\bibfnamefont {A.~R.}\ \bibnamefont {Ganguly}},\ }\bibfield  {title}
  {\enquote {\bibinfo {title} {Deepsd: Generating high resolution climate
  change projections through single image super-resolution},}\ \ }(\bibinfo
  {publisher} {Association for Computing Machinery},\ \bibinfo {year} {2017})\
  pp.\ \bibinfo {pages} {1663--1672}\BibitemShut {NoStop}%
\bibitem [{\citenamefont {Rodrigues}\ \emph {et~al.}(2018)\citenamefont
  {Rodrigues}, \citenamefont {Oliveira}, \citenamefont {Cunha},\ and\
  \citenamefont {Netto}}]{Rodrigues2018}%
  \BibitemOpen
  \bibfield  {author} {\bibinfo {author} {\bibfnamefont {E.~R.}\ \bibnamefont
  {Rodrigues}}, \bibinfo {author} {\bibfnamefont {I.}~\bibnamefont {Oliveira}},
  \bibinfo {author} {\bibfnamefont {R.}~\bibnamefont {Cunha}}, \ and\ \bibinfo
  {author} {\bibfnamefont {M.}~\bibnamefont {Netto}},\ }\bibfield  {title}
  {\enquote {\bibinfo {title} {Deepdownscale: A deep learning strategy for
  high-resolution weather forecast},}\ \ }(\bibinfo {year} {2018})\ pp.\
  \bibinfo {pages} {415--422}\BibitemShut {NoStop}%
\bibitem [{\citenamefont {Onishi}, \citenamefont {Sugiyama},\ and\
  \citenamefont {Matsuda}(2019)}]{Onishi2019}%
  \BibitemOpen
  \bibfield  {author} {\bibinfo {author} {\bibfnamefont {R.}~\bibnamefont
  {Onishi}}, \bibinfo {author} {\bibfnamefont {D.}~\bibnamefont {Sugiyama}}, \
  and\ \bibinfo {author} {\bibfnamefont {K.}~\bibnamefont {Matsuda}},\
  }\bibfield  {title} {\enquote {\bibinfo {title} {Super-resolution simulation
  for real-time prediction of urban micrometeorology},}\ }\href {\doibase
  10.2151/sola.2019-032} {\bibfield  {journal} {\bibinfo  {journal} {SOLA}\
  }\textbf {\bibinfo {volume} {15}},\ \bibinfo {pages} {178--182} (\bibinfo
  {year} {2019})}\BibitemShut {NoStop}%
\bibitem [{\citenamefont {Leinonen}, \citenamefont {Nerini},\ and\
  \citenamefont {Berne}(2020)}]{Leinonen2020}%
  \BibitemOpen
  \bibfield  {author} {\bibinfo {author} {\bibfnamefont {J.}~\bibnamefont
  {Leinonen}}, \bibinfo {author} {\bibfnamefont {D.}~\bibnamefont {Nerini}}, \
  and\ \bibinfo {author} {\bibfnamefont {A.}~\bibnamefont {Berne}},\ }\bibfield
   {title} {\enquote {\bibinfo {title} {Stochastic super-resolution for
  downscaling time-evolving atmospheric fields with a generative adversarial
  network},}\ }\href {\doibase 10.1109/TGRS.2020.3032790} {\bibfield  {journal}
  {\bibinfo  {journal} {IEEE Transactions on Geoscience and Remote Sensing}\ ,\
  \bibinfo {pages} {1--13}} (\bibinfo {year} {2020})}\BibitemShut {NoStop}%
\bibitem [{\citenamefont {Stengel}\ \emph {et~al.}(2020)\citenamefont
  {Stengel}, \citenamefont {Glaws}, \citenamefont {Hettinger},\ and\
  \citenamefont {King}}]{Stengel2020}%
  \BibitemOpen
  \bibfield  {author} {\bibinfo {author} {\bibfnamefont {K.}~\bibnamefont
  {Stengel}}, \bibinfo {author} {\bibfnamefont {A.}~\bibnamefont {Glaws}},
  \bibinfo {author} {\bibfnamefont {D.}~\bibnamefont {Hettinger}}, \ and\
  \bibinfo {author} {\bibfnamefont {R.~N.}\ \bibnamefont {King}},\ }\bibfield
  {title} {\enquote {\bibinfo {title} {Adversarial super-resolution of
  climatological wind and solar data},}\ }\href {\doibase
  10.1073/PNAS.1918964117} {\bibfield  {journal} {\bibinfo  {journal}
  {Proceedings of the National Academy of Sciences}\ }\textbf {\bibinfo
  {volume} {117}},\ \bibinfo {pages} {16805--16815} (\bibinfo {year}
  {2020})}\BibitemShut {NoStop}%
\bibitem [{\citenamefont {Wang}\ \emph
  {et~al.}(2021{\natexlab{b}})\citenamefont {Wang}, \citenamefont {Liu},
  \citenamefont {Foster}, \citenamefont {Chang}, \citenamefont {Kettimuthu},\
  and\ \citenamefont {Kotamarthi}}]{Wang2021GMD}%
  \BibitemOpen
  \bibfield  {author} {\bibinfo {author} {\bibfnamefont {J.}~\bibnamefont
  {Wang}}, \bibinfo {author} {\bibfnamefont {Z.}~\bibnamefont {Liu}}, \bibinfo
  {author} {\bibfnamefont {I.}~\bibnamefont {Foster}}, \bibinfo {author}
  {\bibfnamefont {W.}~\bibnamefont {Chang}}, \bibinfo {author} {\bibfnamefont
  {R.}~\bibnamefont {Kettimuthu}}, \ and\ \bibinfo {author} {\bibfnamefont
  {V.~R.}\ \bibnamefont {Kotamarthi}},\ }\bibfield  {title} {\enquote {\bibinfo
  {title} {Fast and accurate learned multiresolution dynamical downscaling for
  precipitation},}\ }\href {\doibase 10.5194/gmd-14-6355-2021} {\bibfield
  {journal} {\bibinfo  {journal} {Geoscientific Model Development}\ }\textbf
  {\bibinfo {volume} {14}},\ \bibinfo {pages} {6355--6372} (\bibinfo {year}
  {2021}{\natexlab{b}})}\BibitemShut {NoStop}%
\bibitem [{\citenamefont {Wu}\ \emph {et~al.}(2021)\citenamefont {Wu},
  \citenamefont {Teufel}, \citenamefont {Sushama}, \citenamefont {Belair},\
  and\ \citenamefont {Sun}}]{Wu2021}%
  \BibitemOpen
  \bibfield  {author} {\bibinfo {author} {\bibfnamefont {Y.}~\bibnamefont
  {Wu}}, \bibinfo {author} {\bibfnamefont {B.}~\bibnamefont {Teufel}}, \bibinfo
  {author} {\bibfnamefont {L.}~\bibnamefont {Sushama}}, \bibinfo {author}
  {\bibfnamefont {S.}~\bibnamefont {Belair}}, \ and\ \bibinfo {author}
  {\bibfnamefont {L.}~\bibnamefont {Sun}},\ }\bibfield  {title} {\enquote
  {\bibinfo {title} {Deep learning-based super-resolution climate
  simulator-emulator framework for urban heat studies},}\ }\href {\doibase
  10.1029/2021GL094737} {\bibfield  {journal} {\bibinfo  {journal} {Geophysical
  Research Letters}\ }\textbf {\bibinfo {volume} {48}} (\bibinfo {year}
  {2021}),\ 10.1029/2021GL094737}\BibitemShut {NoStop}%
\bibitem [{\citenamefont {Yasuda}\ \emph {et~al.}(2022)\citenamefont {Yasuda},
  \citenamefont {Onishi}, \citenamefont {Hirokawa}, \citenamefont
  {Kolomenskiy},\ and\ \citenamefont {Sugiyama}}]{Yasuda2022}%
  \BibitemOpen
  \bibfield  {author} {\bibinfo {author} {\bibfnamefont {Y.}~\bibnamefont
  {Yasuda}}, \bibinfo {author} {\bibfnamefont {R.}~\bibnamefont {Onishi}},
  \bibinfo {author} {\bibfnamefont {Y.}~\bibnamefont {Hirokawa}}, \bibinfo
  {author} {\bibfnamefont {D.}~\bibnamefont {Kolomenskiy}}, \ and\ \bibinfo
  {author} {\bibfnamefont {D.}~\bibnamefont {Sugiyama}},\ }\bibfield  {title}
  {\enquote {\bibinfo {title} {Super-resolution of near-surface temperature
  utilizing physical quantities for real-time prediction of urban
  micrometeorology},}\ }\href {\doibase
  https://doi.org/10.1016/j.buildenv.2021.108597} {\bibfield  {journal}
  {\bibinfo  {journal} {Building and Environment}\ }\textbf {\bibinfo {volume}
  {209}},\ \bibinfo {pages} {108597} (\bibinfo {year} {2022})}\BibitemShut
  {NoStop}%
\bibitem [{\citenamefont {Weiler}\ and\ \citenamefont
  {Cesa}(2019)}]{Weiler2019}%
  \BibitemOpen
  \bibfield  {author} {\bibinfo {author} {\bibfnamefont {M.}~\bibnamefont
  {Weiler}}\ and\ \bibinfo {author} {\bibfnamefont {G.}~\bibnamefont {Cesa}},\
  }\bibfield  {title} {\enquote {\bibinfo {title} {General e(2)-equivariant
  steerable cnns},}\ }in\ \href
  {https://proceedings.neurips.cc/paper/2019/file/45d6637b718d0f24a237069fe41b0db4-Paper.pdf}
  {\emph {\bibinfo {booktitle} {Advances in Neural Information Processing
  Systems}}},\ Vol.~\bibinfo {volume} {32},\ \bibinfo {editor} {edited by\
  \bibinfo {editor} {\bibfnamefont {H.}~\bibnamefont {Wallach}}, \bibinfo
  {editor} {\bibfnamefont {H.}~\bibnamefont {Larochelle}}, \bibinfo {editor}
  {\bibfnamefont {A.}~\bibnamefont {Beygelzimer}}, \bibinfo {editor}
  {\bibfnamefont {F.}~\bibnamefont {d\textquotesingle Alch\'{e}-Buc}}, \bibinfo
  {editor} {\bibfnamefont {E.}~\bibnamefont {Fox}}, \ and\ \bibinfo {editor}
  {\bibfnamefont {R.}~\bibnamefont {Garnett}}}\ (\bibinfo  {publisher} {Curran
  Associates, Inc.},\ \bibinfo {year} {2019})\BibitemShut {NoStop}%
\bibitem [{\citenamefont {Bronstein}\ \emph {et~al.}(2021)\citenamefont
  {Bronstein}, \citenamefont {Bruna}, \citenamefont {Cohen},\ and\
  \citenamefont {Velickovic}}]{Bronstein2021}%
  \BibitemOpen
  \bibfield  {author} {\bibinfo {author} {\bibfnamefont {M.~M.}\ \bibnamefont
  {Bronstein}}, \bibinfo {author} {\bibfnamefont {J.}~\bibnamefont {Bruna}},
  \bibinfo {author} {\bibfnamefont {T.}~\bibnamefont {Cohen}}, \ and\ \bibinfo
  {author} {\bibfnamefont {P.}~\bibnamefont {Velickovic}},\ }\bibfield  {title}
  {\enquote {\bibinfo {title} {Geometric deep learning: Grids, groups, graphs,
  geodesics, and gauges},}\ }\href {https://arxiv.org/abs/2104.13478}
  {\bibfield  {journal} {\bibinfo  {journal} {CoRR}\ }\textbf {\bibinfo
  {volume} {abs/2104.13478}} (\bibinfo {year} {2021})},\ \Eprint
  {http://arxiv.org/abs/2104.13478} {2104.13478} \BibitemShut {NoStop}%
\bibitem [{\citenamefont {Landau}\ and\ \citenamefont
  {Lifshitz}(1975)}]{Landau1975}%
  \BibitemOpen
  \bibfield  {author} {\bibinfo {author} {\bibfnamefont {L.}~\bibnamefont
  {Landau}}\ and\ \bibinfo {author} {\bibfnamefont {E.}~\bibnamefont
  {Lifshitz}},\ }\href {\doibase
  https://doi.org/10.1016/B978-0-08-025072-4.50005-8} {\emph {\bibinfo {title}
  {The Classical Theory of Fields}}},\ \bibinfo {edition} {fourth edition}\
  ed.,\ \bibinfo {series} {Course of Theoretical Physics}, Vol.~\bibinfo
  {volume} {2}\ (\bibinfo {year} {1975})\BibitemShut {NoStop}%
\bibitem [{\citenamefont {Salmon}(1998)}]{Salmon1998}%
  \BibitemOpen
  \bibfield  {author} {\bibinfo {author} {\bibfnamefont {R.~L.}\ \bibnamefont
  {Salmon}},\ }\href@noop {} {\emph {\bibinfo {title} {Lectures on Geophysical
  Fluid Dynamics}}}\ (\bibinfo  {publisher} {Oxford University Press},\
  \bibinfo {year} {1998})\BibitemShut {NoStop}%
\bibitem [{\citenamefont {LeCun}\ \emph {et~al.}(1989)\citenamefont {LeCun},
  \citenamefont {Boser}, \citenamefont {Denker}, \citenamefont {Henderson},
  \citenamefont {Howard}, \citenamefont {Hubbard},\ and\ \citenamefont
  {Jackel}}]{LeCun1989}%
  \BibitemOpen
  \bibfield  {author} {\bibinfo {author} {\bibfnamefont {Y.}~\bibnamefont
  {LeCun}}, \bibinfo {author} {\bibfnamefont {B.}~\bibnamefont {Boser}},
  \bibinfo {author} {\bibfnamefont {J.~S.}\ \bibnamefont {Denker}}, \bibinfo
  {author} {\bibfnamefont {D.}~\bibnamefont {Henderson}}, \bibinfo {author}
  {\bibfnamefont {R.~E.}\ \bibnamefont {Howard}}, \bibinfo {author}
  {\bibfnamefont {W.}~\bibnamefont {Hubbard}}, \ and\ \bibinfo {author}
  {\bibfnamefont {L.~D.}\ \bibnamefont {Jackel}},\ }\bibfield  {title}
  {\enquote {\bibinfo {title} {Backpropagation applied to handwritten zip code
  recognition},}\ }\href {\doibase 10.1162/neco.1989.1.4.541} {\bibfield
  {journal} {\bibinfo  {journal} {Neural Computation}\ }\textbf {\bibinfo
  {volume} {1}},\ \bibinfo {pages} {541--551} (\bibinfo {year}
  {1989})}\BibitemShut {NoStop}%
\bibitem [{\citenamefont {Krizhevsky}, \citenamefont {Sutskever},\ and\
  \citenamefont {Hinton}(2012)}]{Krizhevsky2012}%
  \BibitemOpen
  \bibfield  {author} {\bibinfo {author} {\bibfnamefont {A.}~\bibnamefont
  {Krizhevsky}}, \bibinfo {author} {\bibfnamefont {I.}~\bibnamefont
  {Sutskever}}, \ and\ \bibinfo {author} {\bibfnamefont {G.~E.}\ \bibnamefont
  {Hinton}},\ }\bibfield  {title} {\enquote {\bibinfo {title} {Imagenet
  classification with deep convolutional neural networks},}\ \ }(\bibinfo
  {publisher} {Curran Associates Inc.},\ \bibinfo {year} {2012})\ pp.\ \bibinfo
  {pages} {1097--1105}\BibitemShut {NoStop}%
\bibitem [{\citenamefont {Cohen}\ and\ \citenamefont
  {Welling}(2016)}]{Cohen2016}%
  \BibitemOpen
  \bibfield  {author} {\bibinfo {author} {\bibfnamefont {T.}~\bibnamefont
  {Cohen}}\ and\ \bibinfo {author} {\bibfnamefont {M.}~\bibnamefont
  {Welling}},\ }\bibfield  {title} {\enquote {\bibinfo {title} {Group
  equivariant convolutional networks},}\ }in\ \href
  {https://proceedings.mlr.press/v48/cohenc16.html} {\emph {\bibinfo
  {booktitle} {Proceedings of The 33rd International Conference on Machine
  Learning}}},\ \bibinfo {series} {Proceedings of Machine Learning Research},
  Vol.~\bibinfo {volume} {48},\ \bibinfo {editor} {edited by\ \bibinfo {editor}
  {\bibfnamefont {M.~F.}\ \bibnamefont {Balcan}}\ and\ \bibinfo {editor}
  {\bibfnamefont {K.~Q.}\ \bibnamefont {Weinberger}}}\ (\bibinfo  {publisher}
  {PMLR},\ \bibinfo {address} {New York, New York, USA},\ \bibinfo {year}
  {2016})\ pp.\ \bibinfo {pages} {2990--2999}\BibitemShut {NoStop}%
\bibitem [{\citenamefont {Cohen}\ and\ \citenamefont
  {Welling}(2017)}]{Cohen2017}%
  \BibitemOpen
  \bibfield  {author} {\bibinfo {author} {\bibfnamefont {T.~S.}\ \bibnamefont
  {Cohen}}\ and\ \bibinfo {author} {\bibfnamefont {M.}~\bibnamefont
  {Welling}},\ }\bibfield  {title} {\enquote {\bibinfo {title} {Steerable
  cnns},}\ }\href {http://arxiv.org/abs/1612.08498} {\bibfield  {journal}
  {\bibinfo  {journal} {5th International Conference on Learning
  Representations, ICLR 2017}\ } (\bibinfo {year} {2017})}\BibitemShut
  {NoStop}%
\bibitem [{\citenamefont {Marcos}\ \emph {et~al.}(2017)\citenamefont {Marcos},
  \citenamefont {Volpi}, \citenamefont {Komodakis},\ and\ \citenamefont
  {Tuia}}]{Marcos2017}%
  \BibitemOpen
  \bibfield  {author} {\bibinfo {author} {\bibfnamefont {D.}~\bibnamefont
  {Marcos}}, \bibinfo {author} {\bibfnamefont {M.}~\bibnamefont {Volpi}},
  \bibinfo {author} {\bibfnamefont {N.}~\bibnamefont {Komodakis}}, \ and\
  \bibinfo {author} {\bibfnamefont {D.}~\bibnamefont {Tuia}},\ }\bibfield
  {title} {\enquote {\bibinfo {title} {Rotation equivariant vector field
  networks},}\ }in\ \href@noop {} {\emph {\bibinfo {booktitle} {Proceedings of
  the IEEE International Conference on Computer Vision (ICCV)}}}\ (\bibinfo
  {year} {2017})\BibitemShut {NoStop}%
\bibitem [{\citenamefont {Worrall}\ \emph {et~al.}(2017)\citenamefont
  {Worrall}, \citenamefont {Garbin}, \citenamefont {Turmukhambetov},\ and\
  \citenamefont {Brostow}}]{Worrall2017}%
  \BibitemOpen
  \bibfield  {author} {\bibinfo {author} {\bibfnamefont {D.~E.}\ \bibnamefont
  {Worrall}}, \bibinfo {author} {\bibfnamefont {S.~J.}\ \bibnamefont {Garbin}},
  \bibinfo {author} {\bibfnamefont {D.}~\bibnamefont {Turmukhambetov}}, \ and\
  \bibinfo {author} {\bibfnamefont {G.~J.}\ \bibnamefont {Brostow}},\
  }\bibfield  {title} {\enquote {\bibinfo {title} {Harmonic networks: Deep
  translation and rotation equivariance},}\ }in\ \href@noop {} {\emph {\bibinfo
  {booktitle} {Proceedings of the IEEE Conference on Computer Vision and
  Pattern Recognition (CVPR)}}}\ (\bibinfo {year} {2017})\BibitemShut {NoStop}%
\bibitem [{\citenamefont {Zhou}\ \emph {et~al.}(2017)\citenamefont {Zhou},
  \citenamefont {Ye}, \citenamefont {Qiu},\ and\ \citenamefont
  {Jiao}}]{Zhou2017}%
  \BibitemOpen
  \bibfield  {author} {\bibinfo {author} {\bibfnamefont {Y.}~\bibnamefont
  {Zhou}}, \bibinfo {author} {\bibfnamefont {Q.}~\bibnamefont {Ye}}, \bibinfo
  {author} {\bibfnamefont {Q.}~\bibnamefont {Qiu}}, \ and\ \bibinfo {author}
  {\bibfnamefont {J.}~\bibnamefont {Jiao}},\ }\bibfield  {title} {\enquote
  {\bibinfo {title} {Oriented response networks},}\ }in\ \href@noop {} {\emph
  {\bibinfo {booktitle} {Proceedings of the IEEE Conference on Computer Vision
  and Pattern Recognition (CVPR)}}}\ (\bibinfo {year} {2017})\BibitemShut
  {NoStop}%
\bibitem [{\citenamefont {Schütt}\ \emph {et~al.}(2017)\citenamefont
  {Schütt}, \citenamefont {Arbabzadah}, \citenamefont {Chmiela}, \citenamefont
  {Müller},\ and\ \citenamefont {Tkatchenko}}]{Schutt2017}%
  \BibitemOpen
  \bibfield  {author} {\bibinfo {author} {\bibfnamefont {K.~T.}\ \bibnamefont
  {Schütt}}, \bibinfo {author} {\bibfnamefont {F.}~\bibnamefont {Arbabzadah}},
  \bibinfo {author} {\bibfnamefont {S.}~\bibnamefont {Chmiela}}, \bibinfo
  {author} {\bibfnamefont {K.~R.}\ \bibnamefont {Müller}}, \ and\ \bibinfo
  {author} {\bibfnamefont {A.}~\bibnamefont {Tkatchenko}},\ }\bibfield  {title}
  {\enquote {\bibinfo {title} {Quantum-chemical insights from deep tensor
  neural networks},}\ }\href {\doibase 10.1038/ncomms13890} {\bibfield
  {journal} {\bibinfo  {journal} {Nature Communications}\ } (\bibinfo {year}
  {2017}),\ 10.1038/ncomms13890}\BibitemShut {NoStop}%
\bibitem [{\citenamefont {Bekkers}\ \emph {et~al.}(2018)\citenamefont
  {Bekkers}, \citenamefont {Lafarge}, \citenamefont {Veta}, \citenamefont
  {Eppenhof}, \citenamefont {Pluim},\ and\ \citenamefont
  {Duits}}]{Bekkers2018}%
  \BibitemOpen
  \bibfield  {author} {\bibinfo {author} {\bibfnamefont {E.~J.}\ \bibnamefont
  {Bekkers}}, \bibinfo {author} {\bibfnamefont {M.~W.}\ \bibnamefont
  {Lafarge}}, \bibinfo {author} {\bibfnamefont {M.}~\bibnamefont {Veta}},
  \bibinfo {author} {\bibfnamefont {K.~A.~J.}\ \bibnamefont {Eppenhof}},
  \bibinfo {author} {\bibfnamefont {J.~P.~W.}\ \bibnamefont {Pluim}}, \ and\
  \bibinfo {author} {\bibfnamefont {R.}~\bibnamefont {Duits}},\ }\bibfield
  {title} {\enquote {\bibinfo {title} {Roto-translation covariant convolutional
  networks for medical image analysis},}\ }in\ \href@noop {} {\emph {\bibinfo
  {booktitle} {Medical Image Computing and Computer Assisted Intervention --
  MICCAI 2018}}},\ \bibinfo {editor} {edited by\ \bibinfo {editor}
  {\bibfnamefont {A.~F.}\ \bibnamefont {Frangi}}, \bibinfo {editor}
  {\bibfnamefont {J.~A.}\ \bibnamefont {Schnabel}}, \bibinfo {editor}
  {\bibfnamefont {C.}~\bibnamefont {Davatzikos}}, \bibinfo {editor}
  {\bibfnamefont {C.}~\bibnamefont {Alberola-L{\'o}pez}}, \ and\ \bibinfo
  {editor} {\bibfnamefont {G.}~\bibnamefont {Fichtinger}}}\ (\bibinfo
  {publisher} {Springer International Publishing},\ \bibinfo {address} {Cham},\
  \bibinfo {year} {2018})\ pp.\ \bibinfo {pages} {440--448}\BibitemShut
  {NoStop}%
\bibitem [{\citenamefont {Sosnovik}, \citenamefont {Moskalev},\ and\
  \citenamefont {Smeulders}(2021)}]{Sosnovik2021}%
  \BibitemOpen
  \bibfield  {author} {\bibinfo {author} {\bibfnamefont {I.}~\bibnamefont
  {Sosnovik}}, \bibinfo {author} {\bibfnamefont {A.}~\bibnamefont {Moskalev}},
  \ and\ \bibinfo {author} {\bibfnamefont {A.~W.}\ \bibnamefont {Smeulders}},\
  }\bibfield  {title} {\enquote {\bibinfo {title} {Scale equivariance improves
  siamese tracking},}\ }in\ \href@noop {} {\emph {\bibinfo {booktitle}
  {Proceedings of the IEEE/CVF Winter Conference on Applications of Computer
  Vision (WACV)}}}\ (\bibinfo {year} {2021})\ pp.\ \bibinfo {pages}
  {2765--2774}\BibitemShut {NoStop}%
\bibitem [{\citenamefont {Kondor}\ and\ \citenamefont
  {Trivedi}(2018)}]{Kondor2018}%
  \BibitemOpen
  \bibfield  {author} {\bibinfo {author} {\bibfnamefont {R.}~\bibnamefont
  {Kondor}}\ and\ \bibinfo {author} {\bibfnamefont {S.}~\bibnamefont
  {Trivedi}},\ }\bibfield  {title} {\enquote {\bibinfo {title} {On the
  generalization of equivariance and convolution in neural networks to the
  action of compact groups},}\ }in\ \href
  {https://proceedings.mlr.press/v80/kondor18a.html} {\emph {\bibinfo
  {booktitle} {Proceedings of the 35th International Conference on Machine
  Learning}}},\ \bibinfo {series} {Proceedings of Machine Learning Research},
  Vol.~\bibinfo {volume} {80},\ \bibinfo {editor} {edited by\ \bibinfo {editor}
  {\bibfnamefont {J.}~\bibnamefont {Dy}}\ and\ \bibinfo {editor} {\bibfnamefont
  {A.}~\bibnamefont {Krause}}}\ (\bibinfo  {publisher} {PMLR},\ \bibinfo {year}
  {2018})\ pp.\ \bibinfo {pages} {2747--2755}\BibitemShut {NoStop}%
\bibitem [{\citenamefont {Cohen}, \citenamefont {Geiger},\ and\ \citenamefont
  {Weiler}(2019)}]{Cohen2019}%
  \BibitemOpen
  \bibfield  {author} {\bibinfo {author} {\bibfnamefont {T.~S.}\ \bibnamefont
  {Cohen}}, \bibinfo {author} {\bibfnamefont {M.}~\bibnamefont {Geiger}}, \
  and\ \bibinfo {author} {\bibfnamefont {M.}~\bibnamefont {Weiler}},\
  }\bibfield  {title} {\enquote {\bibinfo {title} {A general theory of
  equivariant cnns on homogeneous spaces},}\ }in\ \href
  {https://proceedings.neurips.cc/paper/2019/file/b9cfe8b6042cf759dc4c0cccb27a6737-Paper.pdf}
  {\emph {\bibinfo {booktitle} {Advances in Neural Information Processing
  Systems}}},\ Vol.~\bibinfo {volume} {32},\ \bibinfo {editor} {edited by\
  \bibinfo {editor} {\bibfnamefont {H.}~\bibnamefont {Wallach}}, \bibinfo
  {editor} {\bibfnamefont {H.}~\bibnamefont {Larochelle}}, \bibinfo {editor}
  {\bibfnamefont {A.}~\bibnamefont {Beygelzimer}}, \bibinfo {editor}
  {\bibfnamefont {F.}~\bibnamefont {d\textquotesingle Alch\'{e}-Buc}}, \bibinfo
  {editor} {\bibfnamefont {E.}~\bibnamefont {Fox}}, \ and\ \bibinfo {editor}
  {\bibfnamefont {R.}~\bibnamefont {Garnett}}}\ (\bibinfo  {publisher} {Curran
  Associates, Inc.},\ \bibinfo {year} {2019})\BibitemShut {NoStop}%
\bibitem [{\citenamefont {Fuchs}\ \emph {et~al.}(2020)\citenamefont {Fuchs},
  \citenamefont {Worrall}, \citenamefont {Fischer},\ and\ \citenamefont
  {Welling}}]{Fuchs2020}%
  \BibitemOpen
  \bibfield  {author} {\bibinfo {author} {\bibfnamefont {F.}~\bibnamefont
  {Fuchs}}, \bibinfo {author} {\bibfnamefont {D.}~\bibnamefont {Worrall}},
  \bibinfo {author} {\bibfnamefont {V.}~\bibnamefont {Fischer}}, \ and\
  \bibinfo {author} {\bibfnamefont {M.}~\bibnamefont {Welling}},\ }\bibfield
  {title} {\enquote {\bibinfo {title} {Se(3)-transformers: 3d roto-translation
  equivariant attention networks},}\ }in\ \href
  {https://proceedings.neurips.cc/paper/2020/file/15231a7ce4ba789d13b722cc5c955834-Paper.pdf}
  {\emph {\bibinfo {booktitle} {Advances in Neural Information Processing
  Systems}}},\ Vol.~\bibinfo {volume} {33},\ \bibinfo {editor} {edited by\
  \bibinfo {editor} {\bibfnamefont {H.}~\bibnamefont {Larochelle}}, \bibinfo
  {editor} {\bibfnamefont {M.}~\bibnamefont {Ranzato}}, \bibinfo {editor}
  {\bibfnamefont {R.}~\bibnamefont {Hadsell}}, \bibinfo {editor} {\bibfnamefont
  {M.~F.}\ \bibnamefont {Balcan}}, \ and\ \bibinfo {editor} {\bibfnamefont
  {H.}~\bibnamefont {Lin}}}\ (\bibinfo  {publisher} {Curran Associates, Inc.},\
  \bibinfo {year} {2020})\ pp.\ \bibinfo {pages} {1970--1981}\BibitemShut
  {NoStop}%
\bibitem [{\citenamefont {Weiler}\ \emph {et~al.}(2018)\citenamefont {Weiler},
  \citenamefont {Geiger}, \citenamefont {Welling}, \citenamefont {Boomsma},\
  and\ \citenamefont {Cohen}}]{Weiler2018}%
  \BibitemOpen
  \bibfield  {author} {\bibinfo {author} {\bibfnamefont {M.}~\bibnamefont
  {Weiler}}, \bibinfo {author} {\bibfnamefont {M.}~\bibnamefont {Geiger}},
  \bibinfo {author} {\bibfnamefont {M.}~\bibnamefont {Welling}}, \bibinfo
  {author} {\bibfnamefont {W.}~\bibnamefont {Boomsma}}, \ and\ \bibinfo
  {author} {\bibfnamefont {T.~S.}\ \bibnamefont {Cohen}},\ }\bibfield  {title}
  {\enquote {\bibinfo {title} {3d steerable cnns: Learning rotationally
  equivariant features in volumetric data},}\ }in\ \href
  {https://proceedings.neurips.cc/paper/2018/file/488e4104520c6aab692863cc1dba45af-Paper.pdf}
  {\emph {\bibinfo {booktitle} {Advances in Neural Information Processing
  Systems}}},\ Vol.~\bibinfo {volume} {31},\ \bibinfo {editor} {edited by\
  \bibinfo {editor} {\bibfnamefont {S.}~\bibnamefont {Bengio}}, \bibinfo
  {editor} {\bibfnamefont {H.}~\bibnamefont {Wallach}}, \bibinfo {editor}
  {\bibfnamefont {H.}~\bibnamefont {Larochelle}}, \bibinfo {editor}
  {\bibfnamefont {K.}~\bibnamefont {Grauman}}, \bibinfo {editor} {\bibfnamefont
  {N.}~\bibnamefont {Cesa-Bianchi}}, \ and\ \bibinfo {editor} {\bibfnamefont
  {R.}~\bibnamefont {Garnett}}}\ (\bibinfo  {publisher} {Curran Associates,
  Inc.},\ \bibinfo {year} {2018})\BibitemShut {NoStop}%
\bibitem [{\citenamefont {Thomas}\ \emph {et~al.}(2018)\citenamefont {Thomas},
  \citenamefont {Smidt}, \citenamefont {Kearnes}, \citenamefont {Yang},
  \citenamefont {Li}, \citenamefont {Kohlhoff},\ and\ \citenamefont
  {Riley}}]{Thomas2018}%
  \BibitemOpen
  \bibfield  {author} {\bibinfo {author} {\bibfnamefont {N.}~\bibnamefont
  {Thomas}}, \bibinfo {author} {\bibfnamefont {T.}~\bibnamefont {Smidt}},
  \bibinfo {author} {\bibfnamefont {S.~M.}\ \bibnamefont {Kearnes}}, \bibinfo
  {author} {\bibfnamefont {L.}~\bibnamefont {Yang}}, \bibinfo {author}
  {\bibfnamefont {L.}~\bibnamefont {Li}}, \bibinfo {author} {\bibfnamefont
  {K.}~\bibnamefont {Kohlhoff}}, \ and\ \bibinfo {author} {\bibfnamefont
  {P.}~\bibnamefont {Riley}},\ }\bibfield  {title} {\enquote {\bibinfo {title}
  {Tensor field networks: Rotation- and translation-equivariant neural networks
  for 3d point clouds},}\ }\href {http://arxiv.org/abs/1802.08219} {\bibfield
  {journal} {\bibinfo  {journal} {CoRR}\ }\textbf {\bibinfo {volume}
  {abs/1802.08219}} (\bibinfo {year} {2018})},\ \Eprint
  {http://arxiv.org/abs/1802.08219} {arXiv:1802.08219} \BibitemShut {NoStop}%
\bibitem [{\citenamefont {Atz}, \citenamefont {Grisoni},\ and\ \citenamefont
  {Schneider}(2021)}]{Atz2021}%
  \BibitemOpen
  \bibfield  {author} {\bibinfo {author} {\bibfnamefont {K.}~\bibnamefont
  {Atz}}, \bibinfo {author} {\bibfnamefont {F.}~\bibnamefont {Grisoni}}, \ and\
  \bibinfo {author} {\bibfnamefont {G.}~\bibnamefont {Schneider}},\ }\bibfield
  {title} {\enquote {\bibinfo {title} {Geometric deep learning on molecular
  representations},}\ }\href {\doibase 10.1038/s42256-021-00418-8} {\bibfield
  {journal} {\bibinfo  {journal} {Nature Machine Intelligence}\ }\textbf
  {\bibinfo {volume} {3}},\ \bibinfo {pages} {1023--1032} (\bibinfo {year}
  {2021})}\BibitemShut {NoStop}%
\bibitem [{\citenamefont {Chen}, \citenamefont {Tachella},\ and\ \citenamefont
  {Davies}(2021{\natexlab{a}})}]{Chen2021a}%
  \BibitemOpen
  \bibfield  {author} {\bibinfo {author} {\bibfnamefont {D.}~\bibnamefont
  {Chen}}, \bibinfo {author} {\bibfnamefont {J.}~\bibnamefont {Tachella}}, \
  and\ \bibinfo {author} {\bibfnamefont {M.~E.}\ \bibnamefont {Davies}},\
  }\bibfield  {title} {\enquote {\bibinfo {title} {Equivariant imaging:
  Learning beyond the range space},}\ }in\ \href@noop {} {\emph {\bibinfo
  {booktitle} {Proceedings of the IEEE/CVF International Conference on Computer
  Vision (ICCV)}}}\ (\bibinfo {year} {2021})\ pp.\ \bibinfo {pages}
  {4379--4388}\BibitemShut {NoStop}%
\bibitem [{\citenamefont {Chen}, \citenamefont {Tachella},\ and\ \citenamefont
  {Davies}(2021{\natexlab{b}})}]{Chen2021b}%
  \BibitemOpen
  \bibfield  {author} {\bibinfo {author} {\bibfnamefont {D.}~\bibnamefont
  {Chen}}, \bibinfo {author} {\bibfnamefont {J.}~\bibnamefont {Tachella}}, \
  and\ \bibinfo {author} {\bibfnamefont {M.~E.}\ \bibnamefont {Davies}},\
  }\bibfield  {title} {\enquote {\bibinfo {title} {Robust equivariant imaging:
  a fully unsupervised framework for learning to image from noisy and partial
  measurements},}\ }\href {https://arxiv.org/abs/2111.12855} {\bibfield
  {journal} {\bibinfo  {journal} {CoRR}\ }\textbf {\bibinfo {volume}
  {abs/2111.12855}} (\bibinfo {year} {2021}{\natexlab{b}})},\ \Eprint
  {http://arxiv.org/abs/2111.12855} {2111.12855} \BibitemShut {NoStop}%
\bibitem [{\citenamefont {Lee}\ and\ \citenamefont {Mu~Lee}(2021)}]{Lee2021}%
  \BibitemOpen
  \bibfield  {author} {\bibinfo {author} {\bibfnamefont {J.}~\bibnamefont
  {Lee}}\ and\ \bibinfo {author} {\bibfnamefont {K.}~\bibnamefont {Mu~Lee}},\
  }\bibfield  {title} {\enquote {\bibinfo {title} {Structure-resonant
  discriminator for image super-resolution},}\ }in\ \href {\doibase
  10.1109/ICME51207.2021.9428359} {\emph {\bibinfo {booktitle} {2021 IEEE
  International Conference on Multimedia and Expo (ICME)}}}\ (\bibinfo {year}
  {2021})\ pp.\ \bibinfo {pages} {1--6}\BibitemShut {NoStop}%
\bibitem [{\citenamefont {Xie}, \citenamefont {Ding},\ and\ \citenamefont
  {Ji}(2020)}]{Xie2020}%
  \BibitemOpen
  \bibfield  {author} {\bibinfo {author} {\bibfnamefont {Y.}~\bibnamefont
  {Xie}}, \bibinfo {author} {\bibfnamefont {Y.}~\bibnamefont {Ding}}, \ and\
  \bibinfo {author} {\bibfnamefont {S.}~\bibnamefont {Ji}},\ }\bibfield
  {title} {\enquote {\bibinfo {title} {Augmented equivariant attention networks
  for electron microscopy image super-resolution},}\ }\href
  {https://arxiv.org/abs/2011.03633} {\bibfield  {journal} {\bibinfo  {journal}
  {CoRR}\ }\textbf {\bibinfo {volume} {abs/2011.03633}} (\bibinfo {year}
  {2020})},\ \Eprint {http://arxiv.org/abs/2011.03633} {2011.03633}
  \BibitemShut {NoStop}%
\bibitem [{\citenamefont {Ling}, \citenamefont {Kurzawski},\ and\ \citenamefont
  {Templeton}(2016)}]{Ling2016}%
  \BibitemOpen
  \bibfield  {author} {\bibinfo {author} {\bibfnamefont {J.}~\bibnamefont
  {Ling}}, \bibinfo {author} {\bibfnamefont {A.}~\bibnamefont {Kurzawski}}, \
  and\ \bibinfo {author} {\bibfnamefont {J.}~\bibnamefont {Templeton}},\
  }\bibfield  {title} {\enquote {\bibinfo {title} {Reynolds averaged turbulence
  modelling using deep neural networks with embedded invariance},}\ }\href
  {\doibase 10.1017/jfm.2016.615} {\bibfield  {journal} {\bibinfo  {journal}
  {Journal of Fluid Mechanics}\ }\textbf {\bibinfo {volume} {807}},\ \bibinfo
  {pages} {155–166} (\bibinfo {year} {2016})}\BibitemShut {NoStop}%
\bibitem [{\citenamefont {Zhou}, \citenamefont {Han},\ and\ \citenamefont
  {Xiao}(2022)}]{Zhou2022}%
  \BibitemOpen
  \bibfield  {author} {\bibinfo {author} {\bibfnamefont {X.-H.}\ \bibnamefont
  {Zhou}}, \bibinfo {author} {\bibfnamefont {J.}~\bibnamefont {Han}}, \ and\
  \bibinfo {author} {\bibfnamefont {H.}~\bibnamefont {Xiao}},\ }\bibfield
  {title} {\enquote {\bibinfo {title} {Frame-independent vector-cloud neural
  network for nonlocal constitutive modeling on arbitrary grids},}\ }\href
  {\doibase https://doi.org/10.1016/j.cma.2021.114211} {\bibfield  {journal}
  {\bibinfo  {journal} {Computer Methods in Applied Mechanics and Engineering}\
  }\textbf {\bibinfo {volume} {388}},\ \bibinfo {pages} {114211} (\bibinfo
  {year} {2022})}\BibitemShut {NoStop}%
\bibitem [{\citenamefont {Han}, \citenamefont {Zhou},\ and\ \citenamefont
  {Xiao}(2022)}]{Han2022}%
  \BibitemOpen
  \bibfield  {author} {\bibinfo {author} {\bibfnamefont {J.}~\bibnamefont
  {Han}}, \bibinfo {author} {\bibfnamefont {X.-H.}\ \bibnamefont {Zhou}}, \
  and\ \bibinfo {author} {\bibfnamefont {H.}~\bibnamefont {Xiao}},\ }\href
  {https://arxiv.org/abs/2201.01287} {\enquote {\bibinfo {title} {Vcnn-e: A
  vector-cloud neural network with equivariance for emulating reynolds stress
  transport equations},}\ } (\bibinfo {year} {2022}),\ \Eprint
  {http://arxiv.org/abs/2201.01287} {arXiv:2201.01287 [physics.flu-dyn]}
  \BibitemShut {NoStop}%
\bibitem [{\citenamefont {Pawar}\ \emph {et~al.}(2022)\citenamefont {Pawar},
  \citenamefont {San}, \citenamefont {Rasheed},\ and\ \citenamefont
  {Vedula}}]{Pawar2022}%
  \BibitemOpen
  \bibfield  {author} {\bibinfo {author} {\bibfnamefont {S.}~\bibnamefont
  {Pawar}}, \bibinfo {author} {\bibfnamefont {O.}~\bibnamefont {San}}, \bibinfo
  {author} {\bibfnamefont {A.}~\bibnamefont {Rasheed}}, \ and\ \bibinfo
  {author} {\bibfnamefont {P.}~\bibnamefont {Vedula}},\ }\href
  {https://arxiv.org/abs/2201.02928} {\enquote {\bibinfo {title} {Frame
  invariant neural network closures for kraichnan turbulence},}\ } (\bibinfo
  {year} {2022}),\ \Eprint {http://arxiv.org/abs/2201.02928} {arXiv:2201.02928
  [physics.flu-dyn]} \BibitemShut {NoStop}%
\bibitem [{\citenamefont {Wang}, \citenamefont {Walters},\ and\ \citenamefont
  {Yu}(2021)}]{Wang2021}%
  \BibitemOpen
  \bibfield  {author} {\bibinfo {author} {\bibfnamefont {R.}~\bibnamefont
  {Wang}}, \bibinfo {author} {\bibfnamefont {R.}~\bibnamefont {Walters}}, \
  and\ \bibinfo {author} {\bibfnamefont {R.}~\bibnamefont {Yu}},\ }\bibfield
  {title} {\enquote {\bibinfo {title} {Incorporating symmetry into deep
  dynamics models for improved generalization.}}\ \ }(\bibinfo {year}
  {2021})\BibitemShut {NoStop}%
\bibitem [{\citenamefont {Suk}\ \emph {et~al.}(2021)\citenamefont {Suk},
  \citenamefont {de~Haan}, \citenamefont {Lippe}, \citenamefont {Brune},\ and\
  \citenamefont {Wolterink}}]{Suk2021}%
  \BibitemOpen
  \bibfield  {author} {\bibinfo {author} {\bibfnamefont {J.}~\bibnamefont
  {Suk}}, \bibinfo {author} {\bibfnamefont {P.}~\bibnamefont {de~Haan}},
  \bibinfo {author} {\bibfnamefont {P.}~\bibnamefont {Lippe}}, \bibinfo
  {author} {\bibfnamefont {C.}~\bibnamefont {Brune}}, \ and\ \bibinfo {author}
  {\bibfnamefont {J.~M.}\ \bibnamefont {Wolterink}},\ }\bibfield  {title}
  {\enquote {\bibinfo {title} {Equivariant graph neural networks as surrogate
  for computational fluid dynamics in 3d artery models},}\ }in\ \href
  {https://nips.cc/Conferences/2021/Schedule?showEvent=21862} {\emph {\bibinfo
  {booktitle} {Fourth Workshop on Machine Learning and the Physical Sciences
  (NeurIPS 2021).}}}\ (\bibinfo {year} {2021})\BibitemShut {NoStop}%
\bibitem [{\citenamefont {Siddani}, \citenamefont {Balachandar},\ and\
  \citenamefont {Fang}(2021)}]{Siddani2021}%
  \BibitemOpen
  \bibfield  {author} {\bibinfo {author} {\bibfnamefont {B.}~\bibnamefont
  {Siddani}}, \bibinfo {author} {\bibfnamefont {S.}~\bibnamefont
  {Balachandar}}, \ and\ \bibinfo {author} {\bibfnamefont {R.}~\bibnamefont
  {Fang}},\ }\bibfield  {title} {\enquote {\bibinfo {title} {Rotational and
  reflectional equivariant convolutional neural network for data-limited
  applications: Multiphase flow demonstration},}\ }\href {\doibase
  10.1063/5.0066049} {\bibfield  {journal} {\bibinfo  {journal} {Physics of
  Fluids}\ }\textbf {\bibinfo {volume} {33}},\ \bibinfo {pages} {103323}
  (\bibinfo {year} {2021})},\ \Eprint
  {http://arxiv.org/abs/https://doi.org/10.1063/5.0066049}
  {https://doi.org/10.1063/5.0066049} \BibitemShut {NoStop}%
\bibitem [{\citenamefont {Gao}\ \emph {et~al.}(2020)\citenamefont {Gao},
  \citenamefont {Du}, \citenamefont {Li},\ and\ \citenamefont
  {Lin}}]{Gao2020RotEqNet}%
  \BibitemOpen
  \bibfield  {author} {\bibinfo {author} {\bibfnamefont {L.}~\bibnamefont
  {Gao}}, \bibinfo {author} {\bibfnamefont {Y.}~\bibnamefont {Du}}, \bibinfo
  {author} {\bibfnamefont {H.}~\bibnamefont {Li}}, \ and\ \bibinfo {author}
  {\bibfnamefont {G.}~\bibnamefont {Lin}},\ }\bibfield  {title} {\enquote
  {\bibinfo {title} {Roteqnet: Rotation-equivariant network for fluid systems
  with symmetric high-order tensors},}\ }\href
  {https://arxiv.org/abs/2005.04286} {\bibfield  {journal} {\bibinfo  {journal}
  {CoRR}\ }\textbf {\bibinfo {volume} {abs/2005.04286}} (\bibinfo {year}
  {2020})},\ \Eprint {http://arxiv.org/abs/2005.04286} {2005.04286}
  \BibitemShut {NoStop}%
\bibitem [{\citenamefont {Yasuda}(2022)}]{Yasuda2022GH}%
  \BibitemOpen
  \bibfield  {author} {\bibinfo {author} {\bibfnamefont {Y.}~\bibnamefont
  {Yasuda}},\ }\href@noop {} {\enquote {\bibinfo {title}
  {equivariant-sr-2d-fluid},}\ }\bibinfo {howpublished}
  {\url{https://github.com/YukiYasuda2718/equivariant-SR-2D-fluid}} (\bibinfo
  {year} {2022})\BibitemShut {NoStop}%
\bibitem [{\citenamefont {Weiler}, \citenamefont {Hamprecht},\ and\
  \citenamefont {Storath}(2018)}]{Weiler2018CVPR}%
  \BibitemOpen
  \bibfield  {author} {\bibinfo {author} {\bibfnamefont {M.}~\bibnamefont
  {Weiler}}, \bibinfo {author} {\bibfnamefont {F.~A.}\ \bibnamefont
  {Hamprecht}}, \ and\ \bibinfo {author} {\bibfnamefont {M.}~\bibnamefont
  {Storath}},\ }\bibfield  {title} {\enquote {\bibinfo {title} {Learning
  steerable filters for rotation equivariant cnns},}\ }in\ \href {\doibase
  10.1109/CVPR.2018.00095} {\emph {\bibinfo {booktitle} {2018 IEEE/CVF
  Conference on Computer Vision and Pattern Recognition}}}\ (\bibinfo {year}
  {2018})\ pp.\ \bibinfo {pages} {849--858}\BibitemShut {NoStop}%
\bibitem [{\citenamefont {Decelle}, \citenamefont {Martin-Mayor},\ and\
  \citenamefont {Seoane}(2019)}]{Decelle2019}%
  \BibitemOpen
  \bibfield  {author} {\bibinfo {author} {\bibfnamefont {A.}~\bibnamefont
  {Decelle}}, \bibinfo {author} {\bibfnamefont {V.}~\bibnamefont
  {Martin-Mayor}}, \ and\ \bibinfo {author} {\bibfnamefont {B.}~\bibnamefont
  {Seoane}},\ }\bibfield  {title} {\enquote {\bibinfo {title} {Learning a local
  symmetry with neural networks},}\ }\href {\doibase
  10.1103/PhysRevE.100.050102} {\bibfield  {journal} {\bibinfo  {journal}
  {Physical Review E}\ }\textbf {\bibinfo {volume} {100}} (\bibinfo {year}
  {2019}),\ 10.1103/PhysRevE.100.050102}\BibitemShut {NoStop}%
\bibitem [{\citenamefont {Wetzel}\ \emph {et~al.}(2020)\citenamefont {Wetzel},
  \citenamefont {Melko}, \citenamefont {Scott}, \citenamefont {Panju},\ and\
  \citenamefont {Ganesh}}]{Wetzel2020}%
  \BibitemOpen
  \bibfield  {author} {\bibinfo {author} {\bibfnamefont {S.~J.}\ \bibnamefont
  {Wetzel}}, \bibinfo {author} {\bibfnamefont {R.~G.}\ \bibnamefont {Melko}},
  \bibinfo {author} {\bibfnamefont {J.}~\bibnamefont {Scott}}, \bibinfo
  {author} {\bibfnamefont {M.}~\bibnamefont {Panju}}, \ and\ \bibinfo {author}
  {\bibfnamefont {V.}~\bibnamefont {Ganesh}},\ }\bibfield  {title} {\enquote
  {\bibinfo {title} {Discovering symmetry invariants and conserved quantities
  by interpreting siamese neural networks},}\ }\href {\doibase
  10.1103/PhysRevResearch.2.033499} {\bibfield  {journal} {\bibinfo  {journal}
  {Physical Review Research}\ }\textbf {\bibinfo {volume} {2}} (\bibinfo {year}
  {2020}),\ 10.1103/PhysRevResearch.2.033499}\BibitemShut {NoStop}%
\bibitem [{\citenamefont {Ha}\ and\ \citenamefont {Jeong}(2021)}]{Ha2021}%
  \BibitemOpen
  \bibfield  {author} {\bibinfo {author} {\bibfnamefont {S.}~\bibnamefont
  {Ha}}\ and\ \bibinfo {author} {\bibfnamefont {H.}~\bibnamefont {Jeong}},\
  }\bibfield  {title} {\enquote {\bibinfo {title} {Discovering invariants via
  machine learning},}\ }\href {\doibase 10.1103/PhysRevResearch.3.L042035}
  {\bibfield  {journal} {\bibinfo  {journal} {Physical Review Research}\
  }\textbf {\bibinfo {volume} {3}} (\bibinfo {year} {2021}),\
  10.1103/PhysRevResearch.3.L042035}\BibitemShut {NoStop}%
\bibitem [{\citenamefont {Krippendorf}\ and\ \citenamefont
  {Syvaeri}(2021)}]{Krippendorf2021}%
  \BibitemOpen
  \bibfield  {author} {\bibinfo {author} {\bibfnamefont {S.}~\bibnamefont
  {Krippendorf}}\ and\ \bibinfo {author} {\bibfnamefont {M.}~\bibnamefont
  {Syvaeri}},\ }\bibfield  {title} {\enquote {\bibinfo {title} {Detecting
  symmetries with neural networks},}\ }\href {\doibase
  10.1088/2632-2153/abbd2d} {\bibfield  {journal} {\bibinfo  {journal} {Machine
  Learning: Science and Technology}\ }\textbf {\bibinfo {volume} {2}},\
  \bibinfo {pages} {015010} (\bibinfo {year} {2021})}\BibitemShut {NoStop}%
\bibitem [{\citenamefont {Desai}, \citenamefont {Nachman},\ and\ \citenamefont
  {Thaler}(2022)}]{Desai2022}%
  \BibitemOpen
  \bibfield  {author} {\bibinfo {author} {\bibfnamefont {K.}~\bibnamefont
  {Desai}}, \bibinfo {author} {\bibfnamefont {B.}~\bibnamefont {Nachman}}, \
  and\ \bibinfo {author} {\bibfnamefont {J.}~\bibnamefont {Thaler}},\
  }\bibfield  {title} {\enquote {\bibinfo {title} {Symmetry discovery with deep
  learning},}\ }\href {\doibase 10.1103/PhysRevD.105.096031} {\bibfield
  {journal} {\bibinfo  {journal} {Physical Review D}\ }\textbf {\bibinfo
  {volume} {105}} (\bibinfo {year} {2022}),\
  10.1103/PhysRevD.105.096031}\BibitemShut {NoStop}%
\bibitem [{\citenamefont {Du}\ \emph {et~al.}(2018)\citenamefont {Du},
  \citenamefont {Qu}, \citenamefont {He},\ and\ \citenamefont {Guo}}]{Du2018}%
  \BibitemOpen
  \bibfield  {author} {\bibinfo {author} {\bibfnamefont {X.}~\bibnamefont
  {Du}}, \bibinfo {author} {\bibfnamefont {X.}~\bibnamefont {Qu}}, \bibinfo
  {author} {\bibfnamefont {Y.}~\bibnamefont {He}}, \ and\ \bibinfo {author}
  {\bibfnamefont {D.}~\bibnamefont {Guo}},\ }\bibfield  {title} {\enquote
  {\bibinfo {title} {Single image super-resolution based on multi-scale
  competitive convolutional neural network},}\ }\href {\doibase
  10.3390/s18030789} {\bibfield  {journal} {\bibinfo  {journal} {Sensors}\
  }\textbf {\bibinfo {volume} {18}} (\bibinfo {year} {2018}),\
  10.3390/s18030789}\BibitemShut {NoStop}%
\bibitem [{\citenamefont {Zhang}\ \emph {et~al.}(2018)\citenamefont {Zhang},
  \citenamefont {Tian}, \citenamefont {Kong}, \citenamefont {Zhong},\ and\
  \citenamefont {Fu}}]{Zhang2018RDN}%
  \BibitemOpen
  \bibfield  {author} {\bibinfo {author} {\bibfnamefont {Y.}~\bibnamefont
  {Zhang}}, \bibinfo {author} {\bibfnamefont {Y.}~\bibnamefont {Tian}},
  \bibinfo {author} {\bibfnamefont {Y.}~\bibnamefont {Kong}}, \bibinfo {author}
  {\bibfnamefont {B.}~\bibnamefont {Zhong}}, \ and\ \bibinfo {author}
  {\bibfnamefont {Y.}~\bibnamefont {Fu}},\ }\bibfield  {title} {\enquote
  {\bibinfo {title} {Residual dense network for image super-resolution},}\ }in\
  \href@noop {} {\emph {\bibinfo {booktitle} {Proceedings of the IEEE
  Conference on Computer Vision and Pattern Recognition (CVPR)}}}\ (\bibinfo
  {year} {2018})\BibitemShut {NoStop}%
\bibitem [{\citenamefont {Gilmore}(2008)}]{Gilmore2008}%
  \BibitemOpen
  \bibfield  {author} {\bibinfo {author} {\bibfnamefont {R.}~\bibnamefont
  {Gilmore}},\ }\href@noop {} {\emph {\bibinfo {title} {Lie Groups, Physics,
  and Geometry: An Introduction for Physicists, Engineers and Chemists}}}\
  (\bibinfo  {publisher} {Cambridge University Press},\ \bibinfo {year}
  {2008})\BibitemShut {NoStop}%
\bibitem [{\citenamefont {Zee}(2016)}]{Zee2016}%
  \BibitemOpen
  \bibfield  {author} {\bibinfo {author} {\bibfnamefont {A.}~\bibnamefont
  {Zee}},\ }\href@noop {} {\emph {\bibinfo {title} {Group Theory in a Nutshell
  for Physicists}}}\ (\bibinfo  {publisher} {Princeton University Press},\
  \bibinfo {year} {2016})\BibitemShut {NoStop}%
\bibitem [{\citenamefont {Keys}(1981)}]{Key1981}%
  \BibitemOpen
  \bibfield  {author} {\bibinfo {author} {\bibfnamefont {R.}~\bibnamefont
  {Keys}},\ }\bibfield  {title} {\enquote {\bibinfo {title} {Cubic convolution
  interpolation for digital image processing},}\ }\href {\doibase
  10.1109/TASSP.1981.1163711} {\bibfield  {journal} {\bibinfo  {journal} {IEEE
  Transactions on Acoustics, Speech, and Signal Processing}\ }\textbf {\bibinfo
  {volume} {29}},\ \bibinfo {pages} {1153--1160} (\bibinfo {year}
  {1981})}\BibitemShut {NoStop}%
\bibitem [{\citenamefont {Vallis}(2017)}]{Vallis2017}%
  \BibitemOpen
  \bibfield  {author} {\bibinfo {author} {\bibfnamefont {G.~K.}\ \bibnamefont
  {Vallis}},\ }\href@noop {} {\emph {\bibinfo {title} {Atmospheric and Oceanic
  Fluid Dynamics: Fundamentals and Large-Scale Circulation}}},\ \bibinfo
  {edition} {second edition}\ ed.\ (\bibinfo  {publisher} {Cambridge University
  Press},\ \bibinfo {year} {2017})\BibitemShut {NoStop}%
\bibitem [{\citenamefont {Ishioka}(2015)}]{ispack2015}%
  \BibitemOpen
  \bibfield  {author} {\bibinfo {author} {\bibfnamefont {K.}~\bibnamefont
  {Ishioka}},\ }\href {https://www.gfd-dennou.org/arch/ispack/index.htm.en}
  {\enquote {\bibinfo {title} {Ispack 1.0.4},}\ }\bibinfo {howpublished}
  {\url{https://www.gfd-dennou.org/arch/ispack/index.htm.en}} (\bibinfo {year}
  {2015})\BibitemShut {NoStop}%
\bibitem [{\citenamefont {Taira}, \citenamefont {Nair},\ and\ \citenamefont
  {Brunton}(2016)}]{Taira2016}%
  \BibitemOpen
  \bibfield  {author} {\bibinfo {author} {\bibfnamefont {K.}~\bibnamefont
  {Taira}}, \bibinfo {author} {\bibfnamefont {A.~G.}\ \bibnamefont {Nair}}, \
  and\ \bibinfo {author} {\bibfnamefont {S.~L.}\ \bibnamefont {Brunton}},\
  }\bibfield  {title} {\enquote {\bibinfo {title} {Network structure of
  two-dimensional decaying isotropic turbulence},}\ }\href {\doibase
  10.1017/jfm.2016.235} {\bibfield  {journal} {\bibinfo  {journal} {Journal of
  Fluid Mechanics}\ }\textbf {\bibinfo {volume} {795}},\ \bibinfo {pages} {R2}
  (\bibinfo {year} {2016})}\BibitemShut {NoStop}%
\bibitem [{\citenamefont {Paszke}\ \emph {et~al.}(2019)\citenamefont {Paszke},
  \citenamefont {Gross}, \citenamefont {Massa}, \citenamefont {Lerer},
  \citenamefont {Bradbury}, \citenamefont {Chanan}, \citenamefont {Killeen},
  \citenamefont {Lin}, \citenamefont {Gimelshein}, \citenamefont {Antiga},
  \citenamefont {Desmaison}, \citenamefont {Kopf}, \citenamefont {Yang},
  \citenamefont {DeVito}, \citenamefont {Raison}, \citenamefont {Tejani},
  \citenamefont {Chilamkurthy}, \citenamefont {Steiner}, \citenamefont {Fang},
  \citenamefont {Bai},\ and\ \citenamefont {Chintala}}]{pytorch2019}%
  \BibitemOpen
  \bibfield  {author} {\bibinfo {author} {\bibfnamefont {A.}~\bibnamefont
  {Paszke}}, \bibinfo {author} {\bibfnamefont {S.}~\bibnamefont {Gross}},
  \bibinfo {author} {\bibfnamefont {F.}~\bibnamefont {Massa}}, \bibinfo
  {author} {\bibfnamefont {A.}~\bibnamefont {Lerer}}, \bibinfo {author}
  {\bibfnamefont {J.}~\bibnamefont {Bradbury}}, \bibinfo {author}
  {\bibfnamefont {G.}~\bibnamefont {Chanan}}, \bibinfo {author} {\bibfnamefont
  {T.}~\bibnamefont {Killeen}}, \bibinfo {author} {\bibfnamefont
  {Z.}~\bibnamefont {Lin}}, \bibinfo {author} {\bibfnamefont {N.}~\bibnamefont
  {Gimelshein}}, \bibinfo {author} {\bibfnamefont {L.}~\bibnamefont {Antiga}},
  \bibinfo {author} {\bibfnamefont {A.}~\bibnamefont {Desmaison}}, \bibinfo
  {author} {\bibfnamefont {A.}~\bibnamefont {Kopf}}, \bibinfo {author}
  {\bibfnamefont {E.}~\bibnamefont {Yang}}, \bibinfo {author} {\bibfnamefont
  {Z.}~\bibnamefont {DeVito}}, \bibinfo {author} {\bibfnamefont
  {M.}~\bibnamefont {Raison}}, \bibinfo {author} {\bibfnamefont
  {A.}~\bibnamefont {Tejani}}, \bibinfo {author} {\bibfnamefont
  {S.}~\bibnamefont {Chilamkurthy}}, \bibinfo {author} {\bibfnamefont
  {B.}~\bibnamefont {Steiner}}, \bibinfo {author} {\bibfnamefont
  {L.}~\bibnamefont {Fang}}, \bibinfo {author} {\bibfnamefont {J.}~\bibnamefont
  {Bai}}, \ and\ \bibinfo {author} {\bibfnamefont {S.}~\bibnamefont
  {Chintala}},\ }\bibfield  {title} {\enquote {\bibinfo {title} {Pytorch: An
  imperative style, high-performance deep learning library},}\ }in\ \href
  {http://papers.neurips.cc/paper/9015-pytorch-an-imperative-style-high-performance-deep-learning-library.pdf}
  {\emph {\bibinfo {booktitle} {Advances in Neural Information Processing
  Systems 32}}},\ \bibinfo {editor} {edited by\ \bibinfo {editor}
  {\bibfnamefont {H.}~\bibnamefont {Wallach}}, \bibinfo {editor} {\bibfnamefont
  {H.}~\bibnamefont {Larochelle}}, \bibinfo {editor} {\bibfnamefont
  {A.}~\bibnamefont {Beygelzimer}}, \bibinfo {editor} {\bibfnamefont
  {F.}~\bibnamefont {d\textquotesingle Alch\'{e}-Buc}}, \bibinfo {editor}
  {\bibfnamefont {E.}~\bibnamefont {Fox}}, \ and\ \bibinfo {editor}
  {\bibfnamefont {R.}~\bibnamefont {Garnett}}}\ (\bibinfo  {publisher} {Curran
  Associates, Inc.},\ \bibinfo {year} {2019})\ pp.\ \bibinfo {pages}
  {8024--8035}\BibitemShut {NoStop}%
\bibitem [{\citenamefont {Kingma}\ and\ \citenamefont {Ba}(2015)}]{Kingma2015}%
  \BibitemOpen
  \bibfield  {author} {\bibinfo {author} {\bibfnamefont {D.~P.}\ \bibnamefont
  {Kingma}}\ and\ \bibinfo {author} {\bibfnamefont {J.}~\bibnamefont {Ba}},\
  }\bibfield  {title} {\enquote {\bibinfo {title} {Adam: A method for
  stochastic optimization},}\ \ }(\bibinfo {year} {2015})\BibitemShut {NoStop}%
\bibitem [{\citenamefont {Maddox}, \citenamefont {Benton},\ and\ \citenamefont
  {Wilson}(2020)}]{Maddox2020}%
  \BibitemOpen
  \bibfield  {author} {\bibinfo {author} {\bibfnamefont {W.~J.}\ \bibnamefont
  {Maddox}}, \bibinfo {author} {\bibfnamefont {G.~W.}\ \bibnamefont {Benton}},
  \ and\ \bibinfo {author} {\bibfnamefont {A.~G.}\ \bibnamefont {Wilson}},\
  }\bibfield  {title} {\enquote {\bibinfo {title} {Rethinking parameter
  counting in deep models: Effective dimensionality revisited.}}\ }\href
  {https://arxiv.org/abs/2003.02139} {\bibfield  {journal} {\bibinfo  {journal}
  {CoRR}\ }\textbf {\bibinfo {volume} {abs/2003.02139}} (\bibinfo {year}
  {2020})}\BibitemShut {NoStop}%
\bibitem [{\citenamefont {Neyshabur}(2017)}]{Neyshabur2017}%
  \BibitemOpen
  \bibfield  {author} {\bibinfo {author} {\bibfnamefont {B.}~\bibnamefont
  {Neyshabur}},\ }\href {\doibase 10.48550/ARXIV.1709.01953} {\enquote
  {\bibinfo {title} {Implicit regularization in deep learning},}\ } (\bibinfo
  {year} {2017})\BibitemShut {NoStop}%
\bibitem [{\citenamefont {Cohen}\ \emph {et~al.}(2019)\citenamefont {Cohen},
  \citenamefont {Weiler}, \citenamefont {Kicanaoglu},\ and\ \citenamefont
  {Welling}}]{Cohen2019IcosahedralGauge}%
  \BibitemOpen
  \bibfield  {author} {\bibinfo {author} {\bibfnamefont {T.}~\bibnamefont
  {Cohen}}, \bibinfo {author} {\bibfnamefont {M.}~\bibnamefont {Weiler}},
  \bibinfo {author} {\bibfnamefont {B.}~\bibnamefont {Kicanaoglu}}, \ and\
  \bibinfo {author} {\bibfnamefont {M.}~\bibnamefont {Welling}},\ }\bibfield
  {title} {\enquote {\bibinfo {title} {Gauge equivariant convolutional networks
  and the icosahedral {CNN}},}\ }in\ \href
  {https://proceedings.mlr.press/v97/cohen19d.html} {\emph {\bibinfo
  {booktitle} {Proceedings of the 36th International Conference on Machine
  Learning}}},\ \bibinfo {series} {Proceedings of Machine Learning Research},
  Vol.~\bibinfo {volume} {97},\ \bibinfo {editor} {edited by\ \bibinfo {editor}
  {\bibfnamefont {K.}~\bibnamefont {Chaudhuri}}\ and\ \bibinfo {editor}
  {\bibfnamefont {R.}~\bibnamefont {Salakhutdinov}}}\ (\bibinfo  {publisher}
  {PMLR},\ \bibinfo {year} {2019})\ pp.\ \bibinfo {pages}
  {1321--1330}\BibitemShut {NoStop}%
\bibitem [{\citenamefont {Haan}\ \emph {et~al.}(2021)\citenamefont {Haan},
  \citenamefont {Weiler}, \citenamefont {Cohen},\ and\ \citenamefont
  {Welling}}]{Haan2021gauge}%
  \BibitemOpen
  \bibfield  {author} {\bibinfo {author} {\bibfnamefont {P.~D.}\ \bibnamefont
  {Haan}}, \bibinfo {author} {\bibfnamefont {M.}~\bibnamefont {Weiler}},
  \bibinfo {author} {\bibfnamefont {T.}~\bibnamefont {Cohen}}, \ and\ \bibinfo
  {author} {\bibfnamefont {M.}~\bibnamefont {Welling}},\ }\bibfield  {title}
  {\enquote {\bibinfo {title} {Gauge equivariant mesh {\{}cnn{\}}s: Anisotropic
  convolutions on geometric graphs},}\ }in\ \href
  {https://openreview.net/forum?id=Jnspzp-oIZE} {\emph {\bibinfo {booktitle}
  {International Conference on Learning Representations}}}\ (\bibinfo {year}
  {2021})\BibitemShut {NoStop}%
\end{thebibliography}%

\end{document}